\documentclass[fleqn,10pt]{wlscirep}

\usepackage{graphicx}

\usepackage{algorithm}
\usepackage{algpseudocode}

\include{Qcircuit}

\title{Synthesis of Arbitrary Quantum Circuits to Topological Assembly}

\author[1]{Alexandru Paler}
\author[2]{Simon J. Devitt}
\author[3]{Austin G. Fowler}
\affil[1]{Universitatea Transilvania, Facultatea de Matematica si Informatica, Brasov 500091, Romania}
\affil[2]{Center for Emergent Matter Sciences, Riken, Saitama 351-0198, Japan}
\affil[3]{Google Inc., Santa Barbara, California 93117, USA}

\begin{abstract}
Given a quantum algorithm, it is highly nontrivial to devise an efficient sequence of physical gates implementing the algorithm on real hardware and incorporating topological quantum error correction. In this paper, we present a first step towards this goal, focusing on generating correct and simple arrangements of topological structures that correspond to a given quantum circuit and largely neglecting their efficiency. We detail the many challenges that will need to be tackled in the pursuit of efficiency. The software source code can be consulted at \url{https://github.com/alexandrupaler/tqec}.
\end{abstract}

\begin{document}

\flushbottom
\maketitle

\section*{Introduction}
\label{sec:back}

Quantum computers are envisioned to become reality, and a vast amount of research has been devoted to the theoretical foundations of this emerging computing paradigm as well as to investigating and developing initial prototypical hardware to support it. Any practical quantum computer should be able to reliably solve interesting computational problems, but reliability is one of the key engineering challenges.

It has been shown theoretically that it is possible to execute scalable (arbitrarily long) quantum computations on hardware with a failure probability below a certain threshold by using a sufficiently capable error-correcting code. A class of the most promising error-correcting codes, based on \emph{topological cluster states}, enables scalable computations for failure probabilities below 1\%, which is considered an achievable threshold with the current technological state of the art. The advantage of knowing how to reliably implement quantum computations is counterbalanced by the high resources introduced by the error-correction. The challenge to reliably implement quantum computations is partially a problem of optimising quantum error-corrected quantum circuits. However, these have to be firstly obtained (synthesised) from non-error-corrected (non-fault-tolerant) versions.

This work introduces the \emph{synthesis} of \emph{topologically quantum error-corrected (TQEC) circuits}. TQEC circuits are error-corrected quantum circuits operating on information encoded into \emph{topological cluster states}. Synthesis of such circuits requires decomposing (translating) an abstract, high-level circuit description of a computation into a form compatible with TQEC that can be implemented and optimised on real world hardware.

TQEC circuits have a intrinsic visual representation, and their functionality can be described entirely using geometric elements. TQEC synthesis outputs an \emph{assembly} of geometrically abstracted elements which are understood as being protected against errors (faults) by codes using topological considerations. Without affecting the generality of the proposed synthesis method, this work will not present the error-correction mechanisms \cite{fowler2012surface}, and instead focus on establishing a terminology to easily connect the existing theoretical work to future engineering problems.



A very brief introduction to TQEC terminology is required for presenting the proposed synthesis, and this section tries to offer all the necessary details. The introduction starts by showing that TQEC circuits operate at a \emph{logical layer}, which is an abstraction of a physical one. The logical abstraction makes use of topological cluster states, which have the structure of a regular \emph{lattice} of \emph{physical qubits} entangled into a large \emph{graph state}. Therefore, topological cluster states are used to introduce the elements that define the translation between the quantum circuit formalism and the TQEC circuit description. The specification of TQEC circuits is based on the \emph{geometry} of the entities formed in the logical layer.

\subsection*{Defects, Qubits and Gates}
\label{sec:defqub}

\begin{figure}
\centering
\subfloat[]{
\includegraphics[width=.35\columnwidth]{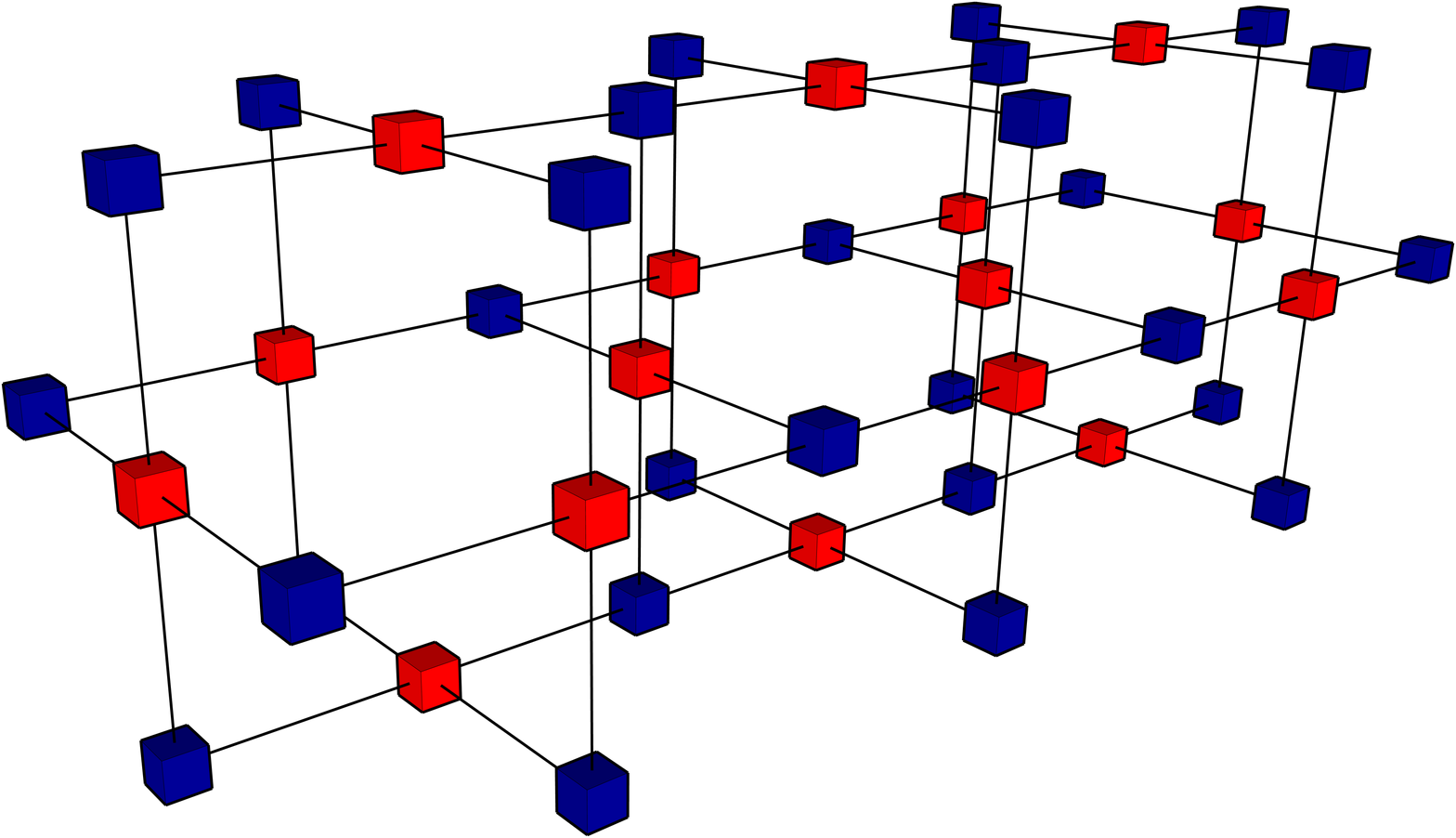}
}\\
\subfloat[]{
\includegraphics[width=.2\columnwidth]{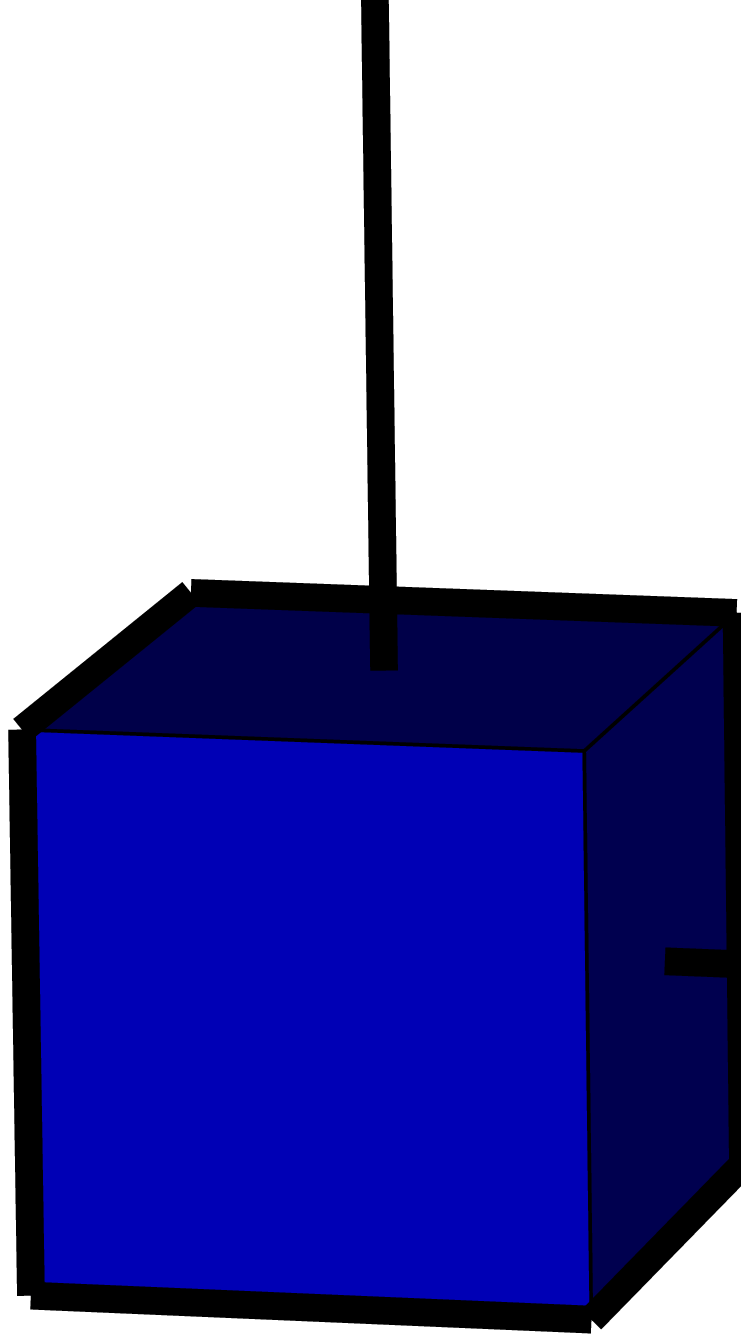}
}\hfil
\subfloat[]{
\includegraphics[width=.2\columnwidth]{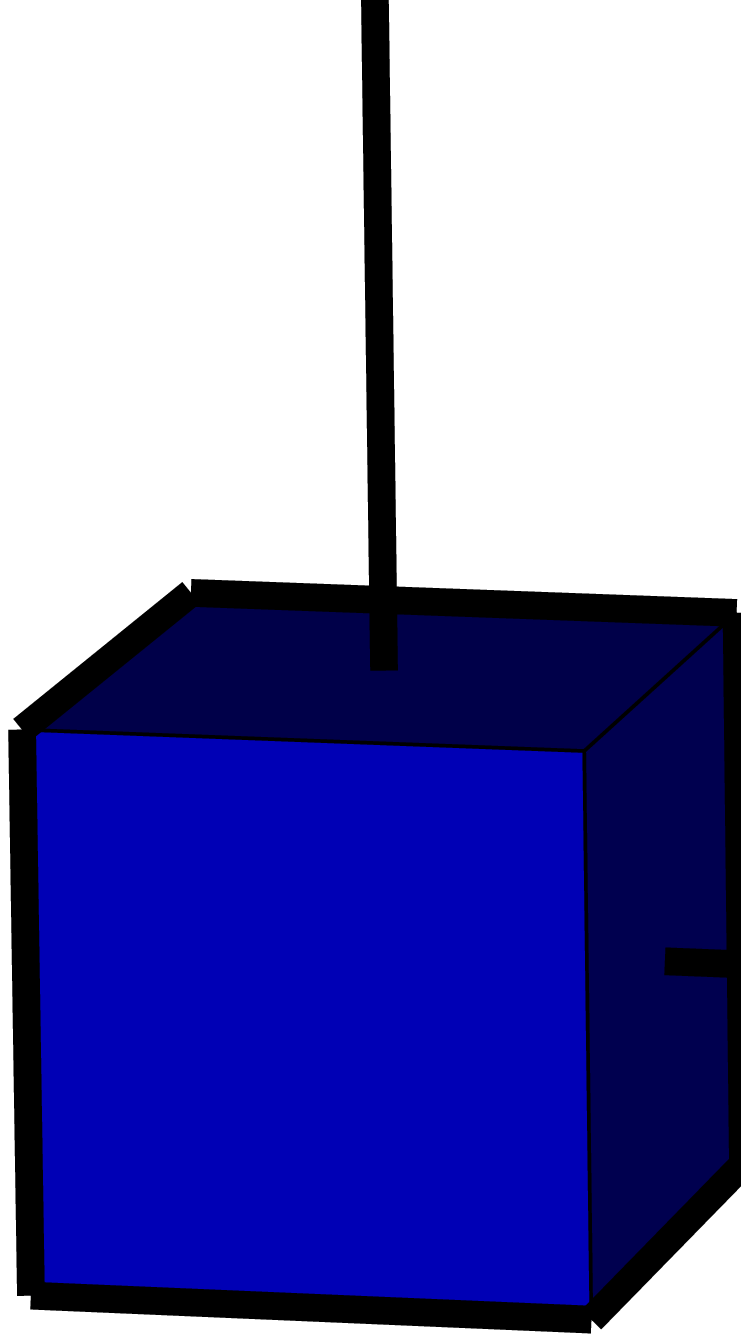}
}\\\
\subfloat[]{
\includegraphics[width=.35\columnwidth]{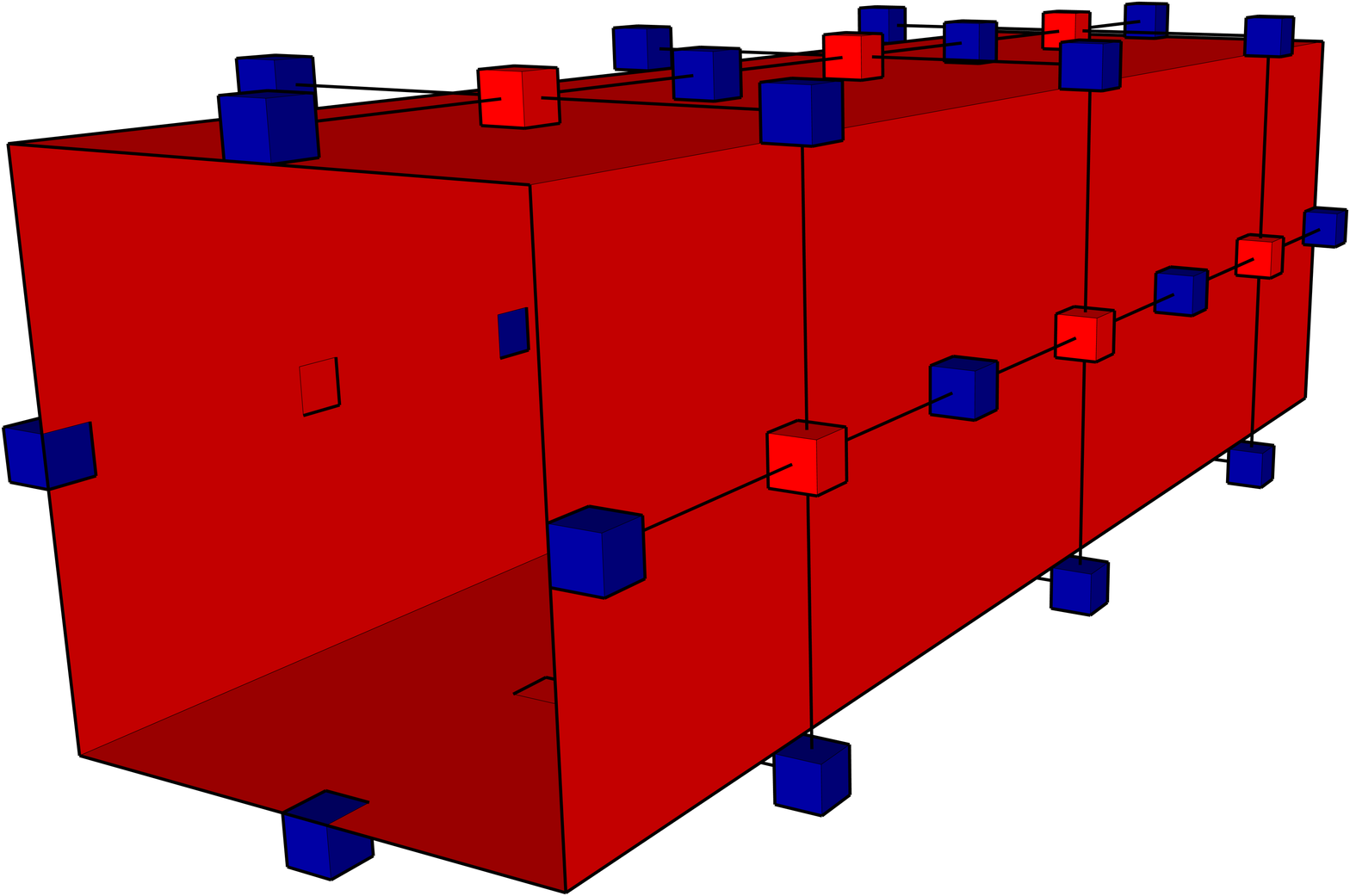}
}\\
\subfloat[]{
\includegraphics[width=.35\columnwidth]{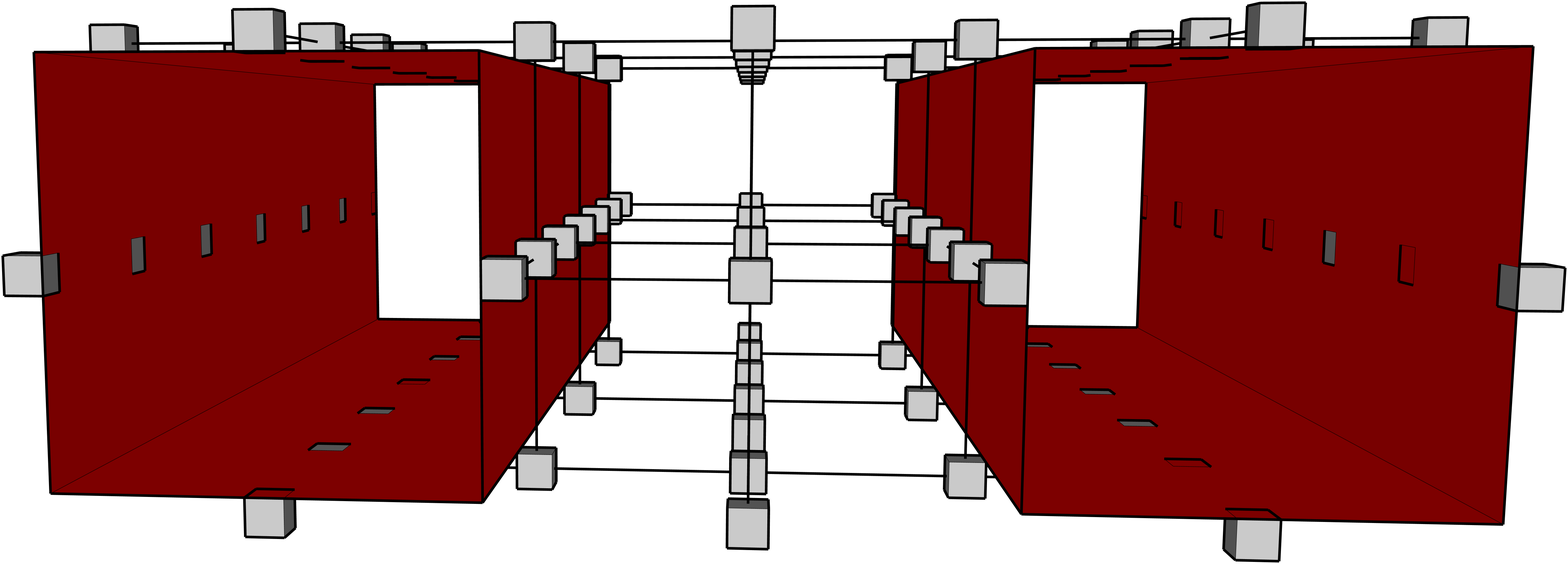}
}\\
\subfloat[]{
\includegraphics[width=.2\columnwidth]{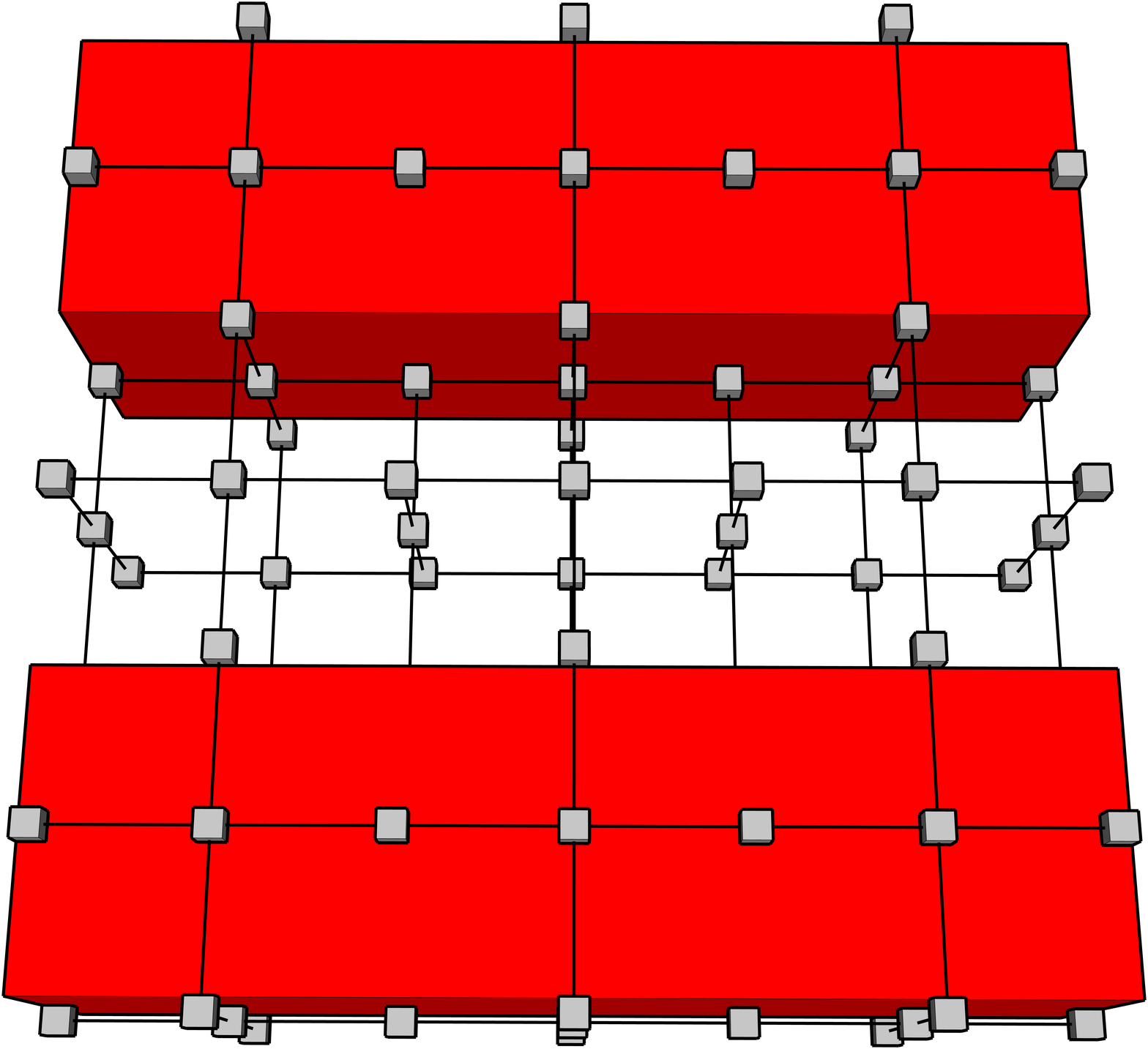}
}\hfil
\subfloat[]{
\includegraphics[width=.2\columnwidth]{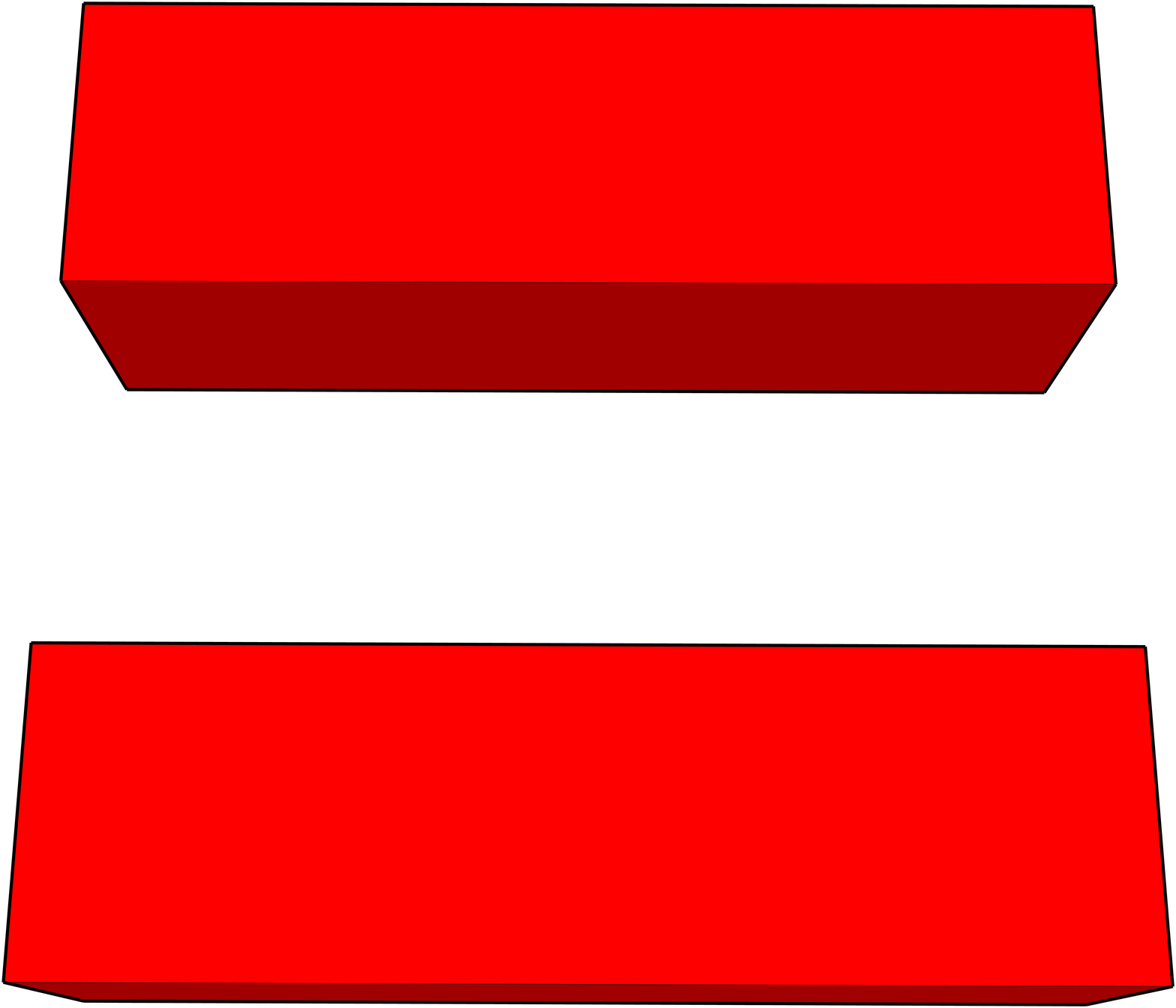}
}
\caption{Construction of defects: a) A lattice of three primal unit cells that will be used to construct a primal defect. For visualisation purposes, no dual unit cell is illustrated in the lattice. The red physical qubits represent primal qubits, and blue ones dual qubits. Removing a region of primal qubits creates a primal defect. Defects have a minimal diameter of one unit cell. b) Rotated view of the previous figure. c) Example of a defect constructed by removing four primal qubits from the graph. d) The boundary of the defect can be abstracted by a cuboid (red). e) Two defect example constructed in a graph consisting of $3 \times 3 \times 1$ unit cells. Physical qubit types are not colour coded. f) Rotated view of the previous figure. g) The cuboids abstracting the defects can be themselves abstracted by segments whose end points are specified in terms of unit cell coordinates. The graph state used to encode logical information is thus not necessary for the specification of a TQEC circuit.}
\label{fig:construction}
\end{figure}

\begin{figure}
\centering
\subfloat[]{
  \label{fig:cell1}
  \includegraphics[width=0.2\columnwidth]{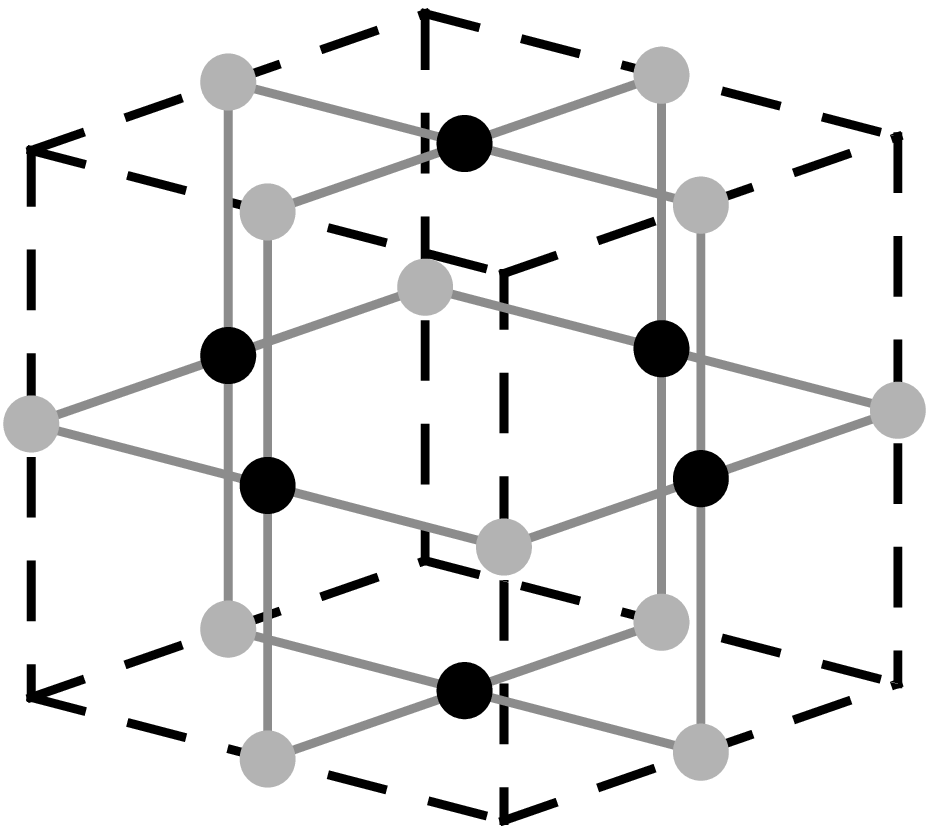}
}
\hfil
\subfloat[]{
  \label{fig:cell2}
  \includegraphics[width=0.2\columnwidth]{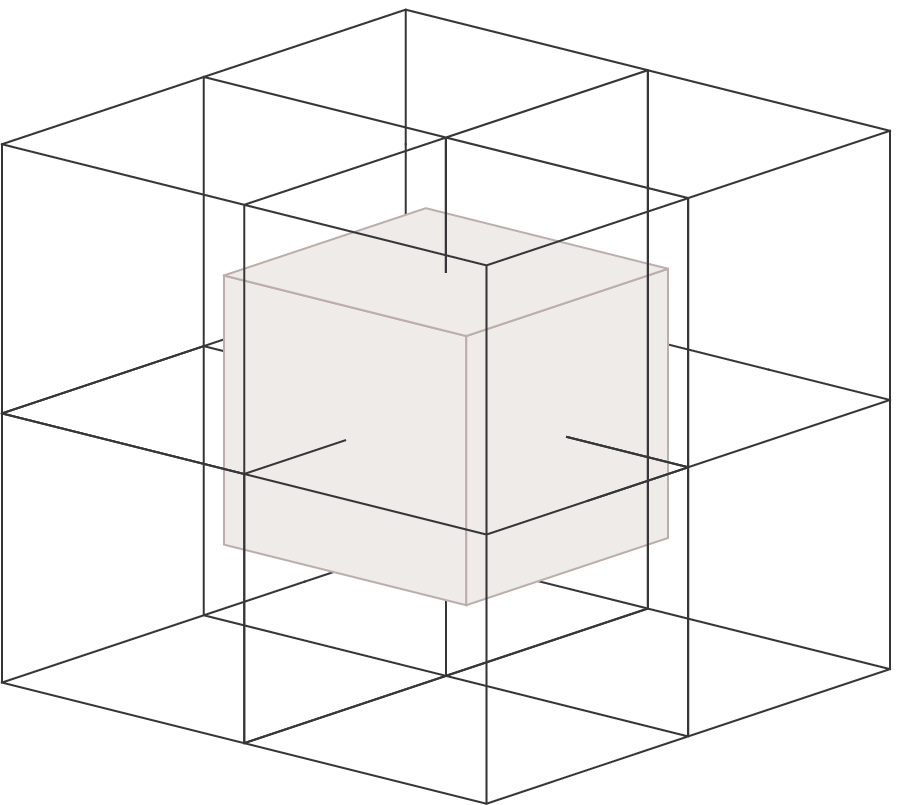}
}
\caption{Unit cells: a) The graph state required for TQEC is constructed using unit cells consisting of 18 physical qubits entangled according to the pattern presented in the figure. The face qubits are marked black, and the edge qubits are gray. Considering that the there are 27 possible positions in the three dimensional space used for illustrating the unit cell, and that the lowest coordinate is (0,0,0), then the cell centre has coordinate (1,1,1). Black qubits have an even coordinate, and grey qubits two even coordinates. b) After stacking eight primal unit cells as in the figure, a unit cell will result in the middle of the $2 \times 2 \times 2$ block. Such cells are called dual, and their centres have even coordinates.}
\label{fig:cell}
\end{figure}

Quantum information encoding into topological cluster states requires firstly the construction of a highly regular graph state, and secondly the removal of specific vertices from a lattice abstracting the state (Fig.~\ref{fig:construction}). Each vertex represents a physical qubit, and edges stand for the entanglement relations between qubits. The removal of vertices results in the definition of \emph{defects}. The lattice consists of two self-similar sublattices (\emph{primal} and \emph{dual}), where both are the result of stacking unit cells (Fig.~\ref{fig:cell1}) along three axes. Their duality is perceived after stacking eight primal unit cells: a dual unit cell results at the geometric centre of a $2 \times 2 \times 2$ (primal) unit cell block (Fig.~\ref{fig:cell2}). The definition of a primal or dual defect depends on which sublattice vertices are removed from. A defect cannot have different types along its path, meaning that if one starts to define a primal defect, primal face qubits from the primal lattice will be removed until the defect ends. Due to practical error-correction reasons, a \emph{logical} qubit is defined by a pair of same-type defects and, as a consequence, there are primal and dual (logical) qubits. Each defect consists of multiple linear \emph{segments}.

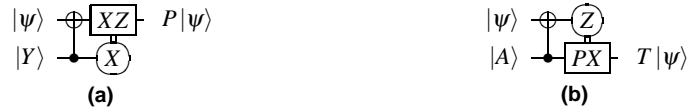
\begin{figure}[ht!]
\small
\centering
\subfloat[]{
	\label{circ:ftv}
	\Qcircuit @C=0.3em @R=.3em {
		\lstick{\ket{\psi}} & \targ & \gate{XZ}  &\qw & \rstick{P\ket{\psi}}\\
		\lstick{\ket{Y} }&\ctrl{-1}&\measure{X} \cwx
	}
}
\hfil
\subfloat[]{
	\label{circ:ftt}
	\Qcircuit @C=0.3em @R=.3em {
		\lstick{\ket{\psi}} & \targ & \measure{Z}\\
		\lstick{\ket{A}}&\ctrl{-1}& \gate{PX} \cwx&\qw &\rstick{T\ket{\psi}}
	}
}
\caption{Teleported rotation gates using the states $\ket{A}$ and $\ket{Y}$. The correction for the $P$-gate is a the Pauli $Y$-gate and can be tracked \cite{paler2014software}, while the correction for the $T$-gate requires a subsequent $P$-gate that must be applied immediately. }
\label{circ:ftcircs}
\end{figure}

Logical qubit types are relevant because the logical CNOT gate is implemented by \emph{braiding} a dual defect around a primal one: the dual qubit always acts as control and the primal qubit as target. However, a CNOT is not sufficient for \emph{computational universality} and more types of gates are required. For practical purposes, the gates supported by a quantum computing architecture form a discrete universal gate set, and TQEC circuits use $\{CNOT, V, P, T\}$. Single qubit \emph{rotation gates} are implemented through teleportation-based mechanisms. Before teleportation, specific ancilla logical states ($\ket{A}$ or $\ket{Y}$) must be prepared. The implementation of the teleported T gate requires an $\ket{A}$ state initialised ancilla, and the teleported P and V gates use an ancilla initialised in $\ket{Y}$ (Fig.~\ref{circ:ftcircs}). The first step when preparing logical ancillae is \emph{state injection}: 1) initialising graph vertices (physical qubits) in those states, and then 2) constructing defects starting from the specially initialised vertices.

\begin{align*}
\ket{A} =\frac{1}{\sqrt{2}}(\ket{0}+e^{i\frac{\pi}{4}}\ket{1}) &&
\ket{Y} =\frac{1}{\sqrt{2}}(\ket{0}+i\ket{1})
\end{align*}

\subsection*{Boxes}

Injected states may have low \emph{fidelity} \cite{fowler2012surface}. State fidelity can be increased by replacing direct injections with the output of a \emph{distillation circuit}. Such circuits take multiple low fidelity instances of injected states and output a single higher fidelity state. 

Distillation procedures are probabilistic, meaning that it could happen that their output does not have the required fidelity. Thus, the \emph{failure probability} of a TQEC rotation gate (T, P or V) is directly related to the failure probability of the box used to output the required ancilla state. In order to lower the failure probability, a larger number of boxes needs to accompany the circuit: if one of the distillations fails, a sufficient number of \emph{spare} boxes is necessary for achieving a targeted lower TQEC gate failure probability.

Distillation circuits are frequently used due to the \emph{ICM} form of TQEC circuits 
 Therefore, it is sufficient to use \emph{boxes} as place holders for distillations in the TQEC circuit's structure. There are two types of boxes, one each for $\ket{A}$ and $\ket{Y}$ distillations.

\subsection*{Pins and Connections}
\label{sec:pins}

\begin{figure}
\centering
\includegraphics[width=0.4\columnwidth]{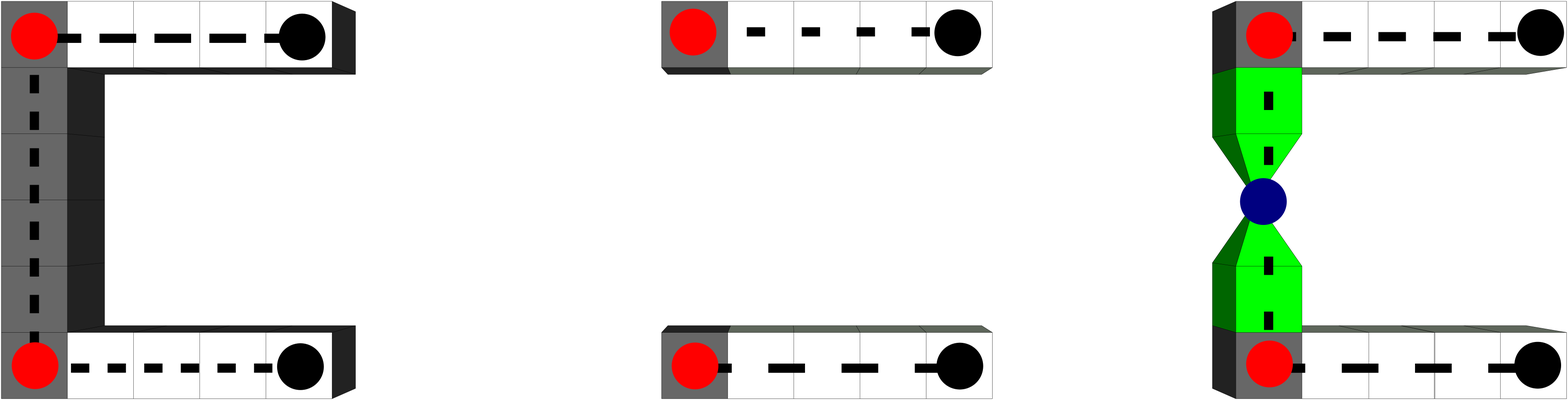}
\caption{The geometries for initialising and measuring logical qubits. The grey segments represent the geometry to construct and join with the defects representing logical qubit segments (white). Each segment in this figure is defined using four unit cells. Dotted lines abstract defect segments, red dots are marking the pins, the black dots represent segment end points. The blue dot marks the graph vertex used for state injection. For primal logical qubits: left) Z-basis initialisation; centre) X-basis initialisation; right) state injection, where from a graph vertex two pyramid-shaped defects are constructed. For dual logical qubits: left) X-basis initialisation; centre) Z-basis initialisation. The measurement of a logical qubit uses the same initialisation geometries. Assuming that in the figures a time axis is represented horizontally from the left to the right, the measurement geometries will be mirrored against the vertical axis.}
\label{fig:minit}
\end{figure}

An arbitrary TQEC circuit has inputs and outputs (state injections are a particular case). Some of these are \emph{configurable} (their defect structure is configurable), while others are not (e.g. teleportation ancillae or state injections). Such inputs/outputs are used to parameterise the computation implemented as a circuit, e.g. input numbers to an adder circuit. The difference between the input/output structures (Fig.~\ref{fig:minit}) is a single defect segment which can be split into two non-disjoint segments sharing a common lattice vertex. Without loss of generality, the shared vertex is the geometric centre of the difference segment (cf. left and centre figures in Fig.~\ref{fig:minit}). While for injected states the defects end right before the injection vertex, for configurable inputs/outputs the defects include the vertex (cf. right and left figures in Fig.~\ref{fig:minit}). The only segments, where one of the end points refers to a vertex, are those in the direct neighbourhood of a state injection or configurable input/output. The other end point of such segments is abstracted using a \emph{pin}. As a result, each injection and configurable input/output determine two pins (\emph{pin pair}).


In general, pins are used for connecting circuits: the output pins of a circuit are \emph{connected} to the input pins of another circuit. For example, the geometry of an adder can be chained multiple times in order to construct a multiplier: the outputs of the first adder will be connected to the inputs of the second one. Another example is when pins are used for connecting distillation boxes to the circuit: box output pins of distillation are connected to the injection pins of the circuit. In particular, TQEC circuit connections are implemented by defects.

\subsection*{Geometric Description}
\label{sec:geomdesc}

TQEC synthesis generates the \emph{geometric description} of a TQEC circuit 
 The encoding (constructing defects) and the manipulation of information (braiding defects) is described independently from the initial lattice by concentrating entirely on the geometry of the defect structures. Geometries are a sufficient description also when considering that logical qubit initialisation and measurement basis are a function of defect configuration and logical qubit type (Fig.~\ref{fig:minit}). Therefore, the positions of defect segments, inputs/outputs, injections and boxes are the elements of a geometric description.

A TQEC geometric description uses three dimensional coordinates because the lattice can be represented in three dimensions. Assuming all the coordinates are positive integers, and observing the lattice structure, it can be concluded that primal unit cell centres have odd coordinates, and dual unit cell centres have even coordinates (Fig.~\ref{fig:cell1}). Vertices on the faces of a primal unit cell (primal face qubits) have an odd and two even coordinates. Consequently, vertices on the faces of a dual unit cell (dual face qubits) have a single even and two odd coordinates.

Defect segments have their end point coordinates specified using unit cell centre coordinates. State injections and configurable inputs/outputs are specified using vertex coordinates. The coordinates of the lattice vertex where a state is injected needs to specified too, because from that point starts the encoding of the logical qubit (the associated physical qubit is measured in the $X$ basis). Pins (e.g. marked by red dots in Fig.~\ref{fig:minit}) are specified using unit cell centre coordinates. Boxes abstract the bounding box of a distillation circuit geometric description. For each of the $\ket{A}$ and $\ket{Y}$ distillation procedures a particular box type with parametrised dimensions is used. The position of each box in the synthesised TQEC circuit is specified using unit cell centre coordinates.

\section*{Results}
\label{sec:synthesis}

The present method allows synthesising arbitrary quantum circuits, supports the placement and the diagrammatic representation of distillation subcircuits, and includes an automated method for connecting two distinct pins from the geometry. The TQEC synthesis presented in this work is more versatile than that of \cite{paler2012synthesis}, because it did not consider configurable inputs/outputs, and included a topological implementation of the logical CNOT which is different from braiding. The current synthesis was implemented and its source code can be \href{https://github.com/alexandrupaler/tqec}{consulted}. The method consists of multiple steps which are described in the following.

\subsection*{Gate Decomposition}
\label{sec:decomp}

Arbitrary quantum circuits can be described by a list of gates parametrised by the qubits they operate on. Each quantum computing architecture uses a specific set of gates to implement universal computations. TQEC supports only the definition of CNOTs and the implementation of T, P and V gates, but it does not directly support the application of a Toffoli gate (widely used in classical reversible circuits \cite{wille2009equivalence}). In order to implement this gate, it has to be first decomposed into a sequence of TQEC supported gates (Fig.~\ref{circ:toffoli}). The same applies for the Hadamard (H) gate, which is decomposed as the P,V,P sequence. 

\subsection*{ICM Conversion}
\label{sec:icm}

TQEC supports the CNOT gate and the rotation gates are implemented using a teleportation circuit \cite{nielsen2010quantum} (Fig.~\ref{circ:ftcircs}) with ancillae initialised to $\ket{A}$ or $\ket{Y}$. The ancillae states are included through state injection into the TQEC circuit.

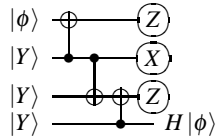
\begin{figure}
\small
\centerline{
\Qcircuit @C=.3em @R=.3em {
		\lstick{\ket{\phi}} & \targ & \qw& \qw &\measure{Z}\\
		\lstick{\ket{Y}} & \ctrl{-1} & \ctrl{1} & \qw & \measure{X}\\
		\lstick{\ket{Y}} & \qw & \targ & \targ& \measure{Z}\\
		\lstick{\ket{Y}} & \qw & \qw & \ctrl{-1} & \rstick{H\ket{\phi}} \qw
	}
}
\caption{The ICM implementation of the Hadamard gate is the result of executing three gate teleportation circuits (corrections are not illustrated).}
\label{circ:hadamard}
\end{figure}

The nature of the teleported gates shows that the structure of an arbitrary TQEC circuit is ICM \cite{paler2015fully}, because the entire circuit consists of qubit (I)nitialisations, followed only by (C)NOT gates and then by (M)easurements. For this reason, during the second step an ICM circuit is generated starting from the output of the first step (gate decomposition). This procedure is detailed in \cite{paler2015fully}.

\subsection*{Generating Circuit Geometry}
\label{sec:gengeom}

\begin{figure}
\centering
\includegraphics[width=0.4\columnwidth]{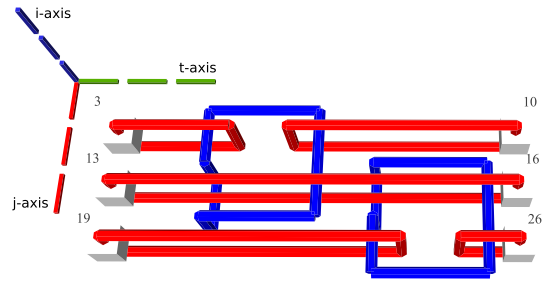}
\caption{The geometry corresponding to the circuit in Fig.~\ref{fig:matrix2}. The light gray cubes numbered 3,13,19 denote configurable circuit inputs, and the ones numbered 10,16,26 are outputs. Primal defects are red, and dual defects blue. The first horizontal defect pair spanned between the spheres 3 and 10 represent the topmost qubit from Fig.~\ref{fig:matrix2}. Each closed dual defect structure represents a CNOT.
}
\label{fig:twocnot}
\end{figure}

The ICM circuit representation synthesised in the previous step is transformed into a three dimensional geometric representation. In order to limit confusion, instead of the x-,y- and z- axis, the i-, j- and t-axis are referenced. A motivation for the choice of the axis names and how a quantum computer will use the geometric representation is offered in the supplementary material.
 The inputs and the outputs of the circuit are aligned along the j-axis, the gate geometries are depicted along the t-axis. The qubit defects are placed along the h-axis. More exactly, as depicted in Fig.~\ref{fig:twocnot}, information is processed from the input towards the outputs, and the inputs are at lowest t-axis coordinate while outputs at the highest t-axis coordinate. The first input has the lowest j-axis coordinate, and the last input the highest. Primal input pins of a qubit have the same j-axis and t-axis coordinates, but different i-axis coordinates.



In the subsequent figures representing TQEC geometries, the three dotted lines indicate the axes: green the t-axis, blue the i-axis, and red the j-axis. The common point where the axes are joined indicates the origin of the three dimensional space.

\subsection*{Distillation Scheduler}
\label{sec:sched}

The injected states required for the construction of teleported rotation gates are the result of distillation procedures \cite{fowler2012surface} abstracted as boxes. A \emph{scheduler} is a distillation box placement algorithm. A \emph{schedule} can be either \emph{heterogeneous}, when both distillation box types are included, or \emph{homogeneous} when a single box type is present. A schedule expresses the geometric placement of distillation boxes so that these are easily connected to the circuit. For the purpose of the current synthesis method, it was considered that all the distillations have to be ready before the circuit starts being executed. However, each distillation circuit has an associated probability of failure, and additional ones are required in order to have enough ready 
 In order to ease the connection to the circuit, the obtained schedules have, similarly to the circuit geometry, a single layer format: the boxes are stacked along the i-axis and aligned along the j-axis.

A schedule is computed starting from a list of circuit pin pairs: a distillation box is scheduled for each pair, to ensure the circuit will implement the designed functionality. The type ($\ket{A}$ or $\ket{Y}$) and the coordinates of the injection are known from the geometry generating step. Thus, the type of the scheduled box is identical to the injection's type, and the geometric position of the box is determined starting from the injection's coordinates. A homogeneous schedule is the result of a same-type injection pin pairs list and its construction is used in a future step of the synthesis.

Each distillation box includes a pair of output pins, too, and box pins need to be connected to circuit pins.
 The simplest way to connect boxes to the circuit is if the circuit and the box pins have a common coordinate. Therefore, in the current version of the scheduler, boxes are placed so that their pins have the j-axis coordinate of their circuit counterparts. Each box is three dimensional and if the j-axis distance between two circuit pins is less then the j-axis dimension of a box, then the boxes will be stacked along the i-axis (see, for example, Fig.~\ref{fig:h1}).

\subsection*{Distillation Failure Simulator}
\label{sec:faild}

Spare boxes need to be scheduled, too, and their number is a function of the box type failure probability. Consequently, there will exist a $\ket{Y}$ spare boxes schedule and a $\ket{A}$ spare boxes schedule. Both are homogeneous schedules obtained after constructing \emph{ghost} injections to the circuit: $\ket{A}$ type and $\ket{Y}$ type ghosts. The ghosts are used only to hint to the scheduler where to place the boxes, and are not used or referenced anywhere in the circuit. Once all the distillation boxes were scheduled, the failure of all the boxes (including spares) is simulated using Algorithm~\ref{alg:faild}.

\subsection*{Connecting Pins}
\label{sec:connect}

There is a high probability to successfully execute a TQEC circuit after the geometry was generated and enough distillation boxes were scheduled. The final step is to connect distillation boxes to the circuit. Knowing which boxes output high fidelity states, the states are used as injected circuit inputs. In a real-world environment, the successful distillation of states to be injected is known after executing the distillation subcircuits, but the presented synthesis method includes a simulation for determining successful ones.

Scheduling was performed so that connecting the boxes to the circuit would be straightforward. The first consequence was that the schedules have a single layer format. Secondly, in the heterogeneous schedule, boxes have the same j-axis coordinate like the circuit pins, meaning that two orthogonal segments (one along the i-axis and one along the t-axis) are sufficient to build a connection.

Additional homogeneous schedules were obtained by using ghost pins nonexistent in the circuit, and by scheduling boxes in an array. A detail not mentioned before 
 is that scheduling is configurable with respect to: 1) which i-axis coordinate is used for the first scheduled box; 2) how j-axis coordinates are iterated (low-to-high or high-to-low); and 3) how i-axis coordinates are iterated. In conjunction with the array arrangement, the configuration influences the corner, the line and the column directions in which the array is filled with boxes. After placing the spare box schedules around the initial schedule (see, for example, Fig.~\ref{fig:toff2}), connecting real circuit pins to the spares requires defects consisting of three orthogonal segments.



\subsection*{Examples}
\label{sec:examples}

The following examples, from a single CNOT circuit to a circuit that implements the Toffoli gate, will illustrate the scheduled distillation boxes, having the smaller boxes for $\ket{Y}$ distillation and the larger ones for $\ket{A}$. For the initial schedules (not including spares), the stacking of boxes is visible (e.g. Fig.~\ref{fig:h1}). The array arrangement of the scheduled spares is also noticeable (e.g. Fig.~\ref{fig:h2}).

\subsection*{CNOT Gate}
\label{sec:excnot}

\begin{figure}
\small
\centering
\subfloat[]{
  \label{fig:geomcirc}
	\raisebox{1.7cm}
	{  \small
	\Qcircuit @C=.4em @R=.4em {
		\lstick{\ket{c_i}}&\targ&\measure{Z}&\ket{0}&&\targ&\qw&&\lstick{\ket{c_o}}\\
		\lstick{\ket{+}}&\ctrl{-1}&\qw&\ctrl{1}&\qw&\ctrl{-1}&\measure{X}&\\
		\lstick{\ket{t_i}}&\qw&\qw&\targ&\qw&\qw&\lstick{\ket{t_o}}&
	}
	}
}
\hfil
\subfloat[]{
	\label{fig:cnot1}
	\includegraphics[width=0.35\columnwidth]{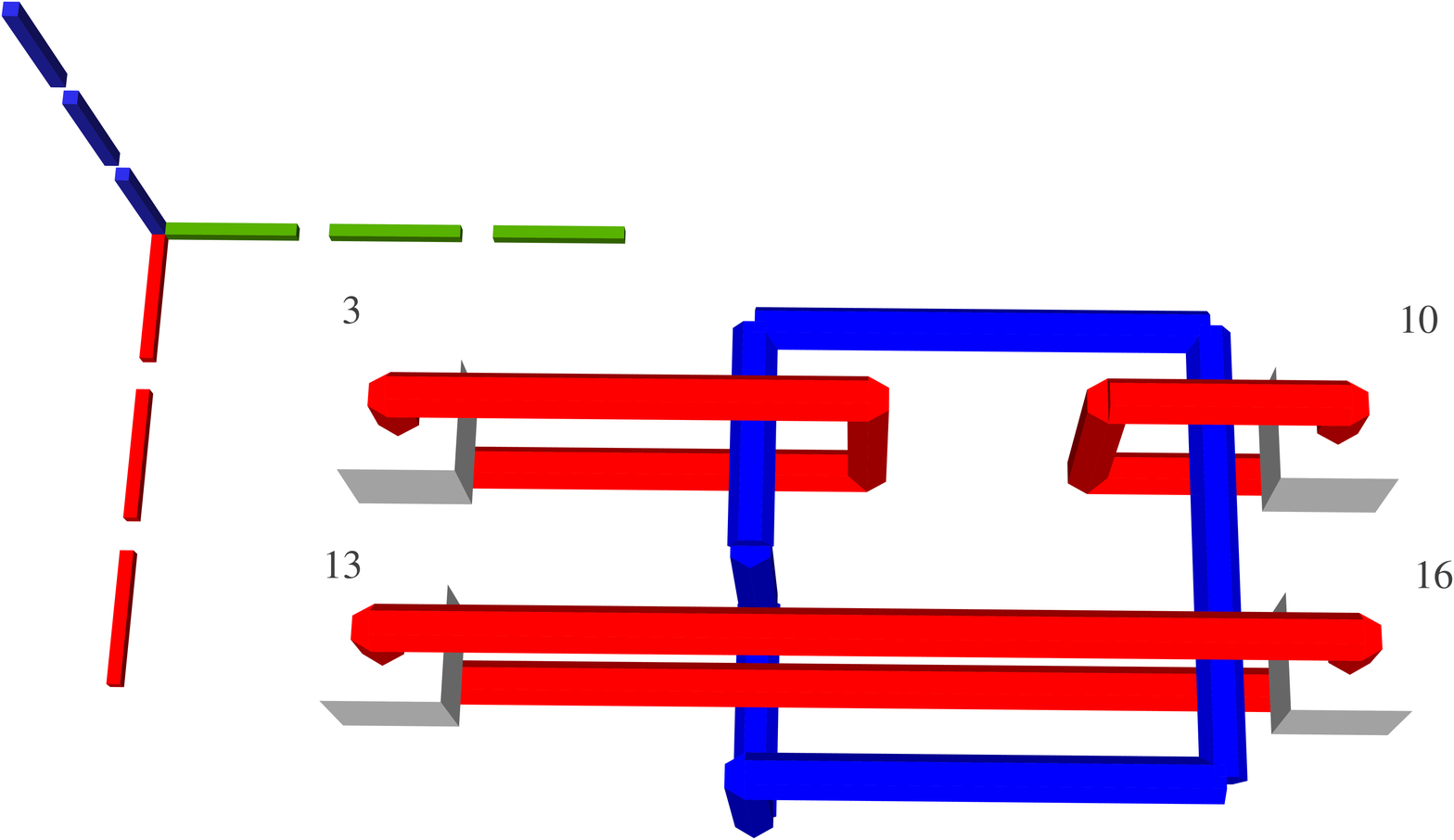}
}
\caption{The primal-primal CNOT circuit identity expressed: a) using the quantum circuit formalism; b) as a TQEC circuit geometric description. The qubit initialised into $\ket{+}$ and measured into the $X$ basis, in a), represents a TQEC dual qubit (blue) in b), and all the other qubits are primal (red).}
\end{figure}

A TQEC circuit consisting of a single CNOT will have two inputs and two outputs corresponding to the control and the target. Both the control and the target qubits will be formed by defect pairs that start/end at the inputs/outputs.
It was mentioned that the braid between a dual defect and a primal defect results in the implementation of a logical CNOT. However, it is preferred to have both the control and the target logical qubits of the same type, and this is achieved by using a circuit identity (Fig.~\ref{fig:geomcirc}). Fig.~\ref{fig:cnot1} depicts the geometry of a single CNOT circuit implemented through the use of the identity.

\subsection*{P Gate}

P gates are implemented by the use of teleportation circuits. A corresponding TQEC circuit will consist of an input, an output an ancilla initialised into the $\ket{Y}$ state (the encoding of an injected state) and a CNOT. The circuit structure is independent of the fidelity of the used $\ket{Y}$ state.

\begin{figure}
\centering
\includegraphics[width=.3\columnwidth]{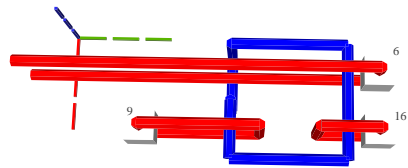}
\caption{TQEC geometry of the teleported P gate (Fig.~\ref{circ:ftcircs}).}
\label{fig:p0}
\end{figure}

\begin{figure}
\centering
\includegraphics[width=.45\columnwidth]{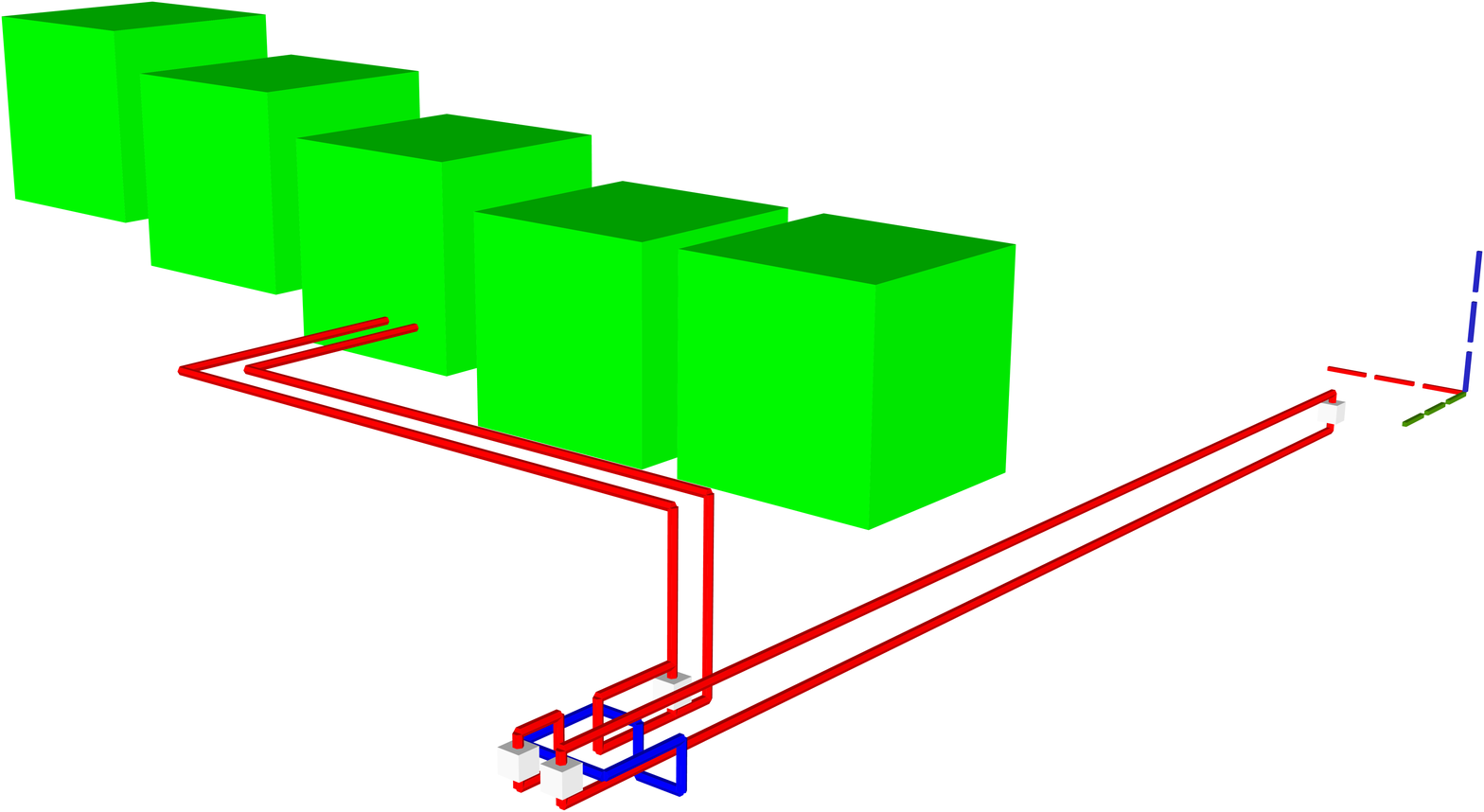}
\caption{Four $\ket{Y}$ spare boxes are required when the injected state fidelity needs to be increased and the distillation boxes have a success rate of 80\%.}
\label{fig:p2}
\end{figure}

Low fidelity injected states need to be distilled and, assuming that distillations have 100\% success rate, a single box would be required (Fig.~\ref{fig:p0}). If the success rate of the box would be lower (e.g. 80\%), additional spare distillation boxes need to be scheduled. The resulting homogeneous schedule is
 placed before the circuit inputs, and after simulating box failures, one of the boxes is connected to the circuit injection ancilla (Fig.~\ref{fig:p2}).

\begin{figure}
\centering
\includegraphics[width=.3\columnwidth]{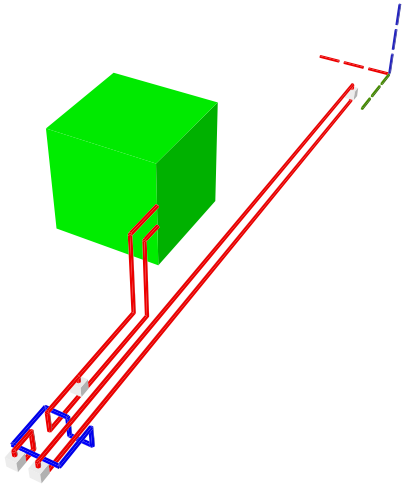}
\caption{If the boxes have 100\% success rate, the schedule of the P gate contains a single $\ket{Y}$ box.}
\label{fig:p1}
\end{figure}

\subsection*{Hadamard Gate}


TQEC circuits are based on topological cluster states, which do not support the direct implementation of a Hadamard gate. One of the solutions is to decompose the gate 
into a series of implementable gates and to synthesise the circuit description afterwards. Topological cluster states are related to the surface quantum error-correction code \cite{fowler2012surface}, which allows a Hadamard implementation without decomposing the gate. The difference between the capability of the topological cluster states and the surface code stems from the graph state structure of the first: the surface code does not use graph states and is thus more flexible to a certain degree.

\begin{figure}
\centering
\includegraphics[width=.3\columnwidth]{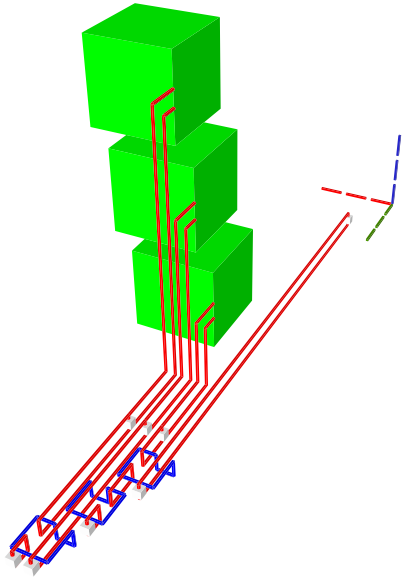}
\caption{The Hadamard gate schedule contains three $\ket{Y}$ boxes if all the injected states have high fidelity and the boxes have 100\% success rates.}
\label{fig:h1}
\end{figure}

\begin{figure}
\centering
\includegraphics[width=.45\columnwidth]{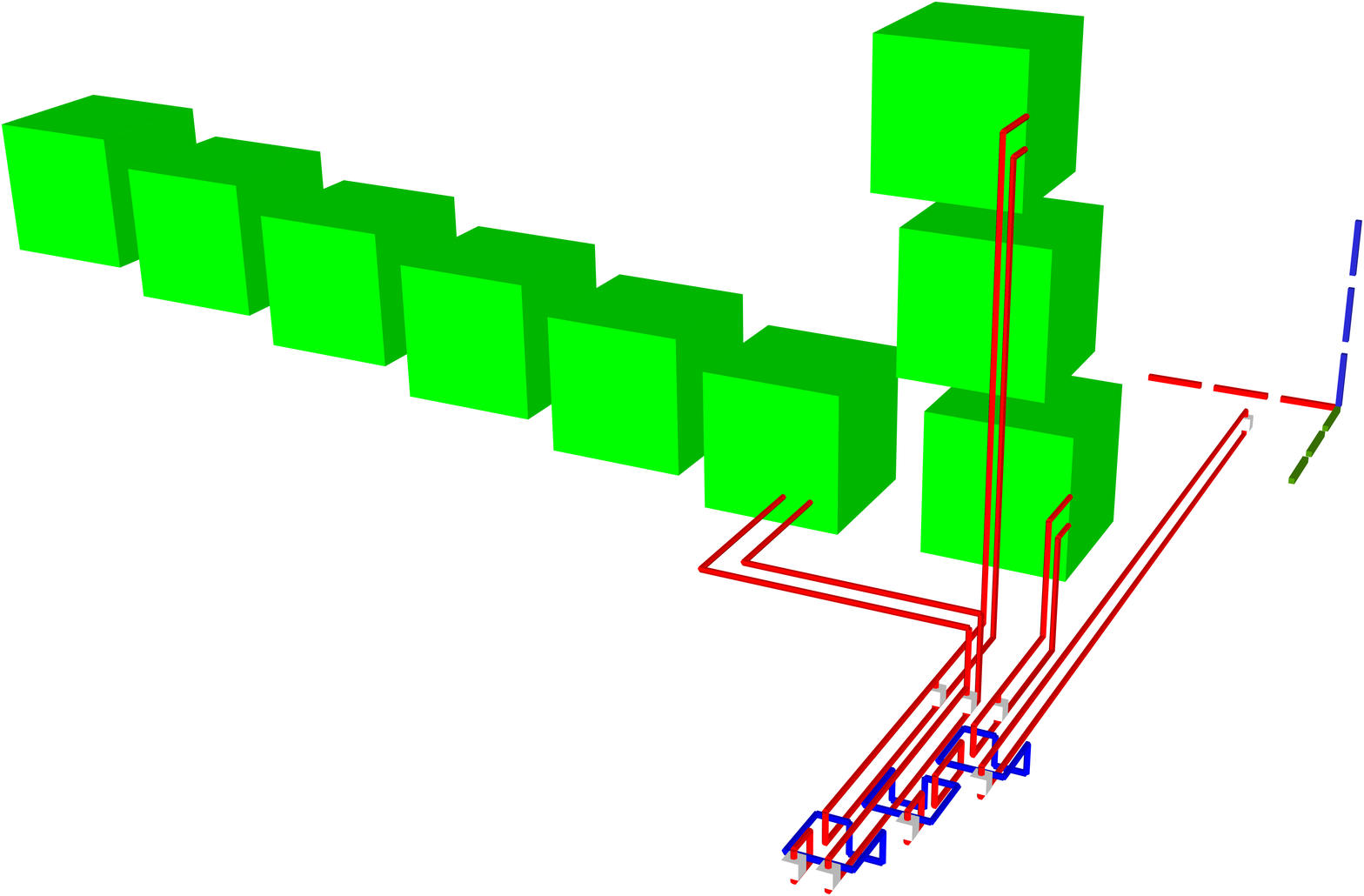}
\caption{The complete schedule, including spares, is illustrated for the case when the injected state fidelity needs to be increased by using distillation boxes with a success rate of 80\%.}
\label{fig:h2}
\end{figure}

A TQEC Hadamard circuit consists of an input, an output, three $\ket{Y}$ state initialised ancillae and three CNOTs (one for each teleported gate implementation). Three distillation boxes need to be scheduled in the case of high fidelity states and 100\% successful distillations (Fig.~\ref{fig:h1}). The additional spare boxes required to compensate a lower box success rate (e.g. 80\%) form a homogeneous schedule positioned on the j-axis before the initial schedule (Fig.~\ref{fig:h2}).

\begin{figure}
\centering
\includegraphics[width=.45\columnwidth]{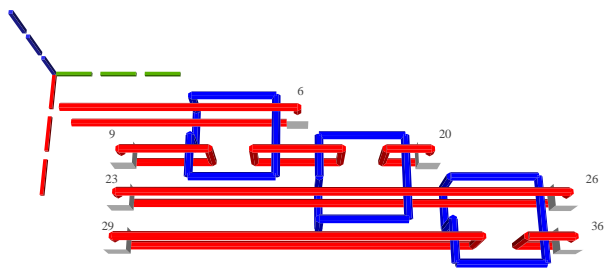}
\caption{TQEC geometry of the circuit from Fig.~\ref{circ:hadamard}.}
\label{fig:h0}
\end{figure}

\subsection*{T Gate}

\begin{figure}
\centering
\includegraphics[width=.45\columnwidth]{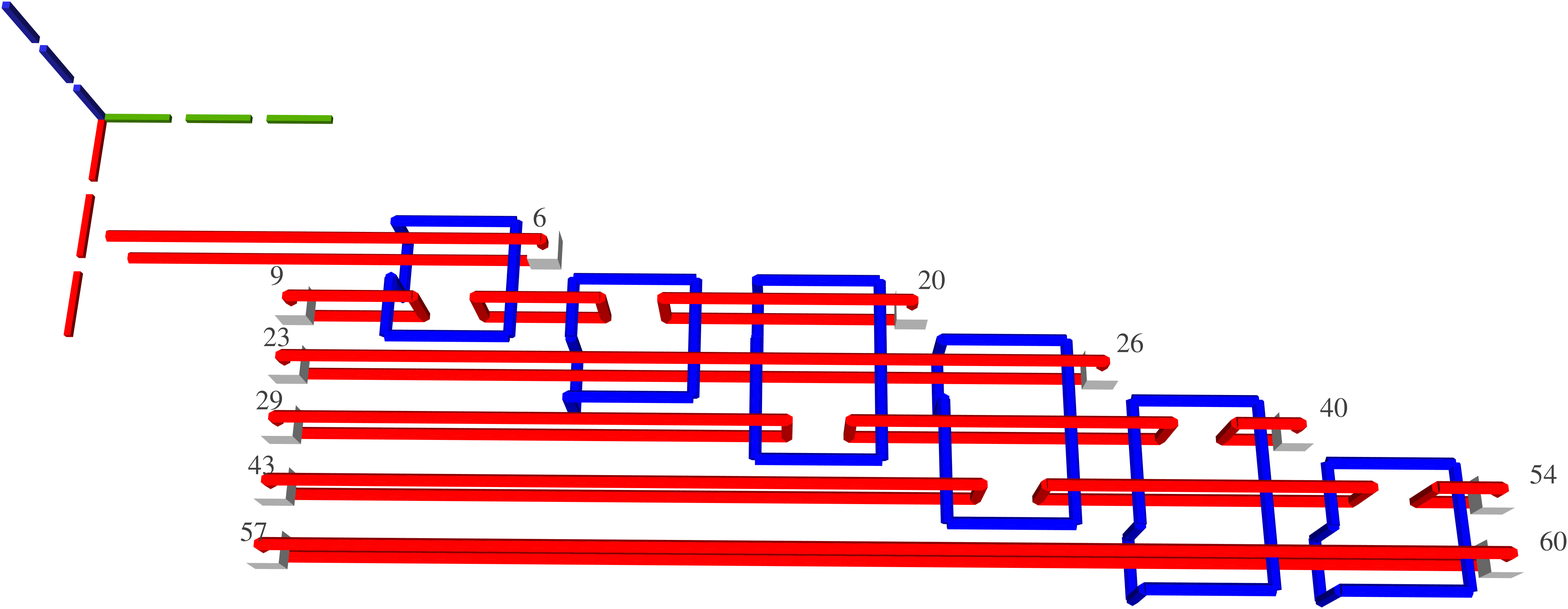}
\caption{TQEC geometry of the circuit from Fig.~\ref{circ:selective}.}
\label{fig:t0}
\end{figure}

Teleported gate applications are probabilistic. 
The P and the Hadamard gates allowed tracking the corrections through the circuit, but this not possible for the T gate: P gate corrections need to be applied right after it. A non-probabilistic T gate would have the same circuit structure like the P gate except that instead of $\ket{Y}$ an $\ket{A}$ is used. Due to their probabilistic nature and correction requirement, each TQEC T gate has to be followed by a selective source/destination circuit, which can be easily transformed into TQEC geometries (Fig.~\ref{fig:t0}).

\begin{figure}
\centering
\includegraphics[width=.45\columnwidth]{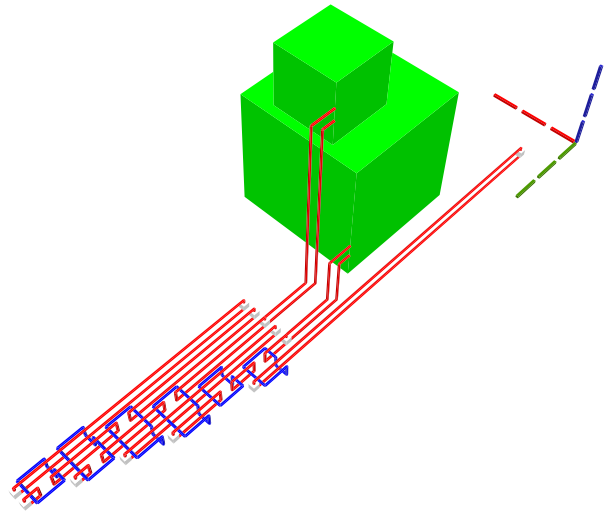}
\caption{The heterogeneous schedule required for the T gate  contains a single $\ket{A}$ and a single $\ket{Y}$ box (used for the correcting P gate). This is illustrated for the situation when all the injected states have high fidelity and the boxes have 100\% success rates.}
\label{fig:t1}
\end{figure}

\begin{figure}
\centering
\includegraphics[width=.45\columnwidth]{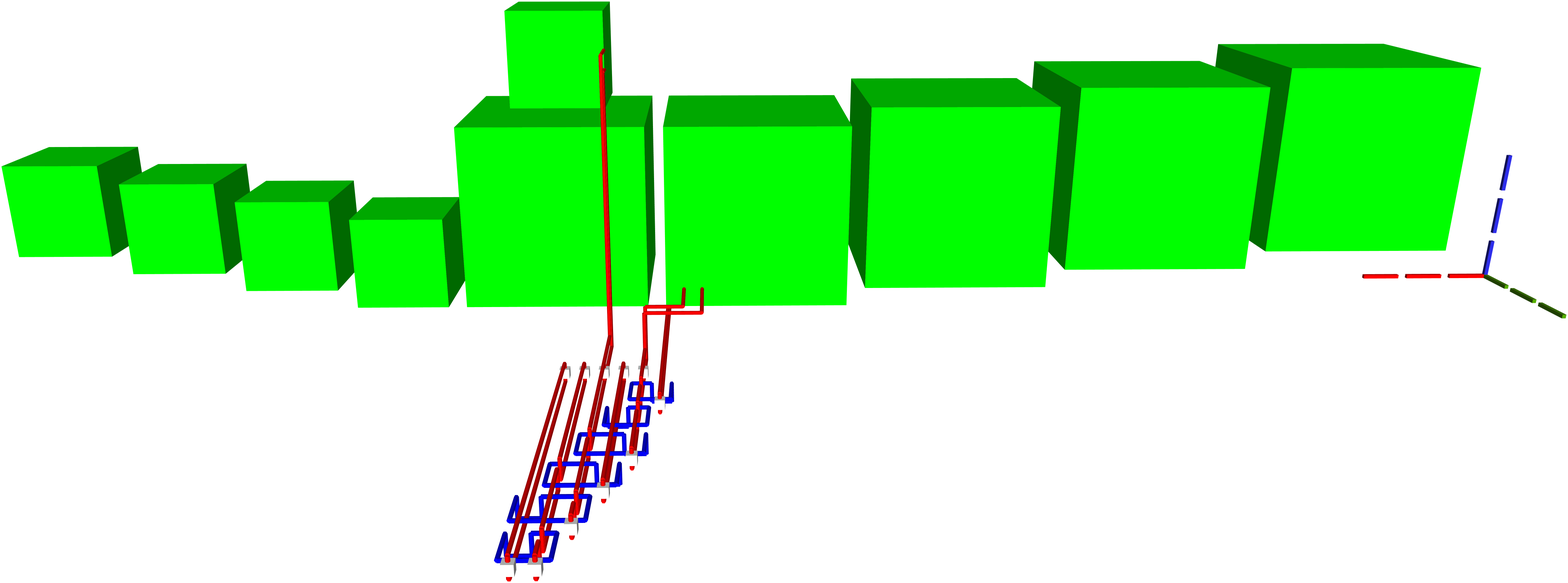}
\caption{All the scheduled boxes, including spares, are illustrated for the situation when the injected state fidelity needs to be increased by using distillation boxes with a success rate of 80\%.}
\label{fig:t2}
\end{figure}

\subsection*{Toffoli Gate}

The TQEC implementation of a Toffoli gate is the result of synthesising its decomposition (Fig.~\ref{circ:toffoli}) which consists of seven T gates, two Hadamards and a single P gate. All the gates in this decomposition will be implemented through teleportations.

\begin{figure*}
\centering
\includegraphics[width=.7\textwidth]{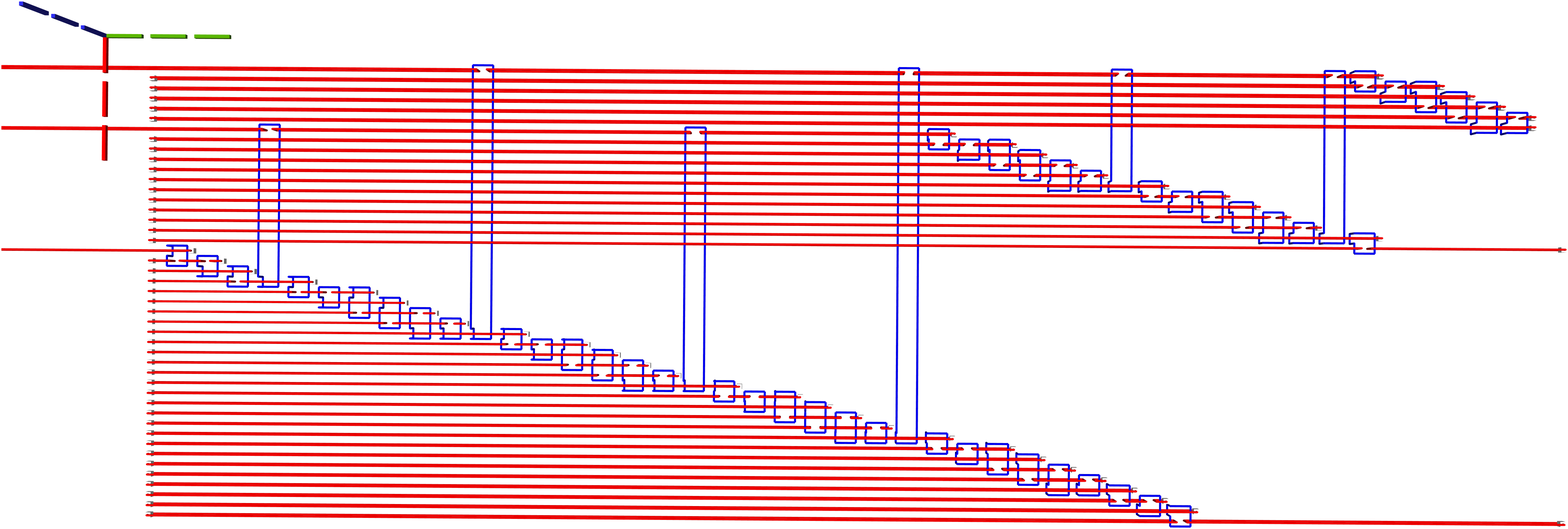}
\caption{TQEC geometry implementing the Toffoli circuit from Fig.~\ref{circ:toffoli}. The T and T$^\dagger$ gates are implemented using selective source/destination subcircuits (Fig.~\ref{circ:selective}) and the Hadamards are decomposed (Fig.~\ref{circ:hadamard}).}
\label{fig:toff0}
\end{figure*}

\begin{figure}
\centering
\includegraphics[width=.8\columnwidth]{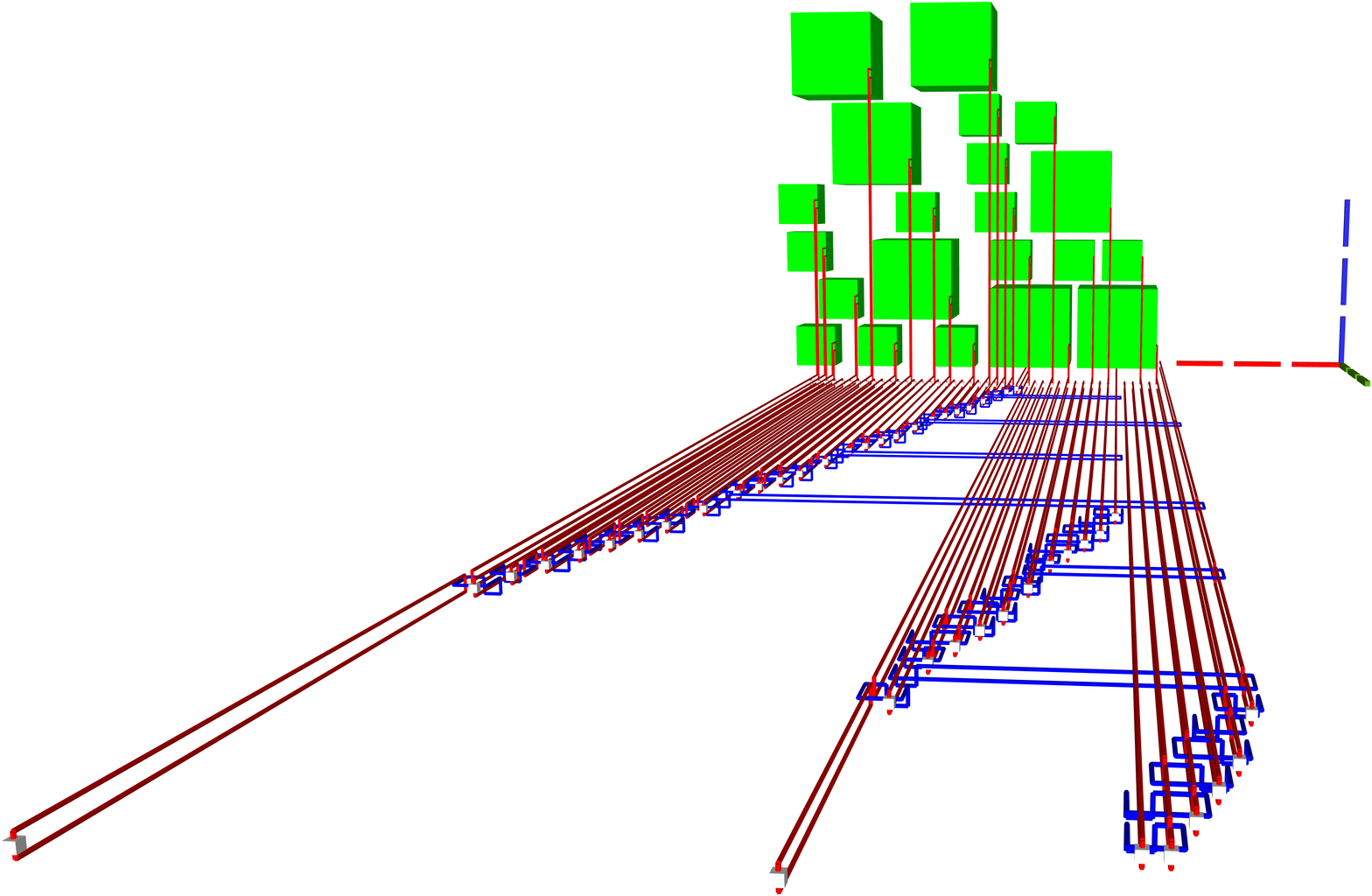}
\caption{The heterogeneous schedule required for the Toffoli gate implementation is illustrated for the situation when all the injected states have high fidelity and the distillation boxes have 100\% success rates. The inputs are at the top of the image, the outputs at the bottom. The scheduled boxes are depicted.}
\label{fig:toff1}
\end{figure}

\begin{figure*}
\centering
\includegraphics[width=.8\textwidth]{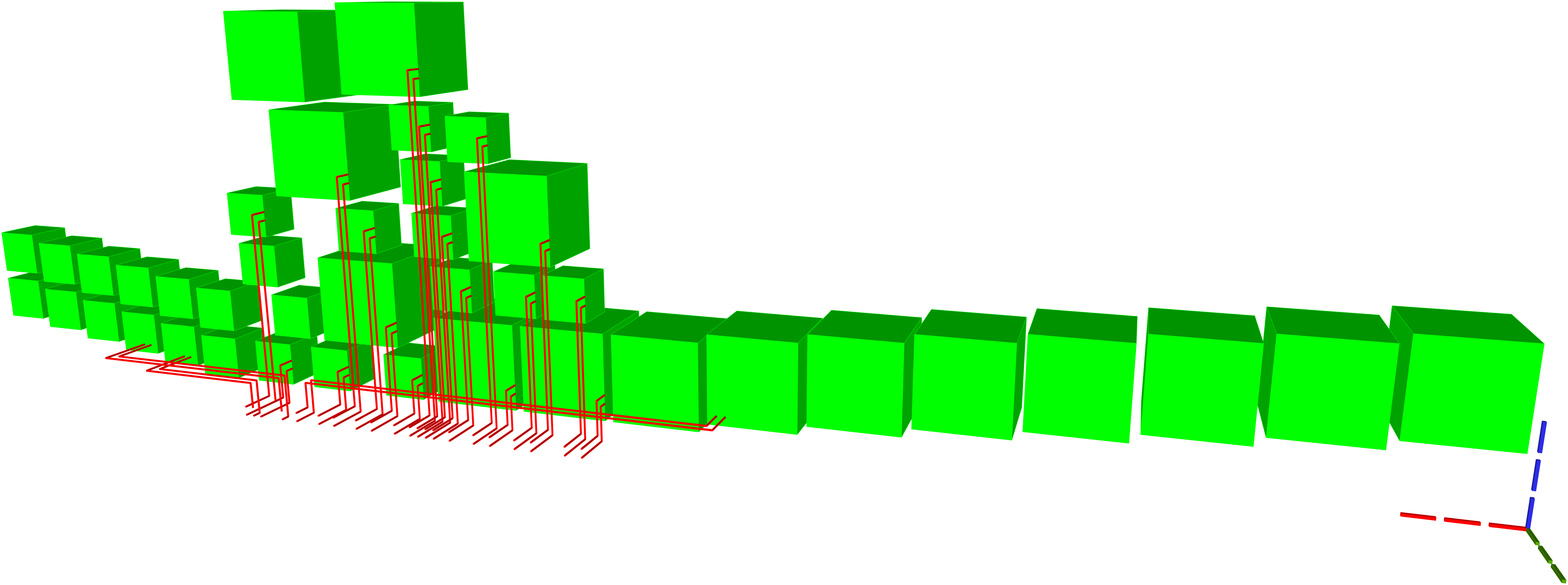}
\caption{All the scheduled boxes, including spares, are illustrated for the situation when the injected state fidelity needs to be increased by using distillation boxes with a success rate of 80\%. The initial schedule (Fig.~\ref{fig:toff1}) contained 21 boxes. The homogeneous schedule on the left side contains 12 $\ket{Y}$ spare boxes, and the schedule on the right contains eight spare $\ket{A}$ boxes. Through simulation it was determined that four $\ket{Y}$ and three $\ket{A}$ from the initial schedule (middle) fail and, as a result, spares are used.}.
\label{fig:toff2}
\end{figure*}

\subsection*{Future Work}
\label{sec:discussion}

TQEC circuit synthesis is the basis of a complete design stack, which is required to include optimisation and \emph{verification} methods. Circuit optimisation will be used, for example, for reducing the bounding box of the generated geometry (including schedules), while through verification the optimisation results will be checked for correctness. This section sketches future work directions which are directly related to the herein presented synthesis method.

\subsubsection*{Geometric Description}

TQEC circuit geometries are assumed to have a single layer format. This is not a requirement, but an implementation decision for the current version, and future synthesis variants could generate geometries extended on multiple layers. Furthermore, it is not required to separate inputs and outputs into two disjoint blocks placed at t-axis opposite coordinates. Inputs and outputs are allowed to have any coordinate, as long as the circuit still implements the designed (correct) functionality.

\begin{figure}
\centering
\subfloat[]{
  \label{fig:rule1}
  \includegraphics[width=0.2\columnwidth]{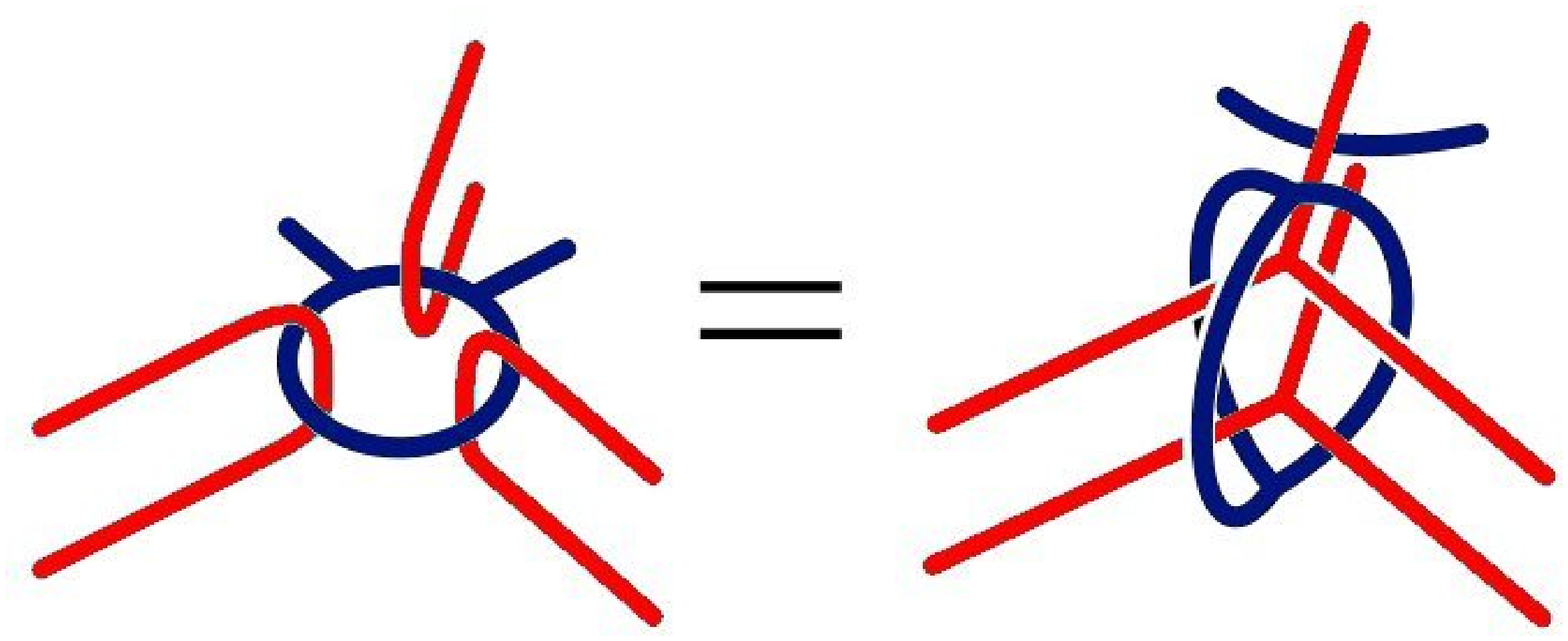}
}
\hfil
\subfloat[]{
  \label{fig:rule2}
  \includegraphics[width=0.2\columnwidth]{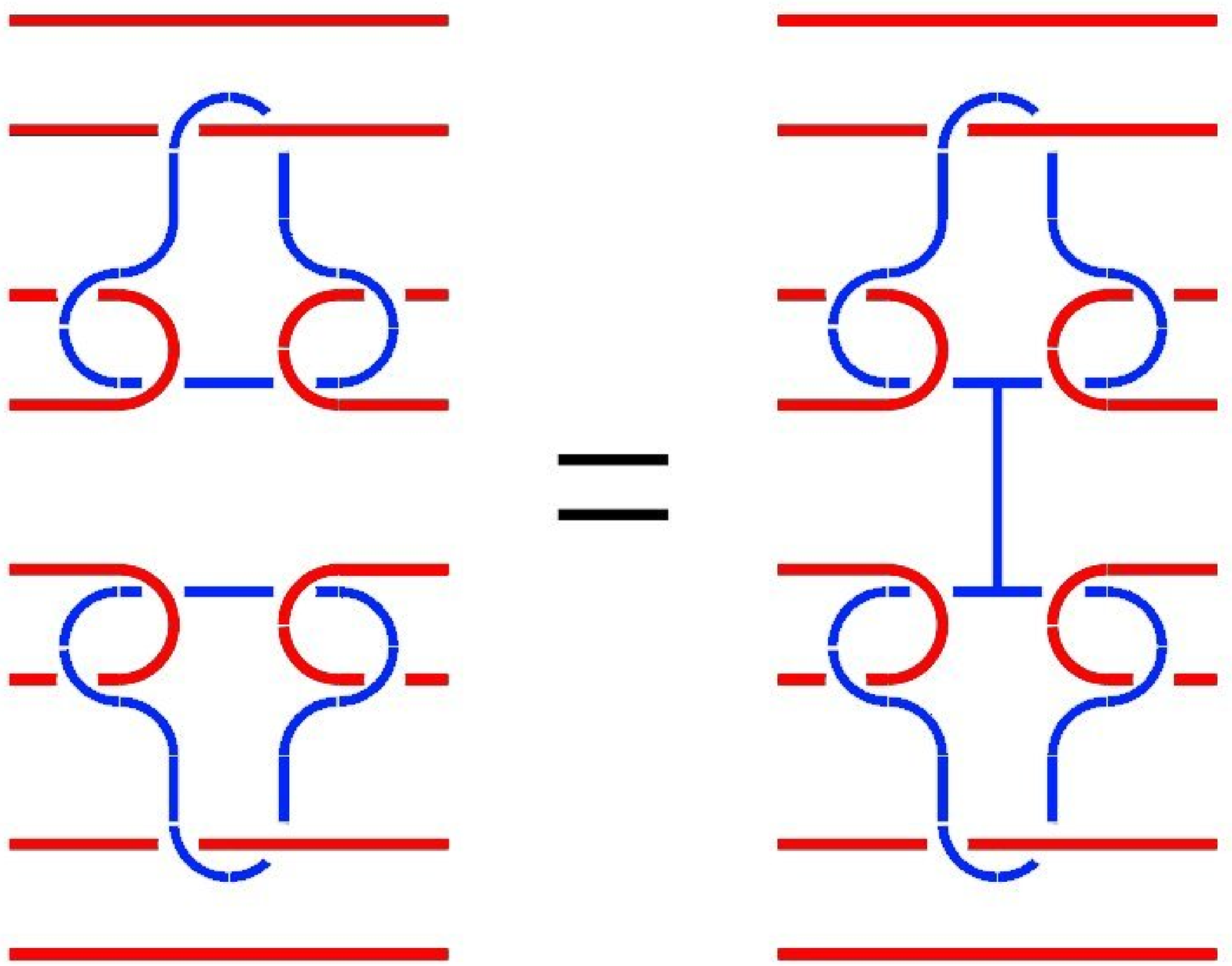}
}
\caption{TQEC geometry transformations: a) the cage rule \cite{raussendorf2007topological}; b) bridging \cite{fowler2012bridge}.}
\label{fig:rules}
\end{figure}

\begin{figure}
\centering
\includegraphics[width=0.2\columnwidth]{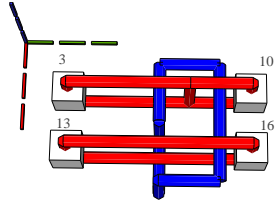}
\caption{The geometry of a primal-primal CNOT after bridging the primal defect structures representing the control qubit. The initial circuit was illustrated in Fig.~\ref{fig:cnot1}.}
\label{fig:bridgecnot}
\end{figure}

TQEC circuit geometries can be manipulated through circuit identities expressed as geometric transformations (Fig.~\ref{fig:rules}). Instead of generating a non-optimised geometry, this can be synthesised from the beginning using, for example, bridging \cite{fowler2012bridge}. That operation does not affect the implemented computation, and has the advantage, for certain types of circuits, of reducing the geometric bounding box of the circuit.
 This is possible by correctly joining certain defects so that no distance exists between them anymore. For example, bridging can be applied to the primal-primal CNOT (Fig.~\ref{fig:cnot1}) at the primal defects representing the control qubit and the result is a shorter geometry along the t-axis (Fig.~\ref{fig:bridgecnot}).

Another topic of interest is to devise automatic geometry compacting algorithms. These will selectively iterate through a set of geometric transformations and determine if the resulting circuit geometry is more optimal with respect to a specified metric. Global optimums will be very difficult to achieve, but the focus will be to firstly research automation, because this was not extensively done up to this point. Geometry optimisation is still performed in the research community by hand, for example \cite{fowler2012bridge}, and any degree of automation will result in a major productivity increase.

\subsubsection*{Distillation Box Scheduling}

The decision to place distillation box schedules at the beginning of the circuit, was taken in order to simplify the synthesis method. Once more, this is not a requirement, and boxes can be placed anywhere in the geometry. A simple optimisation is to move each box along the t-axis right before its output is required. The Toffoli geometry (Fig.~\ref{fig:toff0}) is good example: some $\ket{A}$ and $\ket{Y}$ boxes are connected to qubits which are required only at the end of the circuit. A large portion of the defect lengths is not required because the CNOTs are applied late (considering the t-axis as a time axis). Assuming that the CNOT has a starting t-axis coordinate of 200, and a box placed before an input has coordinate 10, it implies that the defects can be shorter and the box can be placed at, for example, 170. If this solution is chosen, the problem of failing distillation boxes and the scheduling of the required set of spares needs to be addressed in a different manner. 

The presented scheduling introduced a two dimensional placement of the boxes, and schedules are arranged one next to the other along the j-axis. Scheduling boxes anywhere in the geometry will result in a three dimensional scheduler, and the difficulty would be to determine and implement criteria to still maintain the circuit's fault-tolerance: given the per box failure probability, are enough boxes available and ready to be connected to the circuit requiring a high fidelity injected state?

\subsubsection*{Automatic Defect Construction}

Relaxing the assumptions made about geometric construction and box scheduling introduces additional complexities in the automatic construction of connections.
 The current version utilises a maximum of only three segments to connect an arbitrary box to the corresponding injection pins. A more complex connection mechanism is required if multiple layers are used for the geometry and boxes are allowed to be placed anywhere in the three dimensional space. Electronic design methods use routing algorithms to add wires between the placed circuit components. Lee's algorithm or channel routing are among the mostly used starting points in classical circuit design, and these could be very well be adapted for the case of TQEC, where defects are the equivalent of classical circuit wires.

\section*{Discussion}
\label{sec:conclusion}

TQEC circuit engineering is a relatively new field of computer engineering and its initial goal is to help construct the first scalable quantum computer. This will be achieved through developing, improving and using circuit design automation tools. Although steps towards constructing a functional TQEC design stack were previously proposed, this work introduced the complete automated synthesis of TQEC circuits. This is the starting point to the study of optimising geometric descriptions, a challenge which can be solved by expanding the TQEC circuit engineering community and by encouraging collaboration within.

The proposed synthesis method consists of a step sequence which includes gate decomposition, fault-tolerant circuit transformation, generation of TQEC circuit geometries, scheduling of specific ancilla state distillation circuits and, finally, connection of disjoint geometric elements into a single TQEC circuit geometric description. Each step was algorithmically formulated, and the synthesis results of the most commonly used quantum gates were presented and analysed.

Future work building on the synthesis results, including optimised scheduling of distillations and a more compact geometric description, were also presented and discussed.

\section*{Methods}

\subsection*{Gate Decomposition}

Gate decompositions can be computed dynamically using tools like \cite{ross2014optimal} or are known beforehand (e.g. the previous Hadamard and Toffoli decompositions). The current synthesis is able to use both methods, but irrespective of the chosen method, the output of this synthesis step is a circuit where all the non-TQEC gates were replaced with their decomposition into TQEC gate sequences.

\begin{align*}
\small
CNOT =
\begin{pmatrix}
1 & 0 & 0 & 0 \\
0 & 1 & 0 & 0 \\
0 & 0 & 0 & 1 \\
0 & 0 & 1 & 0
\end{pmatrix}&\,\,
V = \begin{pmatrix}
1 & -i \\
-i & 1
\end{pmatrix}\,\,
P = \begin{pmatrix}
1 & 0 \\
0 & i
\end{pmatrix} \,\,
T = \begin{pmatrix}
1 & 0 \\
0 & e^{i\frac{\pi}{4}}
\end{pmatrix}
\end{align*}

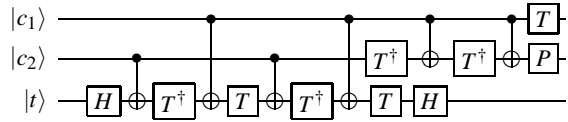
\begin{figure}[t!]
\centerline{
\small{
\Qcircuit @C=.3em @R=.3em {
	\lstick{\ket{c_1}}&\qw&\qw&\qw & \qw & \qw &\qw &\ctrl{2} &\qw&\qw&\qw&\ctrl{2} &\qw&\ctrl{1}&\qw&\ctrl{1}&\gate{T} &\qw\\
	\lstick{\ket{c_2}}&\qw&\qw&\qw& \qw&\ctrl{1}&\qw&\qw&\qw&\ctrl{1}&\qw&\qw&\gate{T^\dagger}&\targ&\gate{T^\dagger}&\targ&\gate{P}&\qw\\
	\lstick{\ket{t}}&\qw&\qw&\qw& \gate{H} & \targ & \gate{T^\dagger} & \targ & \gate{T} & \targ & \gate{T^\dagger} & \targ & \gate{T} & \gate{H} &\qw&\qw&\qw&\qw
	}
}
}
\caption{Toffoli gate implemented by a sequence of CNOT, $T$, $T^\dagger$, $P$ and $H$ gates~\cite[Ch.~4]{nielsen2010quantum}.}
\label{circ:toffoli}
\end{figure}

\subsection*{ICM Conversion}
\begin{figure}[t]
\small
\centering
\subfloat[]{
 \Qcircuit @C=.3em @R=.3em {
		\lstick{\ket{\psi}}& \ctrl{1} & \targ & \qw & \measure{Z} & \measure{X} \\
		\lstick{\ket{0}} & \targ & \qw & \targ & \measure{X}&\measure{Z} \\
		\lstick{\ket{+}} & \qw & \ctrl{-2} & \qw & \gate{P} &\qw\\
		\lstick{\ket{+}} & \qw &\qw &\ctrl{-2} & \gate{I} & \qw 
\gategroup{1}{5}{2}{5}{.4em}{--}\gategroup{1}{6}{2}{6}{.4em}{--}	
	}
}
\hfil
\subfloat[]{
 \Qcircuit @C=.3em @R=.3em {
		\lstick{\ket{\psi_1}}&\ctrl{2}&\qw& \measure{X}&\measure{Z}\\
		\lstick{\ket{\psi_2}}&\qw&\ctrl{1}& \measure{Z}&\measure{X}\\
		\lstick{\ket{0}}&\targ&\targ&\qw&\qw
\gategroup{1}{4}{2}{4}{.4em}{--}\gategroup{1}{5}{2}{5}{.4em}{--}
	}
}
\hfil
\subfloat[]{
 \Qcircuit @C=.3em @R=.3em {
 	\lstick{\ket{\psi}} & \targ & \qw & \qw & \qw & \qw &\measure{Z} & \control \cw & \control \cw \\
	\lstick{\ket{A}} & \ctrl{-1} & \ctrl{1} & \targ & \qw & \qw & \qw &\measure{Z} \cwx & \measure{X}\cwx \\
		\lstick{\ket{0}} & \qw &\targ & \qw & \targ &\qw & \qw & \measure{X}&\measure{Z} \\
		\lstick{\ket{Y}} & \qw &\qw & \ctrl{-2} & \qw &\ctrl{2} &\qw & \measure{X}&\measure{Z}\\
		\lstick{\ket{+}} & \qw &\qw &\qw &\ctrl{-2} & \qw& \ctrl{1} & \measure{Z}&\measure{X}\\
		\lstick{\ket{0}} & \qw &\qw & \qw & \qw & \targ & \targ & \qw & \qw & \qw & \rstick{T\ket{\psi}}
		\gategroup{2}{8}{5}{8}{.4em}{--}\gategroup{2}{9}{5}{9}{.4em}{--}
	}
}
\caption{Teleportations: a) Selective destination; b) Selective source \cite{fowler2012surface} and the combination of the two c), with the $T$ gate teleportation circuit (Fig.~\ref{circ:ftcircs}) to produce deterministic circuitry for a $T$-gate. The first column of measurements is chosen if the corrective P gate making use of the injected $\ket{Y}$ state is required \cite{fowler2012time}.}
\label{circ:selective}
\end{figure}
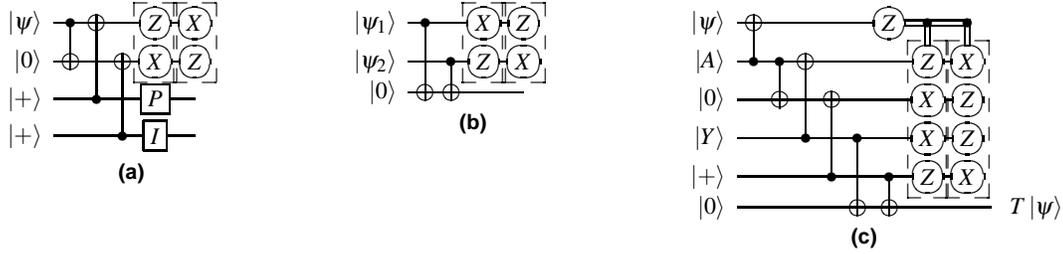

Teleported gate applications are probabilistic, meaning that after measurement the output qubit will either need or not need a correction in the form of another quantum gate. The correction requirement is signalled by the measurement result, and some of the corrections are Pauli gates (X,Y,Z), while others are non-Pauli. Pauli gate corrections are not required to be applied directly, because their effect can be tracked through the circuit \cite{paler2014software}, and applied only at the end of the computation (at circuit outputs). However, tracking is not possible for non-Pauli corrections, and this is the case for the T gate which requires P gate corrections. The selective source and destination mechanism can be used instead of dynamically modifying the circuit structure to probabilistically accommodate P gates (Fig.~\ref{circ:selective}).

\subsection*{Generating Circuit Geometry}
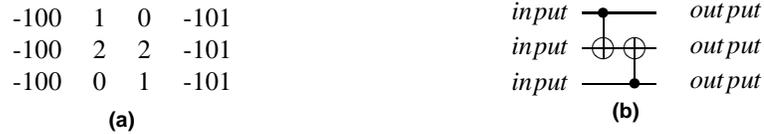
\begin{figure}
\centering
\subfloat[]{
  \label{fig:matrix1}
  \begin{tabular}{cccc}
  -100&1&0&-101\\
  -100&2&2&-101\\
  -100&0&1&-101
  \end{tabular}
}
\hfil
\subfloat[]{
  \raisebox{0.6cm}{
  \label{fig:matrix2}
  \Qcircuit @C=.4em @R=.8em {
		\lstick{input}&\ctrl{1}&\qw&\qw&&\rstick{output}\\
		\lstick{input}&\targ&\targ&\qw&&\rstick{output}\\
		\lstick{input}&\qw&\ctrl{-1}&\qw&&\rstick{output}
	}
}
}
\caption{A matrix-like representation of the circuit is used to generate a circuit's geometry obtained after marking the inputs/outputs accordingly with different values (e.g. -100 and -101), and the controls and the targets with separate values (e.g. 1 and 2). The matrix representation of a) is equivalent to the circuit from b). For each line in the matrix, a defect pair is defined, and a CNOT will have the control/target drawn at the coordinates indicated by the row and the column corresponding to its matrix encoded position. The geometry of a single CNOT circuit is presented in Fig.~\ref{fig:cnot1}.}
\label{fig:matrix}
\end{figure}

When considering the t-axis horizontal and the j-axis vertical, the gates are aligned horizontally and the qubits vertically. A \emph{matrix-like representation} of the circuit, inspired by the quantum circuit formalism, is used: qubits are the lines of the matrix and each CNOT occupies a certain column. Each initialisation and measurement basis is encoded by an integer (e.g. -99 for $\ket{A}$ initialisation, and -98 for $\ket{A}$ measurement), circuit input outputs marked distinctively (e.g. -100 for input, -101 for output), and the controls and the targets have separate values (e.g. 1 and 2). The geometry is constructed by traversing the matrix column wise, thus drawing the entire sequence of CNOTS starting from circuit inputs towards outputs (Fig.~\ref{fig:matrix}).

Due to the direct mapping between the ICM representation, the matrix representation of the circuit and the geometry, the resulting geometry has a single layer format and does not include any \emph{optimisation} like bridging \cite{fowler2012bridge}.

\subsection*{Distillation Scheduler}

Algorithm~\ref{alg:sched} is sketching the scheduling procedure which takes as inputs a list $L$ of injection pin pairs and a list $Dim$ of box dimensions. The pseudo code of the algorithms uses partially an object oriented syntax. For example, $l.coord.j$ gets the j-axis coordinate from the coordinates attribute $coord$ of object $l$. The algorithm uses an object $R$ which represents a two dimensional \emph{region} representing the layer (parallel to the j- and i-axis) where the boxes will be placed in the three dimensional space. For each injection pin pair ($l \in  L$) three parameters are required to schedule a box of type $l.type$: the j-axis coordinate of the pins ($j$ at Line~\ref{algsched:5}) and the necessary jspan and ispan of the box to be scheduled ($reqj$ and $reqi$ at Lines~\ref{algsched:6},~\ref{algsched:7}). The parameters are used to compute the set $A$ of possible sub regions of $R$ where the box could be placed. The sub region $rmin$ with the lowest i-axis coordinate is chosen (Line~\ref{algsched:9}), the sub region is removed from $R$ in order to not allow other boxes to be placed there, and finally the box is scheduled at $posj, rmin.i$.

\begin{algorithm}
\caption{Distillation Scheduler}\label{alg:sched}
\begin{algorithmic}[1]
\State{Input: $L$ list of injection pin pairs}
\State{Input: $Dim$ list of box dimensions}
\State{Region $R$}
\ForAll{$l \in L$}
\State{$posj \gets l.coord.j$}\label{algsched:5}
\State{$reqj \gets Dim[l.type].jspan$}\label{algsched:6}
\State{$reqi \gets Dim[l.type].ispan$}\label{algsched:7}
\State{$A \gets$ all subregions of $R$ having $jspan=reqj$, $ispan=reqi$ and placed at the j-coordinate $posj$};
\State{$rmin \gets$ the subregion from $A$ having the lowest i-axis coordinate}\label{algsched:9}
\State{Remove the region $rmin$ from $R$}
\State{Schedule box of type $l.type$ at coordinates $posj, rmin.i$}
\EndFor
\end{algorithmic}
\end{algorithm}

\subsection*{Distillation Failure Simulator}

Algorithm~\ref{alg:additional} generates ghost pins having a large enough distance along the j-axis, so that boxes are not stacked but linearly arranged in a row. The algorithm will generate $n$ pin pairs of type $type$, places the first pair at the j-axis coordinate $sw$ and subsequent boxes with a j-axis distance of $\mathit{offj}$ between. Multiple rows (e.g. $m$) of boxes (similar to stacking) can be obtained by generating ghosts at the same coordinates. After repeating the algorithm $m$ times, the output list $L$ (Lines~\ref{algadd:6},~\ref{algadd:11}) of ghost pin pairs will contain $m$ times the same $n$ pins. The pins of a pair have the same j-axis coordinate, and the t-axis and i-axis coordinates are not computed because they are not relevant (Lines~\ref{algadd:9},~\ref{algadd:10}) for the scheduler (Algorithm~\ref{alg:sched}). After scheduling the $m\times n$ spare boxes, the homogeneous schedule is organised as an array. This arrangement simplifies the connection between boxes and real (not ghost) circuit pins, because the defect segments will be constructed using straightforward rules.

\begin{algorithm}
\caption{Construct Homogeneous Schedule}\label{alg:additional}
\begin{algorithmic}[1]
\State{Input: $n$ number of boxes to schedule}
\State{Input: $type$ type of boxes to schedule}
\State{Input: $sj$ starting j-axis coordinate}
\State{Input: $Dim$ list of box type dimensions}
\State{Input: $\mathit{offj}$ j-axis offset between boxes}
\State{Input: $L$ existing list of pins}\label{algadd:6}
\ForAll{$0 \leq idx<n$}
\State{Ghost pin pair $g_1, g_2$}
\State{$g_1.j \gets sj + idx*Dim[type].jspan + \mathit{offj}$}\label{algadd:9}
\State{$g_2 \gets g_1$}\label{algadd:10}
\State{$L = L \cup (g_1, g_2)$}\label{algadd:11}
\EndFor
\State{Call Algorithm~\ref{alg:sched} with parameters $L,Dim$}
\end{algorithmic}
\end{algorithm}

The next problem is to determine which succcessful box should be connected to which circuit injection pin pair (ghosts are not considered). All the boxes of type $t$ are stored in a queue $B_t$ and it is assumed that the per box failure probability is $p$. All the type $t$ circuit pin pairs are stored in the $Q_t$ queue. The solution is to take the first unused successful box for every pin pair $q \in Q_t$. The algorithm runs until the queue $B_t$ is empty (Line~\ref{algfaild:5}). The failure of each queued box is randomly determined based on $p$ (Lines~\ref{algfaild:7}--\ref{algfaild:11}), and a successful box will be connected to the current circuit pin pair $q$ (Line~\ref{algfaild:9}).

\begin{algorithm}
\caption{Distillation Failure Simulator}\label{alg:faild}
\begin{algorithmic}[1]
\State{Input: $B_t$ queue of boxes of type $t$}
\State{Input: $p$ failure probability}
\State{Input: $Q_t$ queue of injection pin pairs of type $t$}
\ForAll{$q \in Q_t$}
\While{$B_t$ not empty}\label{algfaild:5}
\State{$b \gets B_t.pop()$}
\State{$rnd \gets$ random number $\in [0,1]$}\label{algfaild:7}
\If{$rnd<p$}
\State{Connect $b$ to $q$}\label{algfaild:9}
\State{break}
\EndIf\label{algfaild:11}
\EndWhile
\EndFor
\end{algorithmic}
\end{algorithm}

\subsection*{Connecting Pins}
Algorithm~\ref{alg:connect} captures the rules of connecting the distillation boxes to the circuit. Pairs of box and circuit pins are resulting after the distillatin failure simulation. 
The connection algorithm takes as input the set of pin pairs computed by Algorithm~\ref{alg:faild} and determines the defect segments necessary to connect the pins. Because pins are specified using unit cell coordinates, and by the way the schedules were obtained, it is sufficient to compute the Manhattan distances between box and pin coordinates (Lines~\ref{algcon:4},~\ref{algcon:7},~\ref{algcon:9}). For example, the t-axis distance between box pin $b$ and circuit pin $c$ is $c.coord.t-b.coord.t$. The distances are used to determine segment end points (e.g. $ep_1$ and $ep_2$). The first end point has the coordinates of the box pin (Line~\ref{algcon:3}), and subsequent end points are computed by adding the corresponding distances. A segment is determined by its end point coordinates, and segments of zero length are not constructed.


\begin{algorithm}
\caption{Connect Boxes to Circuit}\label{alg:connect}
\begin{algorithmic}[1]
\State{Input: $P=\{(b,c) \mid b \text{ box pin}; c \text{ circuit pin}\}$}
\ForAll{$(b,c) \in P$}
\State{$ep_1 \gets b.coord$}\label{algcon:3}
\State{$ep_1.t = ep_1.t + (c.coord.t - b.coord.t)$}\label{algcon:4}
\State{Construct segment $b.coord, ep_1$}
\State{$ep_2 \gets ep_1$}
\State{$ep_2.i = ep_2.i + (c.coord.i - b.coord.i)$}\label{algcon:7}
\State{Construct segment $ep_1, ep_2$}
\State{$ep_2.j = ep_2.j + (c.coord.j - b.coord.j)$}\label{algcon:9}
\State{Construct segment $ep_2, c.coord$}
\EndFor
\end{algorithmic}
\end{algorithm}


\begin{thebibliography}{10}
\expandafter\ifx\csname url\endcsname\relax
  \def\url#1{\texttt{#1}}\fi
\expandafter\ifx\csname urlprefix\endcsname\relax\def\urlprefix{URL }\fi
\providecommand{\bibinfo}[2]{#2}
\providecommand{\eprint}[2][]{\url{#2}}

\bibitem{fowler2012surface}
\bibinfo{author}{Fowler, A.~G.}, \bibinfo{author}{Mariantoni, M.},
  \bibinfo{author}{Martinis, J.~M.} \& \bibinfo{author}{Cleland, A.~N.}
\newblock \bibinfo{title}{Surface codes: Towards practical large-scale quantum
  computation}.
\newblock \emph{\bibinfo{journal}{Physical Review A}}
  \textbf{\bibinfo{volume}{86}}, \bibinfo{pages}{032324}
  (\bibinfo{year}{2012}).

\bibitem{paler2014software}
\bibinfo{author}{Paler, A.}, \bibinfo{author}{Devitt, S.},
  \bibinfo{author}{Nemoto, K.} \& \bibinfo{author}{Polian, I.}
\newblock \bibinfo{title}{Software-based pauli tracking in fault-tolerant
  quantum circuits}.
\newblock In \emph{\bibinfo{booktitle}{Design, Automation and Test in Europe
  Conference and Exhibition (DATE), 2014}}, \bibinfo{pages}{1--4}
  (\bibinfo{organization}{IEEE}, \bibinfo{year}{2014}).

\bibitem{paler2012synthesis}
\bibinfo{author}{Paler, A.}, \bibinfo{author}{Devitt, S.},
  \bibinfo{author}{Nemoto, K.} \& \bibinfo{author}{Polian, I.}
\newblock \bibinfo{title}{Synthesis of topological quantum circuits}.
\newblock In \emph{\bibinfo{booktitle}{Proceedings of the 2012 IEEE/ACM
  International Symposium on Nanoscale Architectures}},
  \bibinfo{pages}{181--187} (\bibinfo{organization}{ACM},
  \bibinfo{year}{2012}).

\bibitem{wille2009equivalence}
\bibinfo{author}{Wille, R.}, \bibinfo{author}{Gro{\ss}e, D.},
  \bibinfo{author}{Miller, D.~M.} \& \bibinfo{author}{Drechsler, R.}
\newblock \bibinfo{title}{Equivalence checking of reversible circuits}.
\newblock In \emph{\bibinfo{booktitle}{Multiple-Valued Logic, 2009. ISMVL'09.
  39th International Symposium on}}, \bibinfo{pages}{324--330}
  (\bibinfo{organization}{IEEE}, \bibinfo{year}{2009}).

\bibitem{nielsen2010quantum}
\bibinfo{author}{Nielsen, M.~A.} \& \bibinfo{author}{Chuang, I.~L.}
\newblock \emph{\bibinfo{title}{Quantum computation and quantum information}}
  (\bibinfo{publisher}{Cambridge university press}, \bibinfo{year}{2010}).

\bibitem{paler2015fully}
\bibinfo{author}{Paler, A.}, \bibinfo{author}{Polian, I.},
  \bibinfo{author}{Nemoto, K.} \& \bibinfo{author}{Devitt, S.~J.}
\newblock \bibinfo{title}{A fully fault-tolerant representation of quantum
  circuits}.
\newblock In \emph{\bibinfo{booktitle}{Reversible Computation}},
  \bibinfo{pages}{139--154} (\bibinfo{publisher}{Springer},
  \bibinfo{year}{2015}).

\bibitem{fowler2012bridge}
\bibinfo{author}{Fowler, A.~G.} \& \bibinfo{author}{Devitt, S.~J.}
\newblock \bibinfo{title}{A bridge to lower overhead quantum computation}.
\newblock \emph{\bibinfo{journal}{arxiv}} \textbf{\bibinfo{volume}{1209}}
  (\bibinfo{year}{2012}).

\bibitem{raussendorf2007topological}
\bibinfo{author}{Raussendorf, R.}, \bibinfo{author}{Harrington, J.} \&
  \bibinfo{author}{Goyal, K.}
\newblock \bibinfo{title}{Topological fault-tolerance in cluster state quantum
  computation}.
\newblock \emph{\bibinfo{journal}{New Journal of Physics}}
  \textbf{\bibinfo{volume}{9}}, \bibinfo{pages}{199} (\bibinfo{year}{2007}).

\bibitem{ross2014optimal}
\bibinfo{author}{Ross, N.~J.} \& \bibinfo{author}{Selinger, P.}
\newblock \bibinfo{title}{Optimal ancilla-free clifford+ t approximation of
  z-rotations}.
\newblock \emph{\bibinfo{journal}{arXiv preprint arXiv:1403.2975}}
  (\bibinfo{year}{2014}).

\bibitem{fowler2012time}
\bibinfo{author}{Fowler, A.~G.}
\newblock \bibinfo{title}{Time-optimal quantum computation}.
\newblock \emph{\bibinfo{journal}{arXiv preprint arXiv:1210.4626}}
  (\bibinfo{year}{2012}).

\bibitem{fowler2009topological}
\bibinfo{author}{Fowler, A.~G.} \& \bibinfo{author}{Goyal, K.}
\newblock \bibinfo{title}{Topological cluster state quantum computing}.
\newblock \emph{\bibinfo{journal}{Quantum Information \& Computation}}
  \textbf{\bibinfo{volume}{9}}, \bibinfo{pages}{721--738}
  (\bibinfo{year}{2009}).

\bibitem{paler2014mapping}
\bibinfo{author}{Paler, A.}, \bibinfo{author}{Devitt, S.~J.},
  \bibinfo{author}{Nemoto, K.} \& \bibinfo{author}{Polian, I.}
\newblock \bibinfo{title}{Mapping of topological quantum circuits to physical
  hardware}.
\newblock \emph{\bibinfo{journal}{Scientific reports}}
  \textbf{\bibinfo{volume}{4}} (\bibinfo{year}{2014}).

\bibitem{childs2005unified}
\bibinfo{author}{Childs, A.~M.}, \bibinfo{author}{Leung, D.~W.} \&
  \bibinfo{author}{Nielsen, M.~A.}
\newblock \bibinfo{title}{Unified derivations of measurement-based schemes for
  quantum computation}.
\newblock \emph{\bibinfo{journal}{Physical Review A}}
  \textbf{\bibinfo{volume}{71}}, \bibinfo{pages}{032318}
  (\bibinfo{year}{2005}).

\end{thebibliography}

\section*{Additional Information}

\subsection*{Author contributions statement}
A.P., S.J.D. and A.G.F. conceived the idea. A.P. was responsible for algorithmic simulations. S.J.D. and A.G.F. were responsible for results verification. All authors were responsible for drafting of the manuscript.

\subsection*{Competing financial interests}
The authors declare no competing financial interests.

\clearpage

{\huge Supplemental Material}

\section*{Appendix: Geometrical Description Volume}
\label{apx:volume}

Bounding boxes were initially introduced when referring to the distillation subcircuits, and mentioned in the context of optimising a synthesised geometric description (Sec.~\ref{sec:discussion}). However, the metric of bounding boxes was not introduced and discussed accordingly. This section presents the \emph{equivalent volume} \cite{fowler2012bridge} of a geometric description and the considerations backing its definition.

\begin{figure}[h]
  \centering
  \includegraphics[width=0.6\columnwidth]{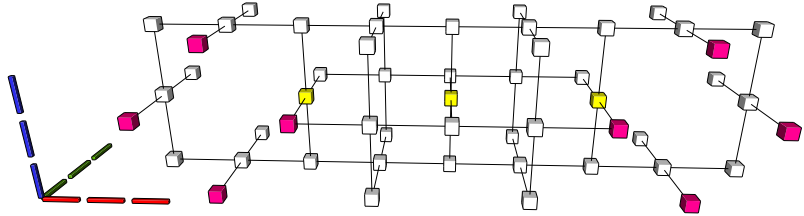}
  \caption{A lattice of $3 \times 1 \times 1$ primal unit cells where two single cell defects were defined. The ring of qubits marked pink (length = 4) and the chain of qubits marked yellow (length = 3) are used for error-correction.}
  \label{fig:chains}
\end{figure}

TQEC circuits use a quantum error-correction procedure based on topological cluster states (Sec.~\ref{sec:back}) in which information is encoded by constructing defects and operated on by braiding the defects. Error-correction is based on logical operators which are defined as rings of specific physical qubits around defects and specific physical qubit chains connecting defects. The overall code distance is the minimum value between the shortest ring and the shortest chain. The shortest ring is found at the defect with the smallest diameter, and the shortest chain exists between the closest defects. Fig.~\ref{fig:chains} depicts the way rings and chains are determined when considering the lattice layout: each unit cell on the defect margin contributes a physical qubits to the ring diameter $d_f$, each unit cell along the distance between two defects contributes a physical qubit to the chain length $d_d$. The perimeter $p_f$ of a ring is $4d_f$, where $d_f$ equals the number of unit cells along the diameter, because the defects are considered to have quadratic cross-sections. In order to have a code distance of $p_f$, the shortest chain has to have a length greater than $p_f$: $d_d=4d_f + 1$ is achieved when two defects are at least $4d_f$ unit cells apart. Therefore, in general, it can be argued, that in order to use an error-correction with distance $4d$, the defects are required to have a diameter of $d$ unit cells and need to be at least $4d$ unit cells apart. Because the code distance is a function of physical failure rates, diameters and defect distances need to be increased for high failure rates.

The general error-detection procedure is detailed in \cite{fowler2009topological, fowler2012surface} and for the purpose of this section it suffices to mention that it uses the parities of lattice unit cells: there are six face qubits on each unit cell, and their X basis measurement parity is even in the absence of errors. An error is detected on odd parity cells, and correction is performed after choosing pairs of odd parity cells as end points of even parity unit cell chains.

\begin{figure}[t!]
\centering
\subfloat[]{
  \includegraphics[width=0.2\columnwidth]{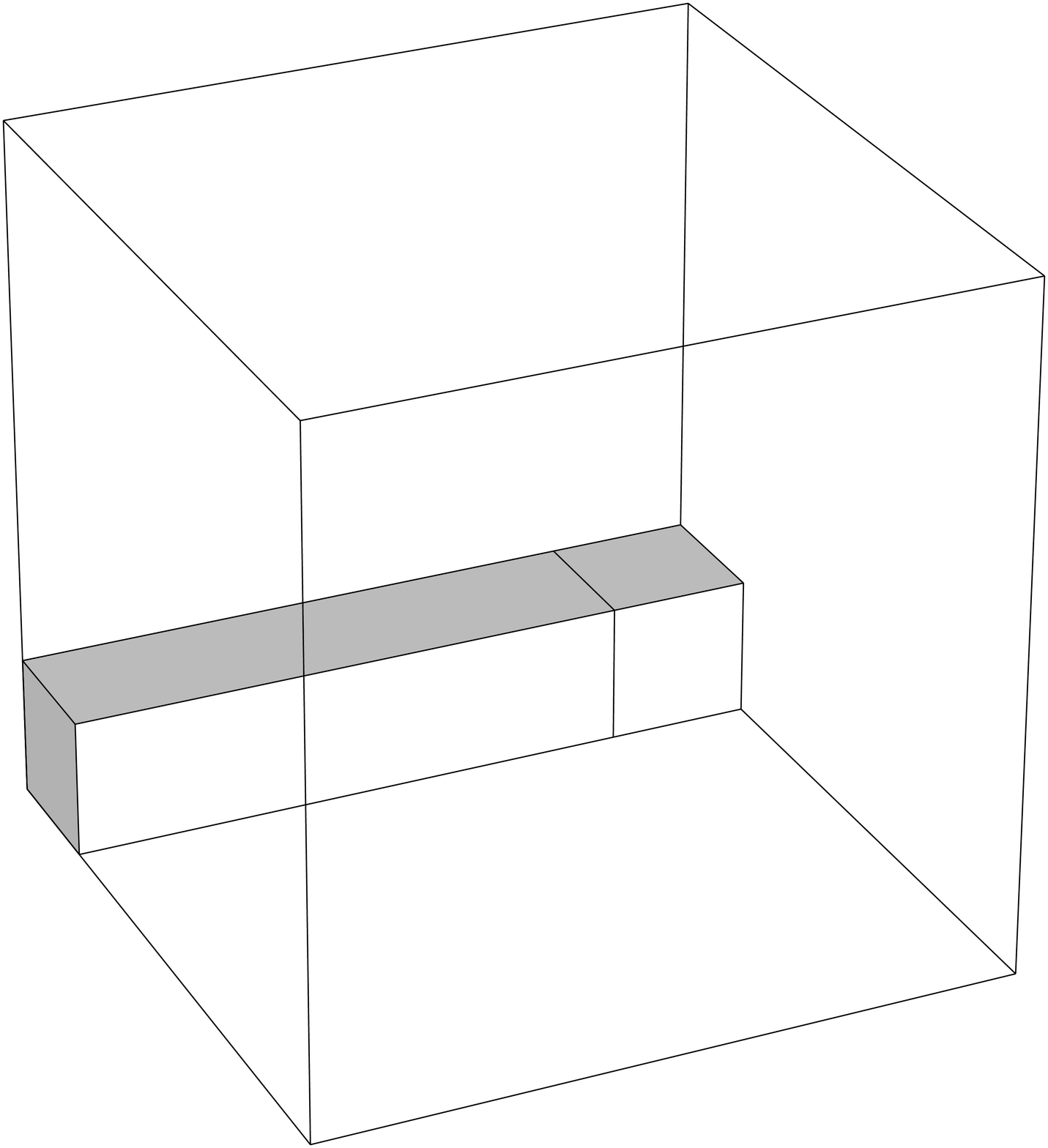}
}
\hfil
\subfloat[]{
  \includegraphics[width=0.2\columnwidth]{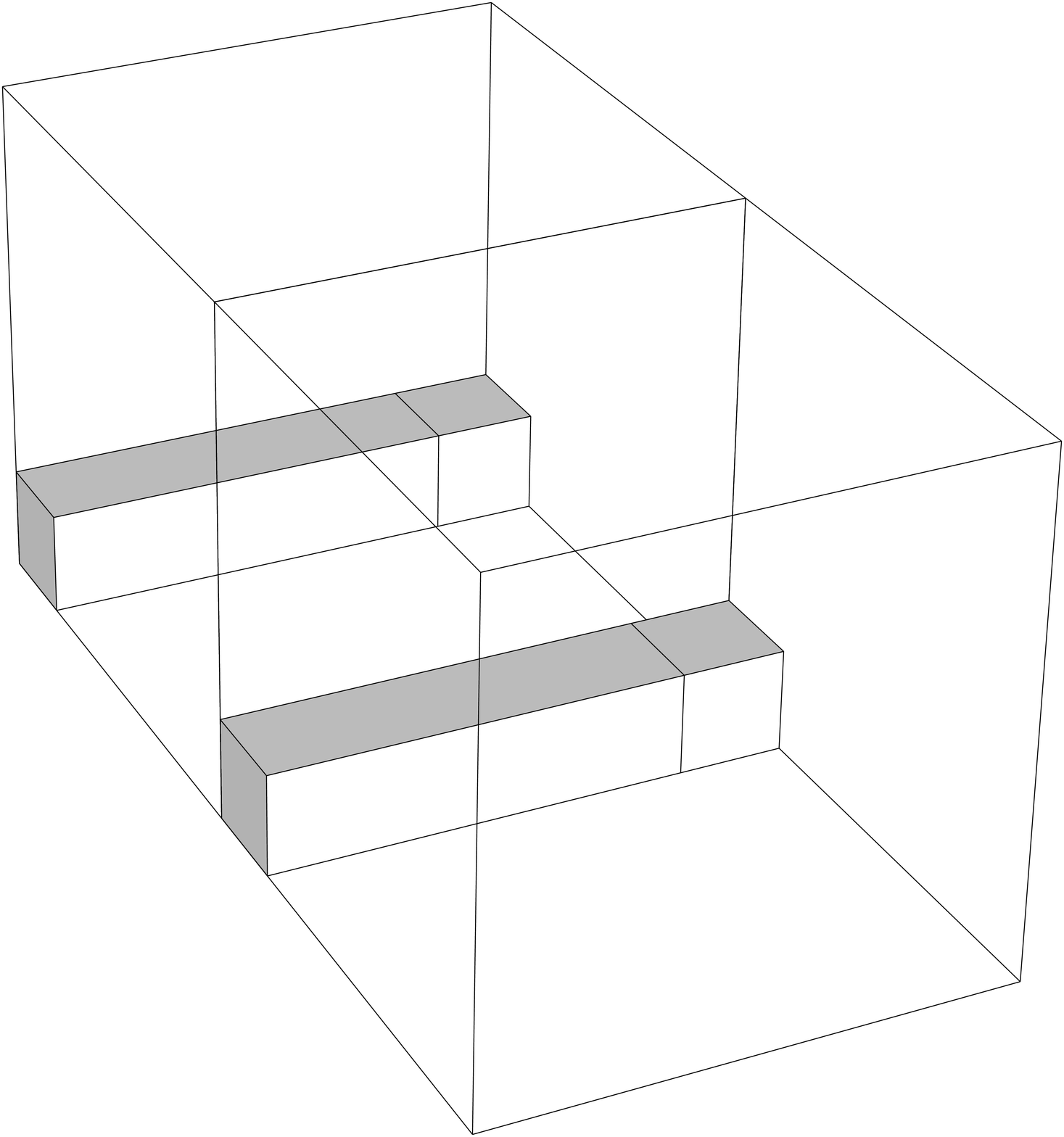}
}
\\
\subfloat[]{
  \includegraphics[width=0.2\columnwidth]{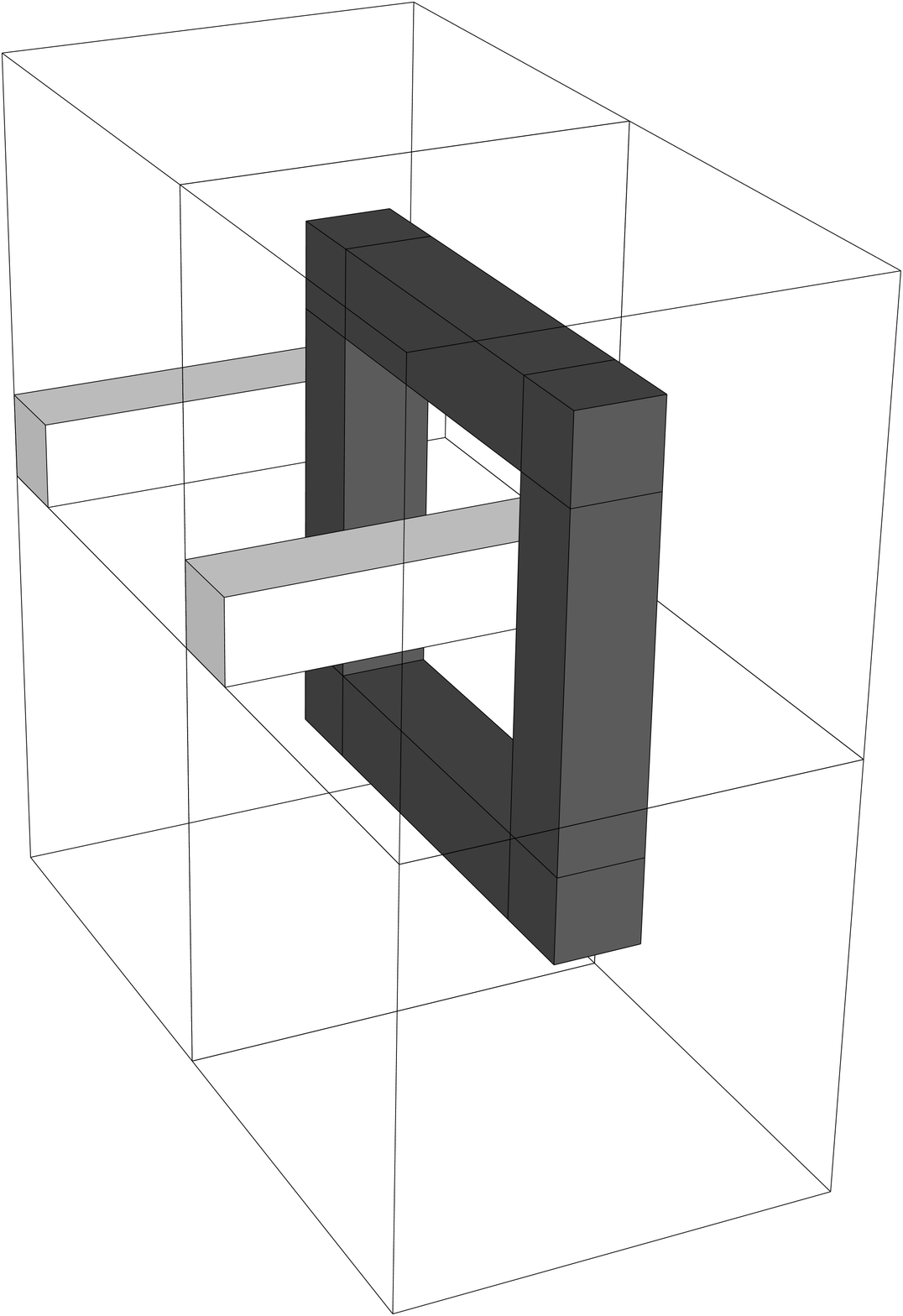}
}
\hfil
\subfloat[]{
  \includegraphics[width=0.2\columnwidth]{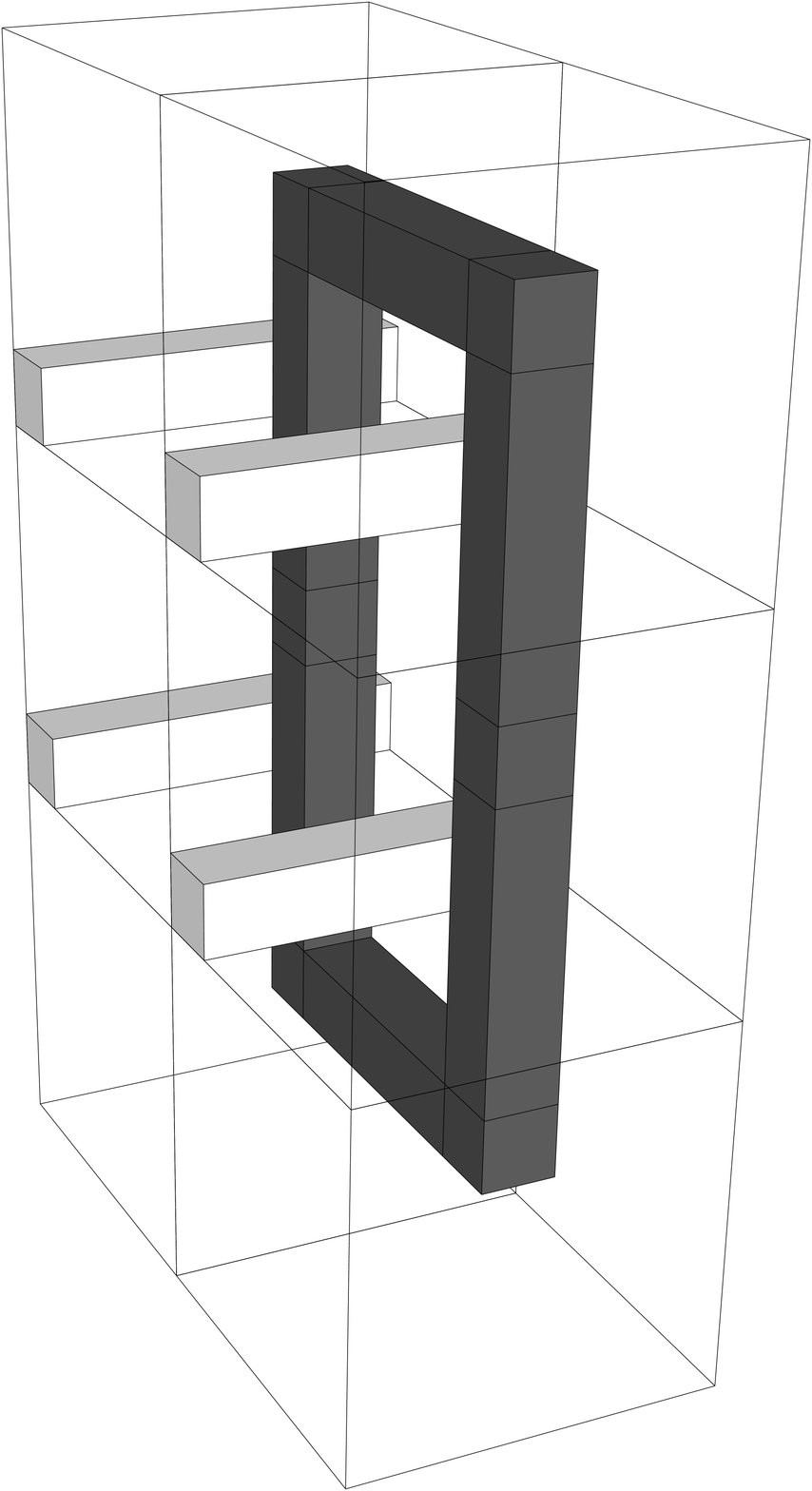}
}
\\
\subfloat[]{
  \includegraphics[width=0.2\columnwidth]{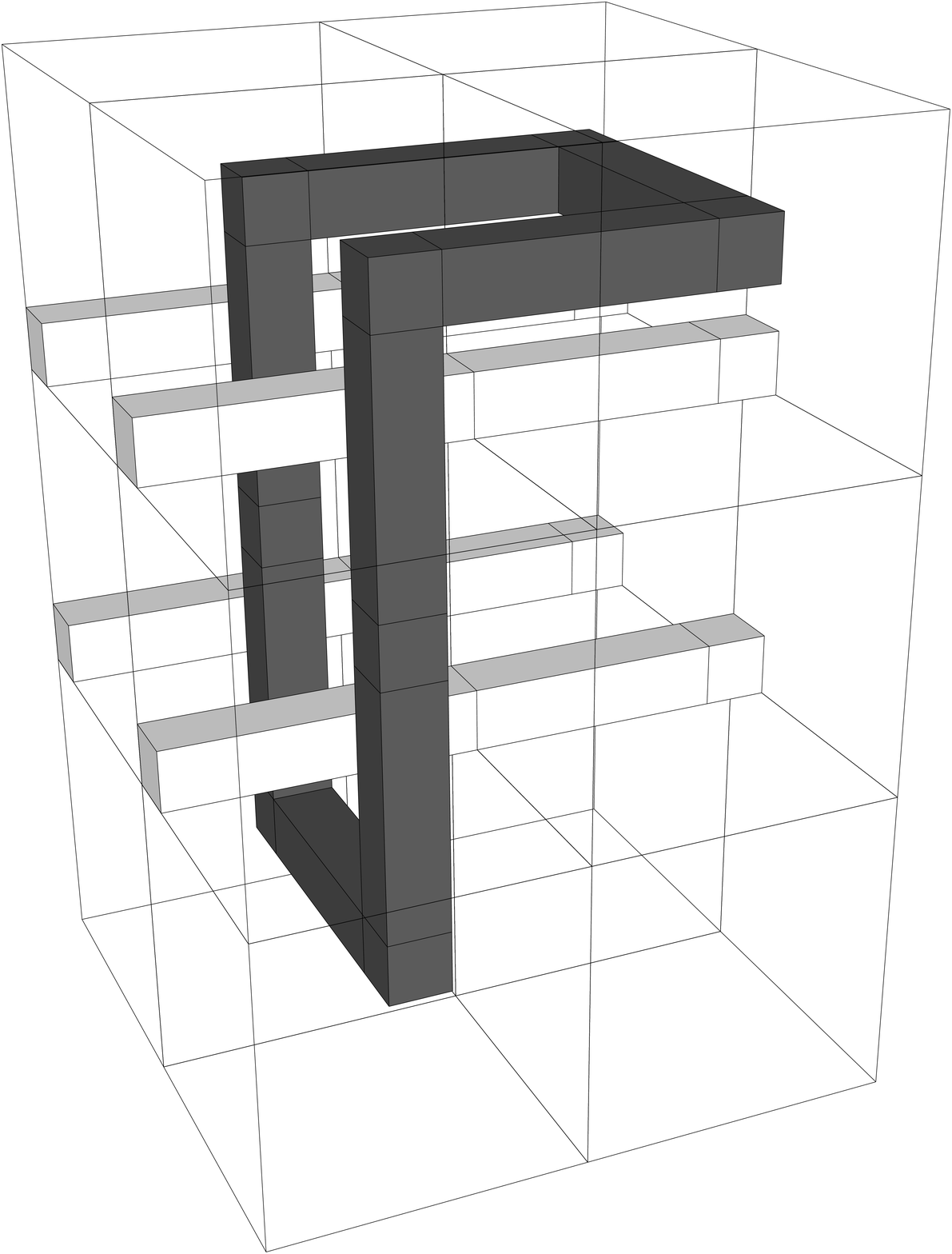}
}
\hfil
\subfloat[]{
  \includegraphics[width=0.2\columnwidth]{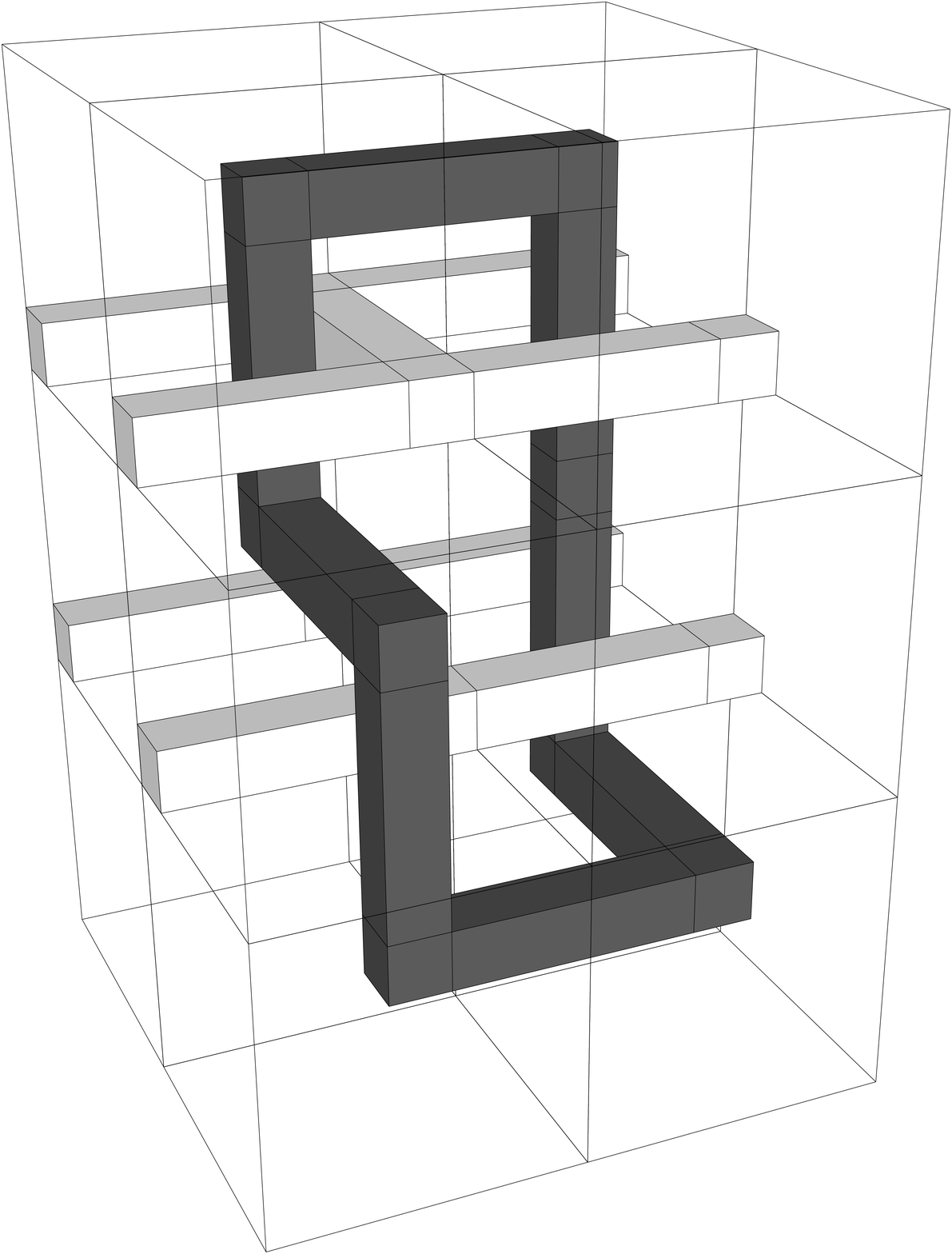}
}
\caption{A volume unit consists of $5^3$ primal cubes (very light gray): a) a five boxes long defect occupies a volume unit; b) two primal defects require two volume units; c) a dual circular defect (dark gray) around a single primal defect; d) a dual circular defect around two primal defects; d) the necessary volume is doubled after deforming the dual defect and extending the primal defects; e) twelve volume units are required for the bridged CNOT.}
\label{fig:volume}
\end{figure}

Apart from the previous discussion, the minimal distance (three) is achieved when the diameter is one unit cell wide, and each defect is two unit cells apart from another. Therefore, the definition of a logical qubit requires at least four unit cells from the i,j-plane of a three dimensional lattice. Considering the configurable geometries at input/outputs (Sec.~\ref{sec:pins}), the minimal length of a defect connecting an input to an output will be four unit cells long. However, such a construction is not fault-tolerant \cite{fowler2009topological}, because errors on the qubits measured in the Z basis cannot be detected and could propagate in an uncontrolled manner (\emph{cascade}). Fault-tolerance is achieved after increasing the defect diameters and using the error-correction method presented in \cite{fowler2009topological}: consider the chain of pairs of neighbouring five-sided dual unit cells sharing Z-measured qubits nearest to the defect border. Thus, the minimal defect diameter required to implement fault-tolerant TQEC is of two unit cells ($2\times 2$ unit cells, $d_f=2$). The code distance will be $d=8$, requiring at least eight unit cells between any two defects ($d_d=9$). Thus, the $1/4$ ratio between diameter and distance is again obtained.

It is simpler to use a volume measure which is code distance independent, instead of counting the number of lattice unit cells required to execute a geometrically described TQEC computation. For this reason, a cubic \emph{volume unit}, a standardised lattice with fixed dimensions, is introduced (Fig.~\ref{fig:volume}). The volume unit is obtained by tiling cubes instead of unit cells, where a cube with sides of length $d$ contains $d^3$ primal unit cells. Therefore, the complete lattice required for executing a TQEC circuit is the result of tiling volume units, which contain an exact number of cubes, which in turn are three dimensional tilings of unit cells. Volume units and cubes are intermediate abstractions between a complete lattice and unit cells.

A volume unit contains $(1+4)^3=5^3$ cubes having the necessary diameter for defect construction so that distance $4d$ error-correction is possible. Cubes are formed of primal unit cells but, the definition could be extended, after observing that a cube of dual unit cells is obtained in the middle of a $2 \times 2 \times 2$ cube arrangement. Tiling cubes has the same effect as tiling unit cells (Fig.~\ref{fig:cell}).

In a volume unit there are enough cubes so that a primal defect having three non-parallel segments could be defined at the margins. Two volume units are the minimum for defining two defects parallel in the same i,j plane. It is not possible to construct a circular dual defect around any of the defects without increasing the lattice to $1 \times 2 \times 2$ volume units. A dual defect around both primal defects requires a lattice of $1\times 2 \times 3$ volume units. Implementing the bridge CNOT (cf. Fig.~\ref{fig:bridgecnot} and \ref{fig:volume}) requires a lattice of 12 volume units due to the $2 \times 2 \times 3$ necessary arrangement.

As a conclusion, the geometric description captures the computation (ICM circuit) being implemented into TQEC circuits, and is independent of the code distance. However, in the previous sections it was always mentioned that segment end point coordinates are specified using unit cell coordinates. For geometries implemented using a distance four error-correction this is true, as each defect is only a single unit cell wide. For codes with larger distances, those definitions do not hold for unit cells. However, if one replaces every occurrence of unit cell with the term cube, the geometric description is valid for arbitrary error-correction procedures using topological cluster states. A single aspect remains to be discussed: the effect of code distance scaling on defect construction. For the distance four code, a cube includes a single unit cell. Scaling the code distance by two, a cube will contain eight unit cells, and each cube face will consist of four unit cell faces. Therefore, when a segment end point is specified having the coordinate of cube and code distance is scaled by two, four unit cell face qubits from the same cube face are measured in the Z basis.

\section*{Appendix: Using a Geometrical Description}
\label{apx:use}

The geometric description of a synthesised TQEC circuit is a high-level description of the operations a fault-tolerant quantum computer needs to execute. This section explains why the i- and j-axis coordinates can be understood as hardware entity addresses, and the t-axis coordinate as the time when the entities are to be used. The execution of a geometrically described computation starts by mapping the description to an empty lattice \cite{paler2014mapping}, determining the bases of physical qubit measurement bases and then operating the quantum hardware. There are multiple justifications why the latter operation is performed by gradually constructing and measuring the lattice. 

The first explanation is related to the complexity of TQEC circuits. Circuit depth is a measure of complexity for classical and quantum circuits, and it represents the longest path, expressed usually in terms of number of gates, between any input and output of the circuit. The depth of a geometry is not considered an appropriate measure of the underlying ICM circuit complexity. This is because, ICM circuits are Clifford+T \cite{paler2015fully} instances, and the relevant complexity in executing such circuits is generated by the number of T gates. It could be argued that the depth of an ICM circuit can be defined as the maximum number of T gates between an input and an output. The T gates on any circuit path introduce a temporal ordering: the sequence in which gates have to be executed. In terms of ICM circuit, this would correspond to an ordered sequence of (M)easurements because gates are implemented through teleportations; gate execution is finalised only after corresponding qubits were measured. Moreover, in ICM circuits, the T gate ordering is equivalent to the ordering of the probabilistic P gate corrections necessary for teleported T gates. If any of these corrections is left out, the executed computation is not correct. The correction mechanism was briefly mentioned in Sec.~\ref{sec:geomdesc} and illustrated using the circuits from Fig.~\ref{circ:ftcircs}.

TQEC geometric descriptions were synthesised so that inputs have the lowest and outputs have the highest t-axis coordinate, while braids and measurements have intermediary coordinates. For example, in Fig.~\ref{fig:toff0}, the measurements corresponding to selective source and destination subcircuits follow on the t-axis immediately after the braids. It can be also observed that the ordering of the measurements has a geometric representation: for the gates on the same wire in Fig.~\ref{circ:toffoli}, the corresponding measurements have increasing t-and j-axis coordinates. The j-coordinate increases because of the additional ancillae introduced during ICM transformation, while the t-coordinate is just a reflection of the gate ordering.

The second justification for which the lattice is gradually constructed and measured is related to a quantum hardware optimisation problem. If the complete lattice would be constructed and measured afterwards, a very large number of quantum hardware entities corresponding to physical qubits would be required. Anyway, as previously explained, there is an order in which the measurements have to be executed, meaning that on average each hardware entity would be required to maintain the states of the physical qubits for a long time. This requirement has a threefold problematic: firstly, it is very difficult to build hardware able to correctly maintain states for long times; secondly, even if the hardware would support this, the cost of building such a high number of entities would be prohibitive; and thirdly why should the physical qubits be initiated that soon if their states are unused for a long time?

\begin{figure}[t!]
\centering
\subfloat[]{
  \includegraphics[width=0.3\columnwidth]{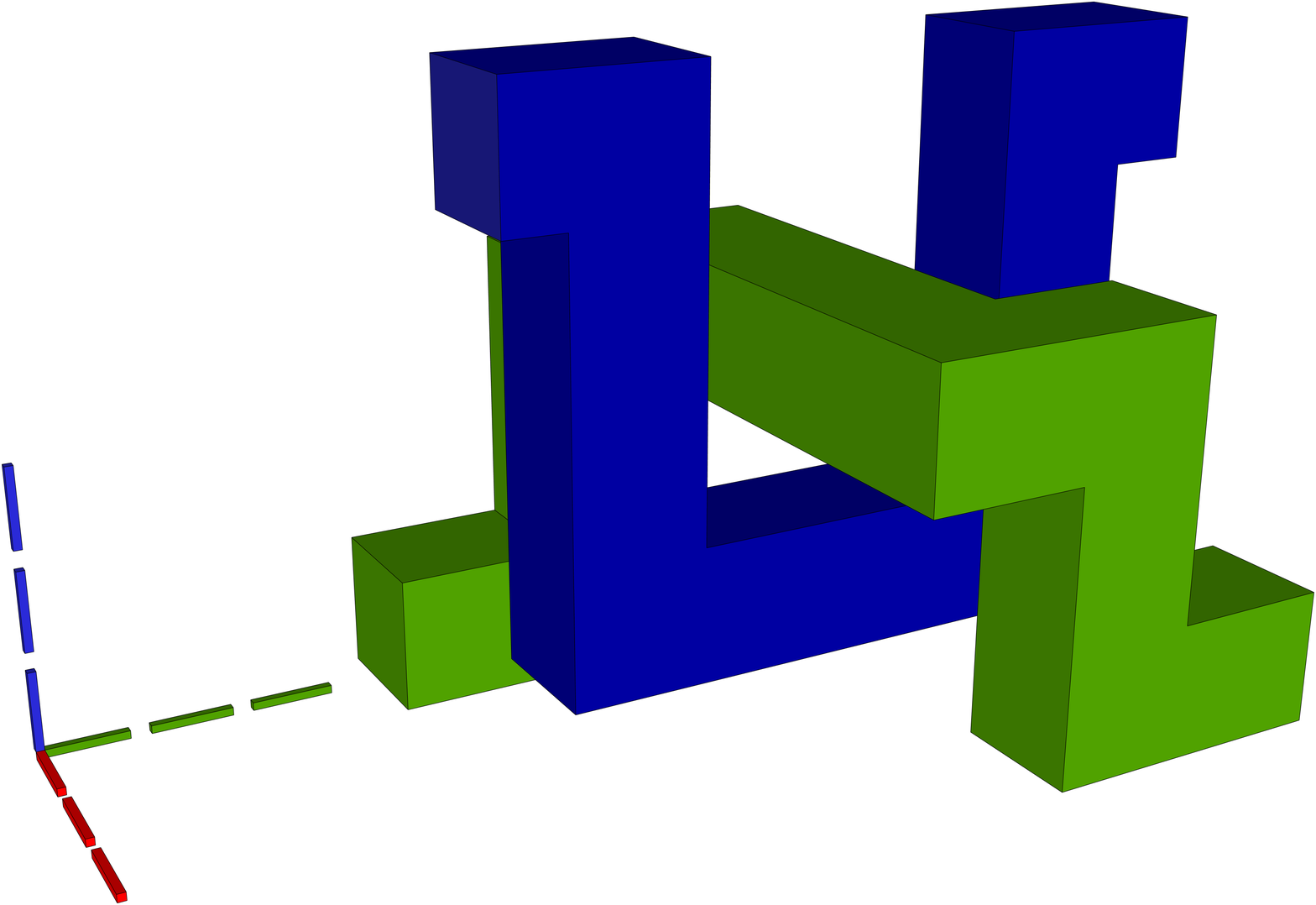}
}
\hfil
\subfloat[]{
  \includegraphics[width=0.3\columnwidth]{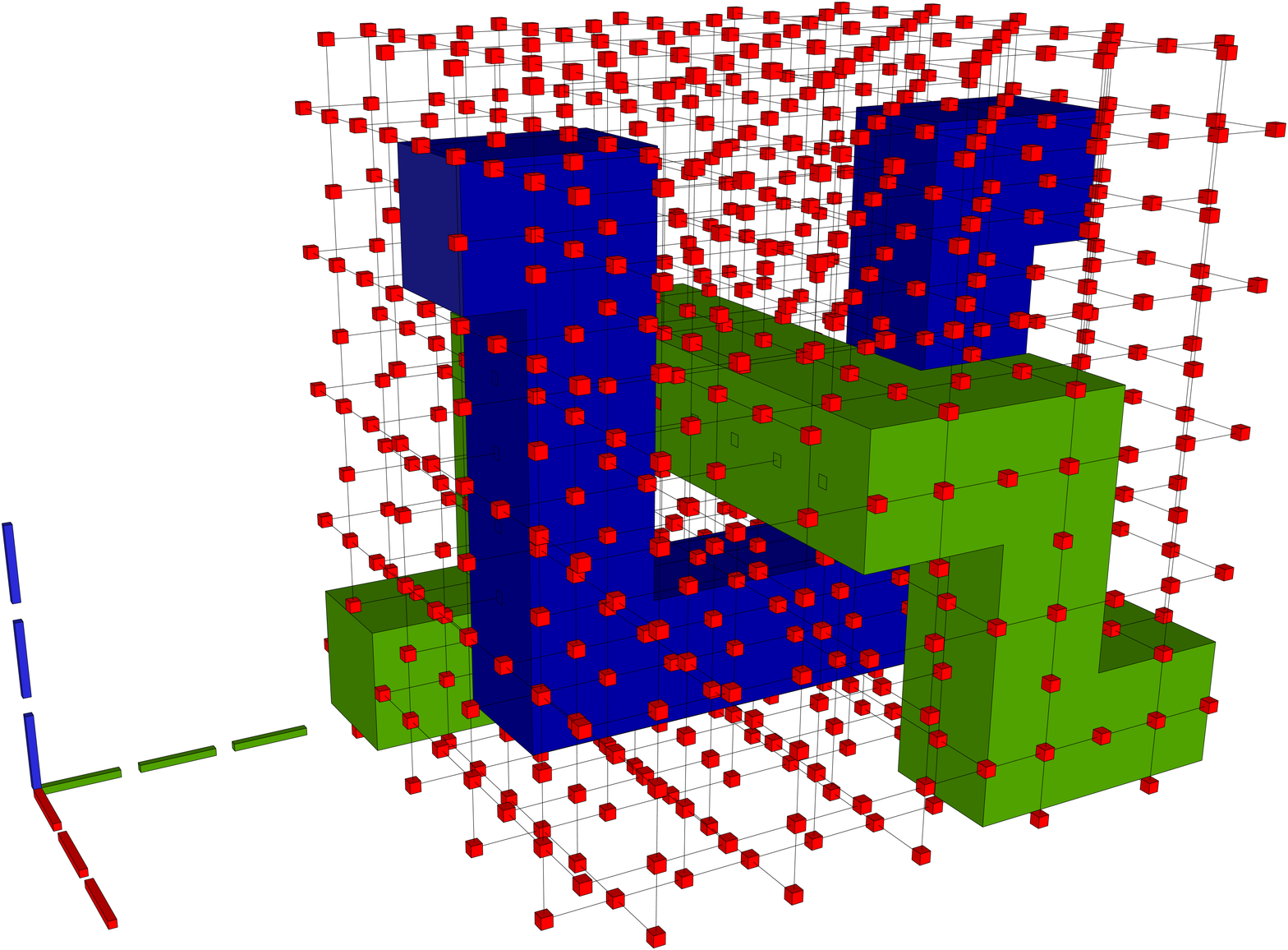}
}
\hfil
\subfloat[]{
  \includegraphics[width=0.3\columnwidth]{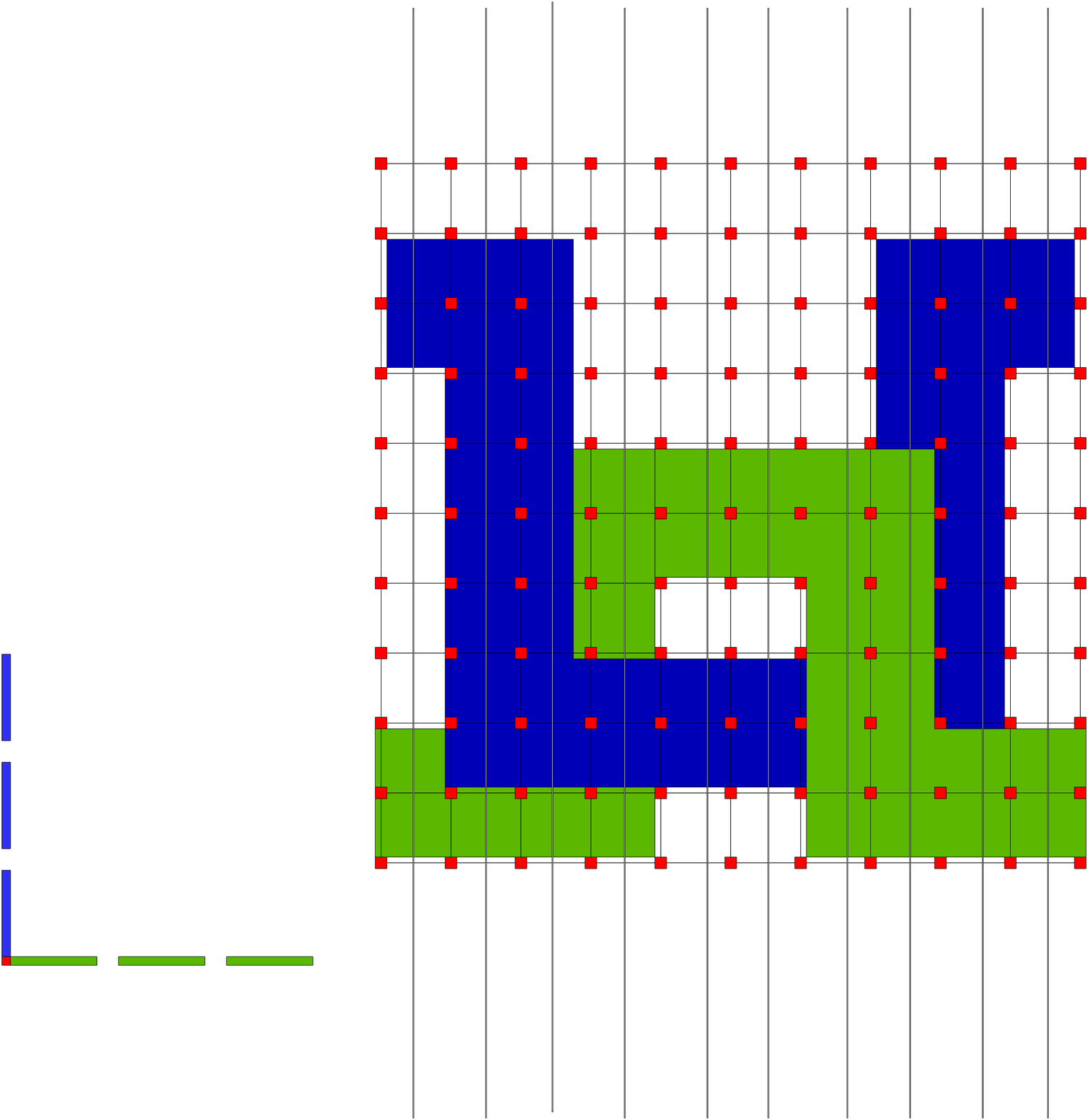}
}
\\
\subfloat[]{
  \label{fig:sliceprimal}
  \includegraphics[width=0.15\columnwidth]{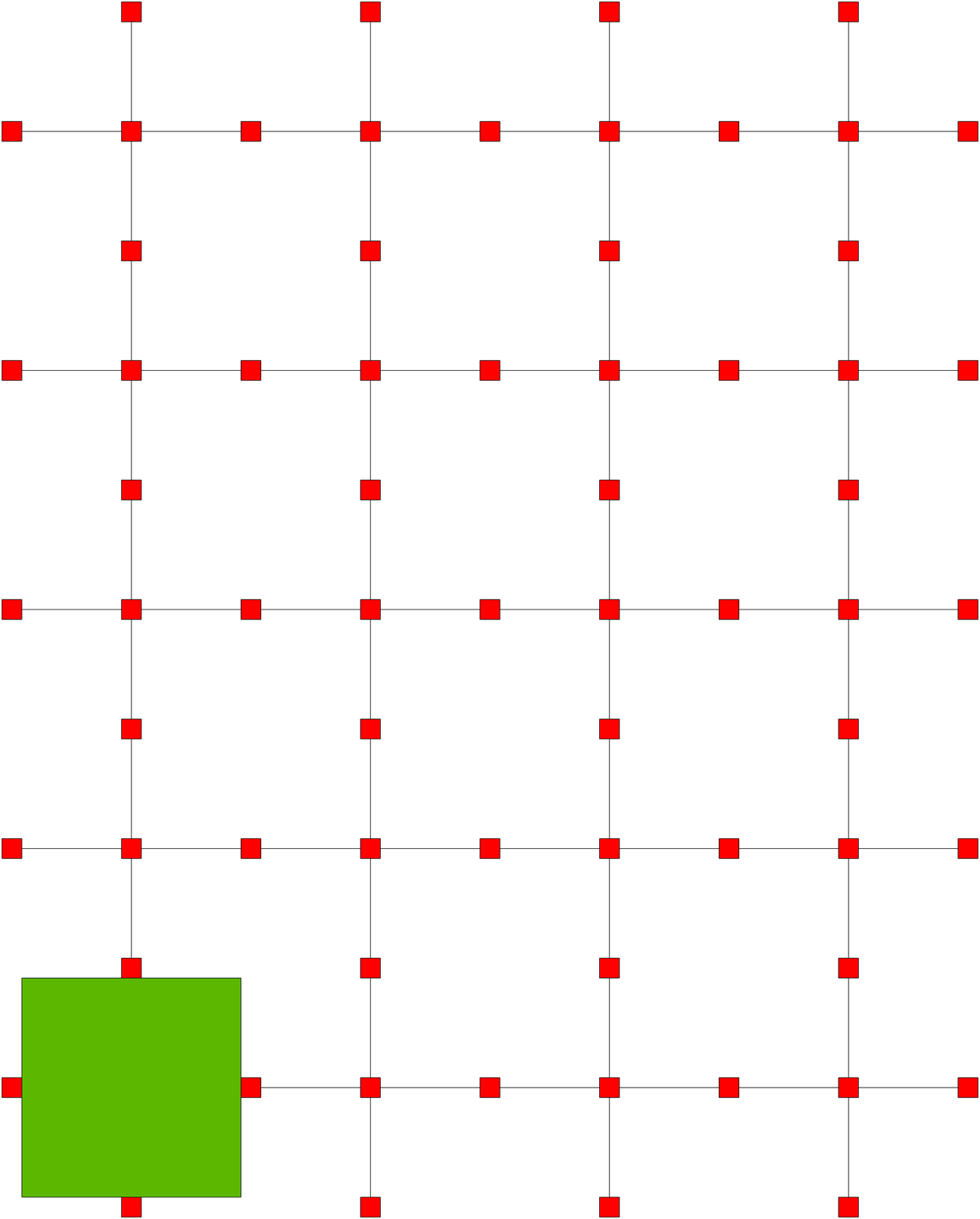}
}
\hfil
\subfloat[]{
  \label{fig:slicedual}
  \includegraphics[width=0.15\columnwidth]{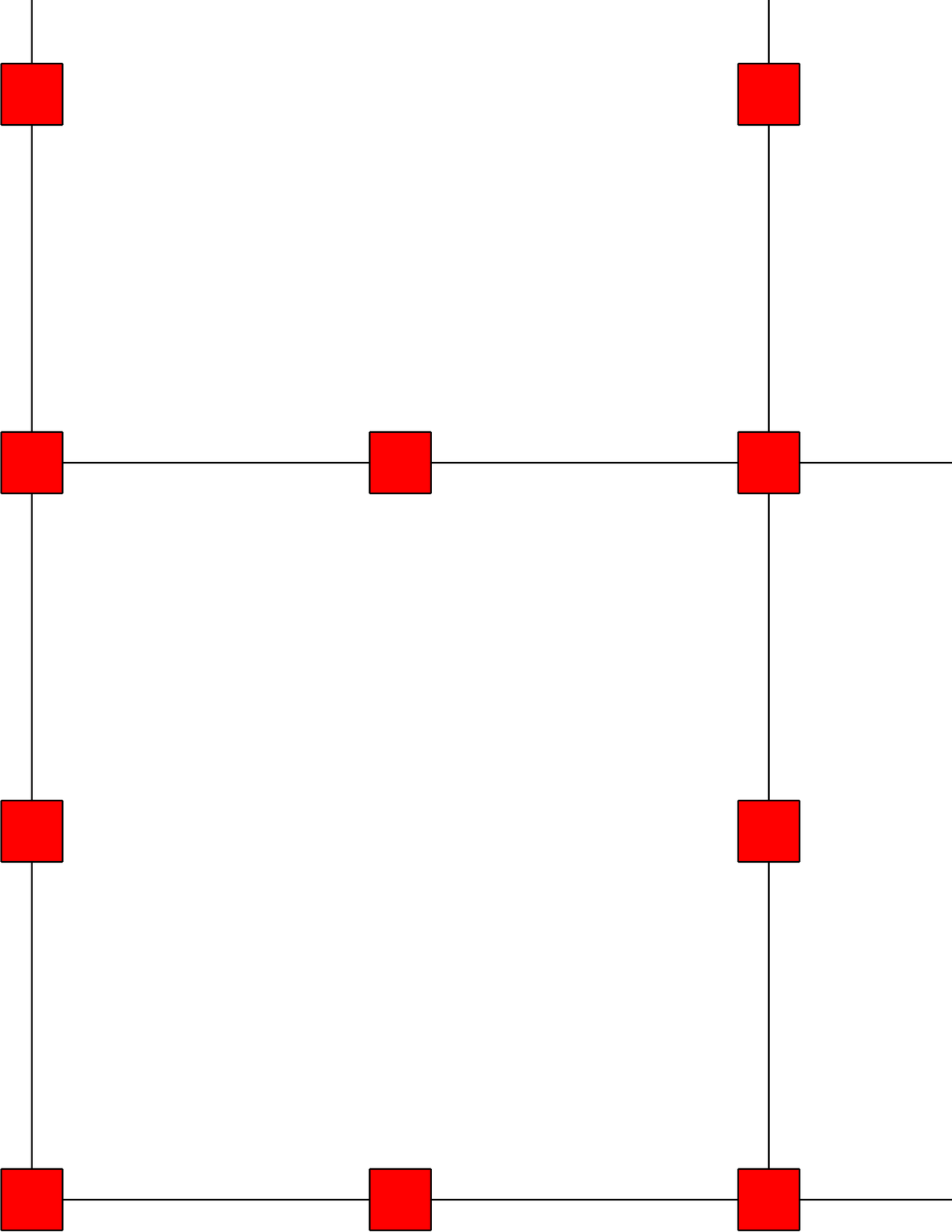}
}
\hfil
\subfloat[]{
  \includegraphics[width=0.15\columnwidth]{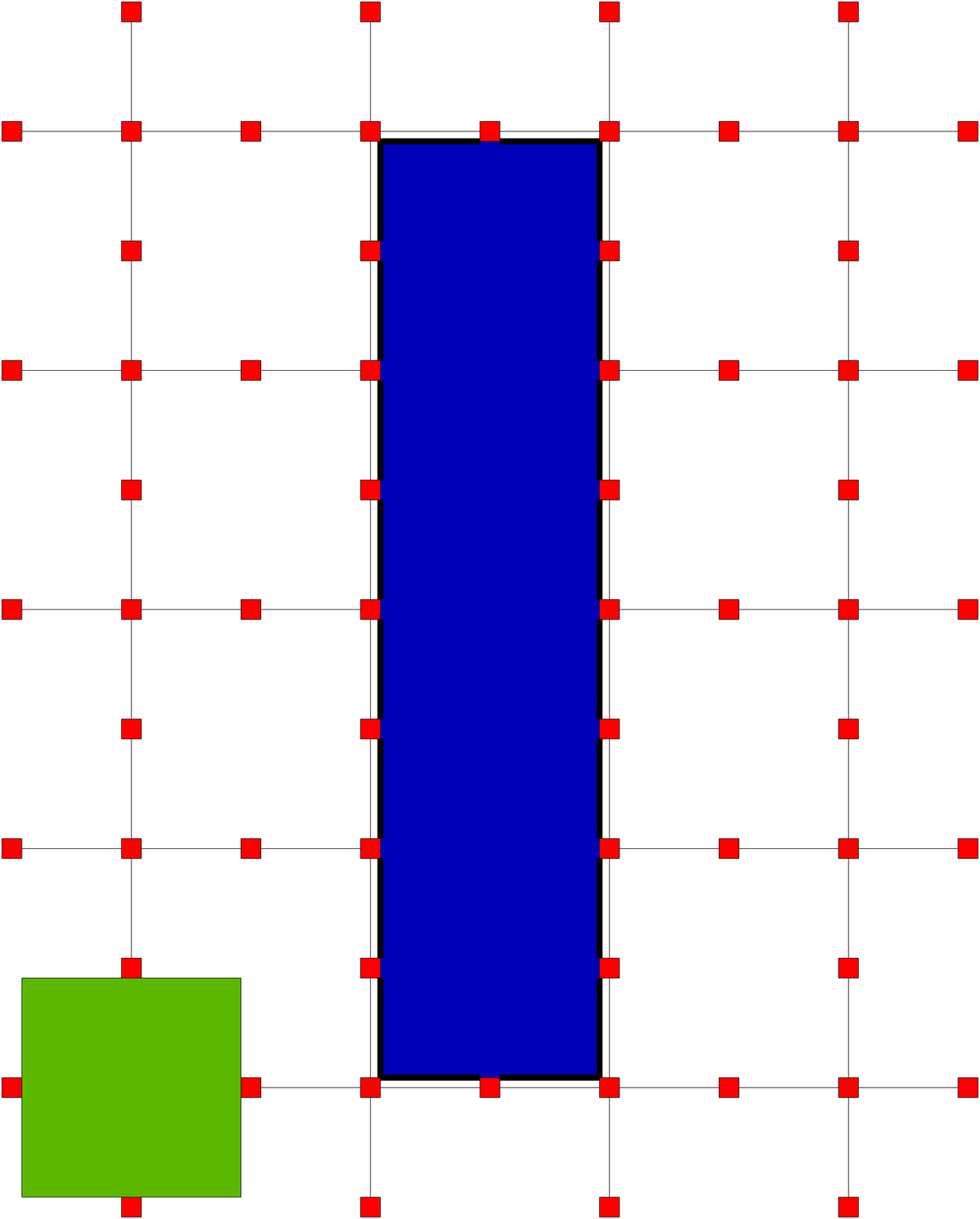}
}
\\
\subfloat[]{
  \includegraphics[width=0.15\columnwidth]{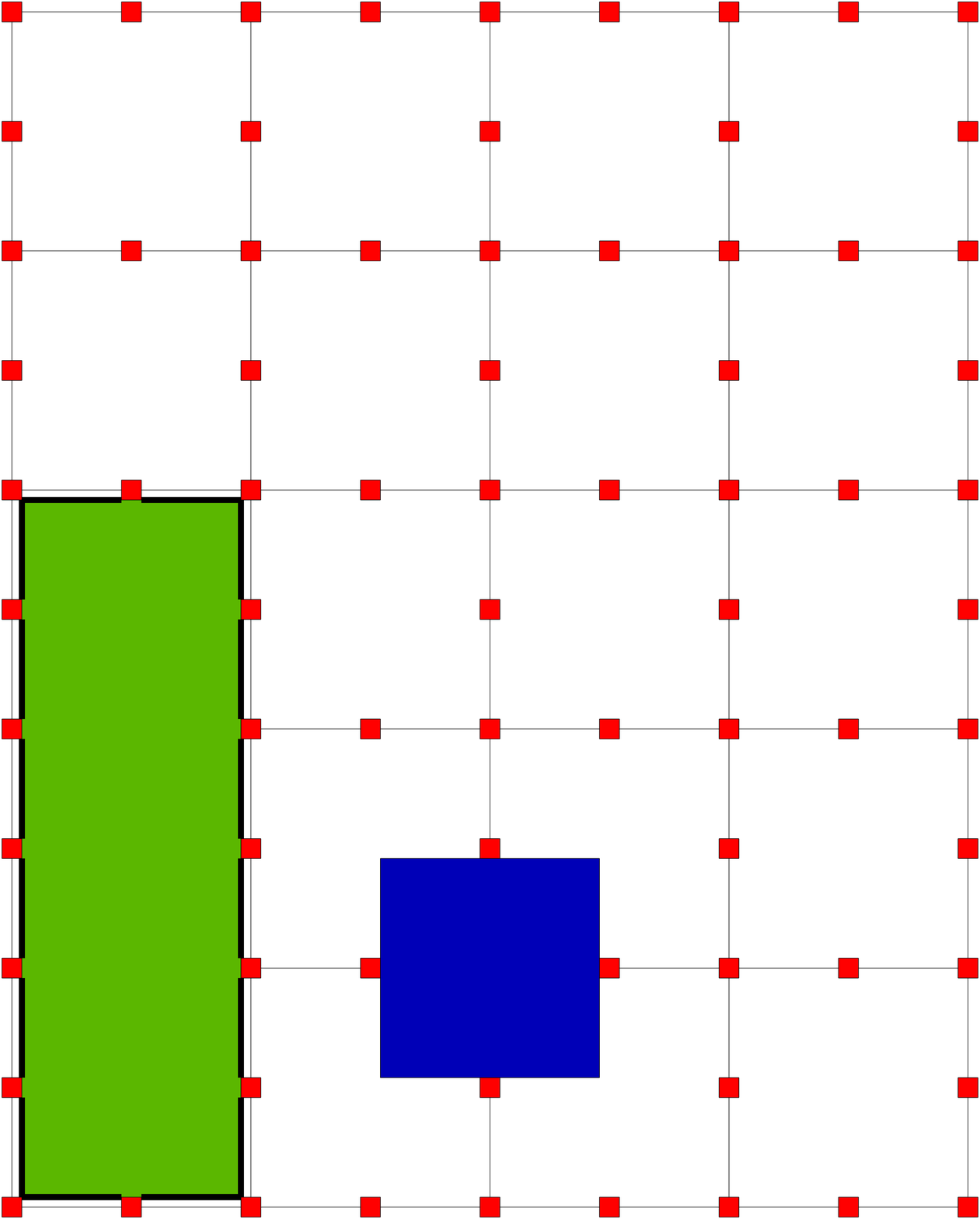}
}
\hfil
\subfloat[]{
  \includegraphics[width=0.15\columnwidth]{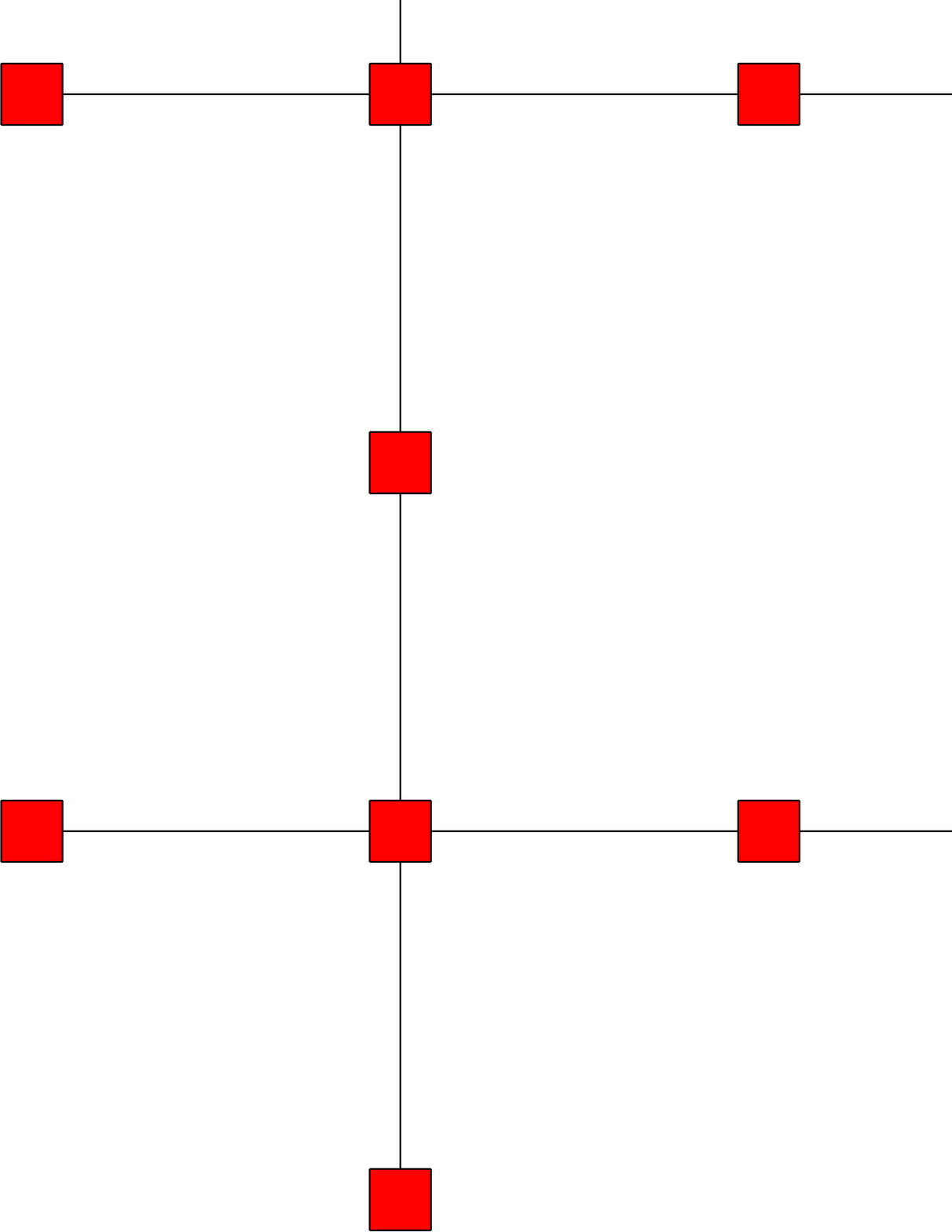}
}
\hfil
\subfloat[]{
  \includegraphics[width=0.15\columnwidth]{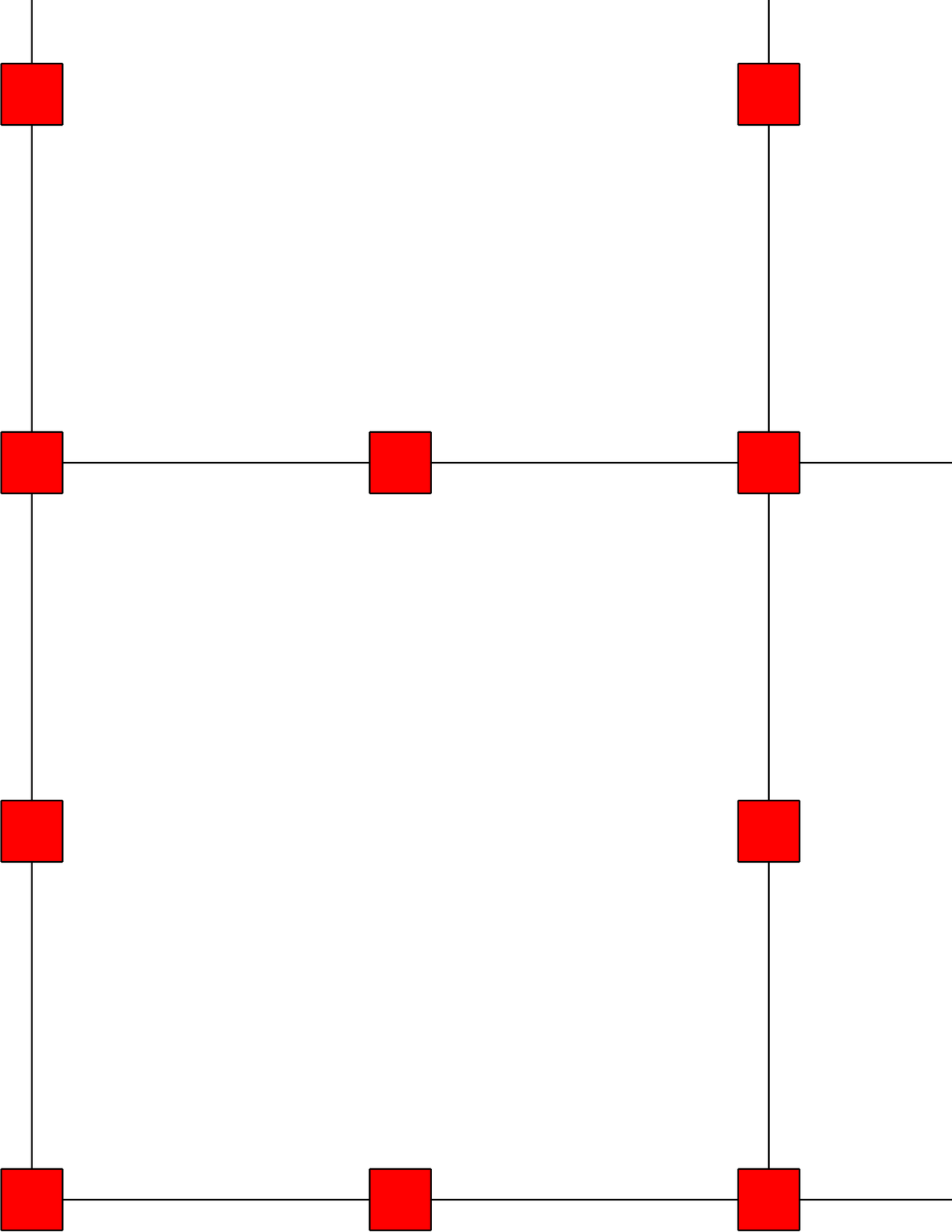}
}
\\
\subfloat[]{
  \includegraphics[width=0.15\columnwidth]{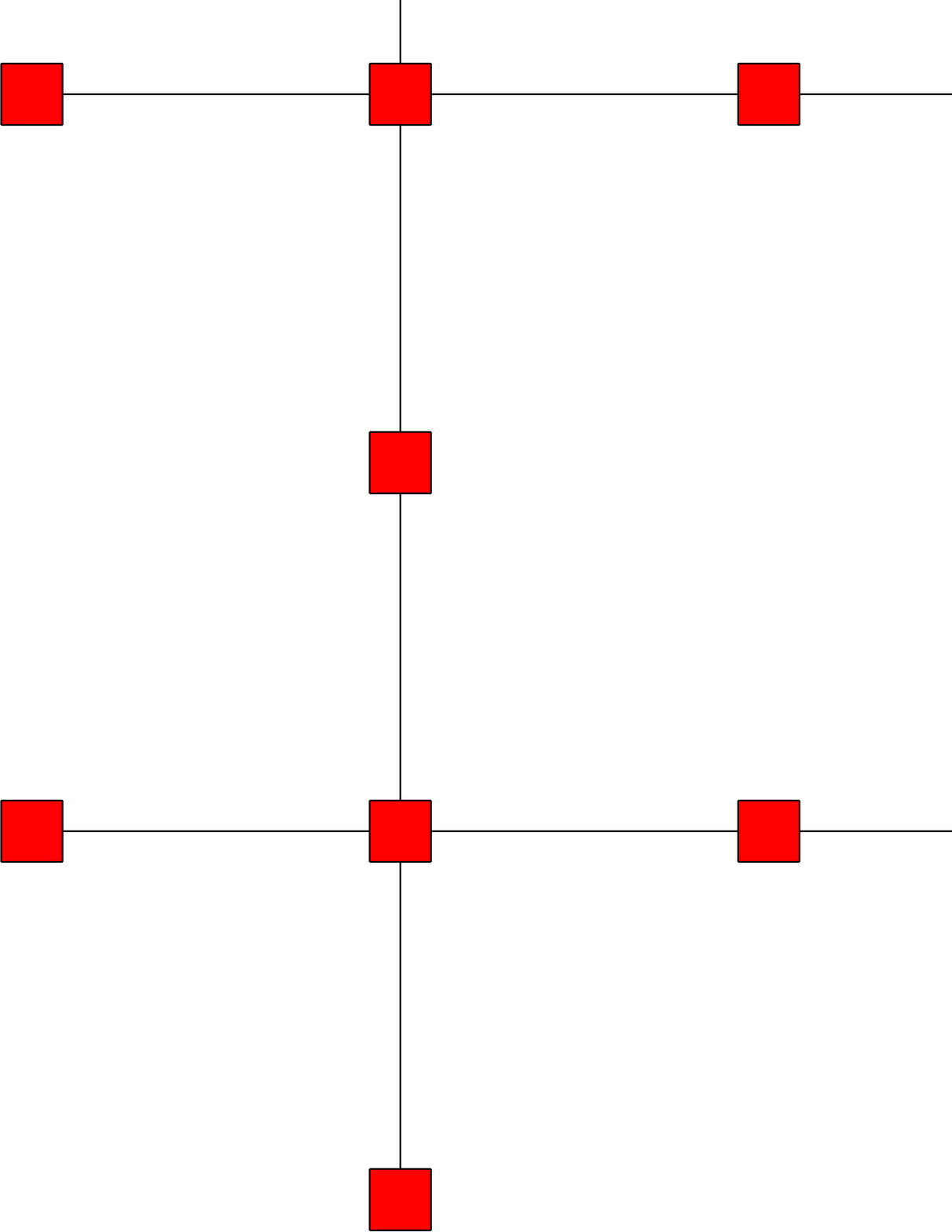}
}
\hfil
\subfloat[]{
  \includegraphics[width=0.15\columnwidth]{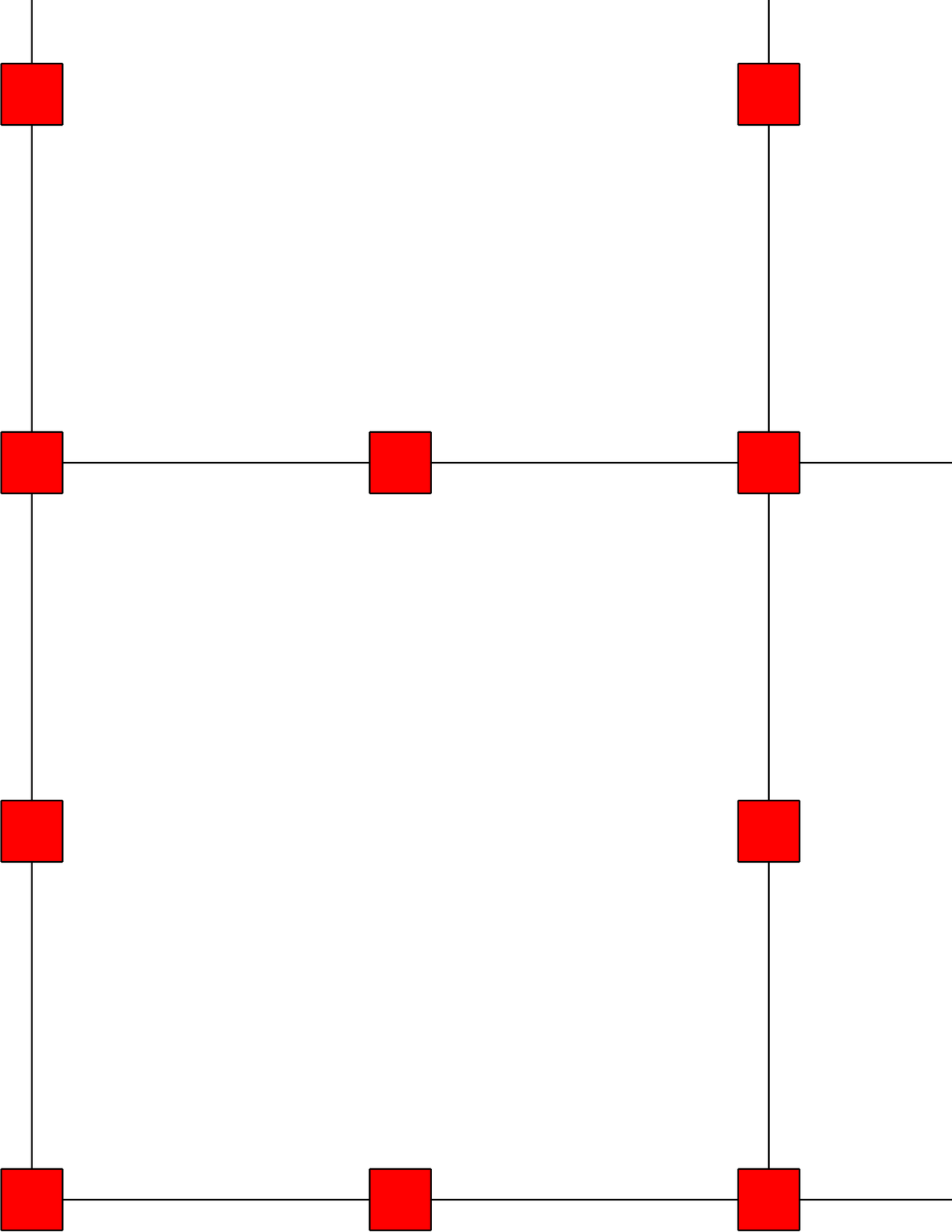}
}
\hfil
\subfloat[]{
  \includegraphics[width=0.15\columnwidth]{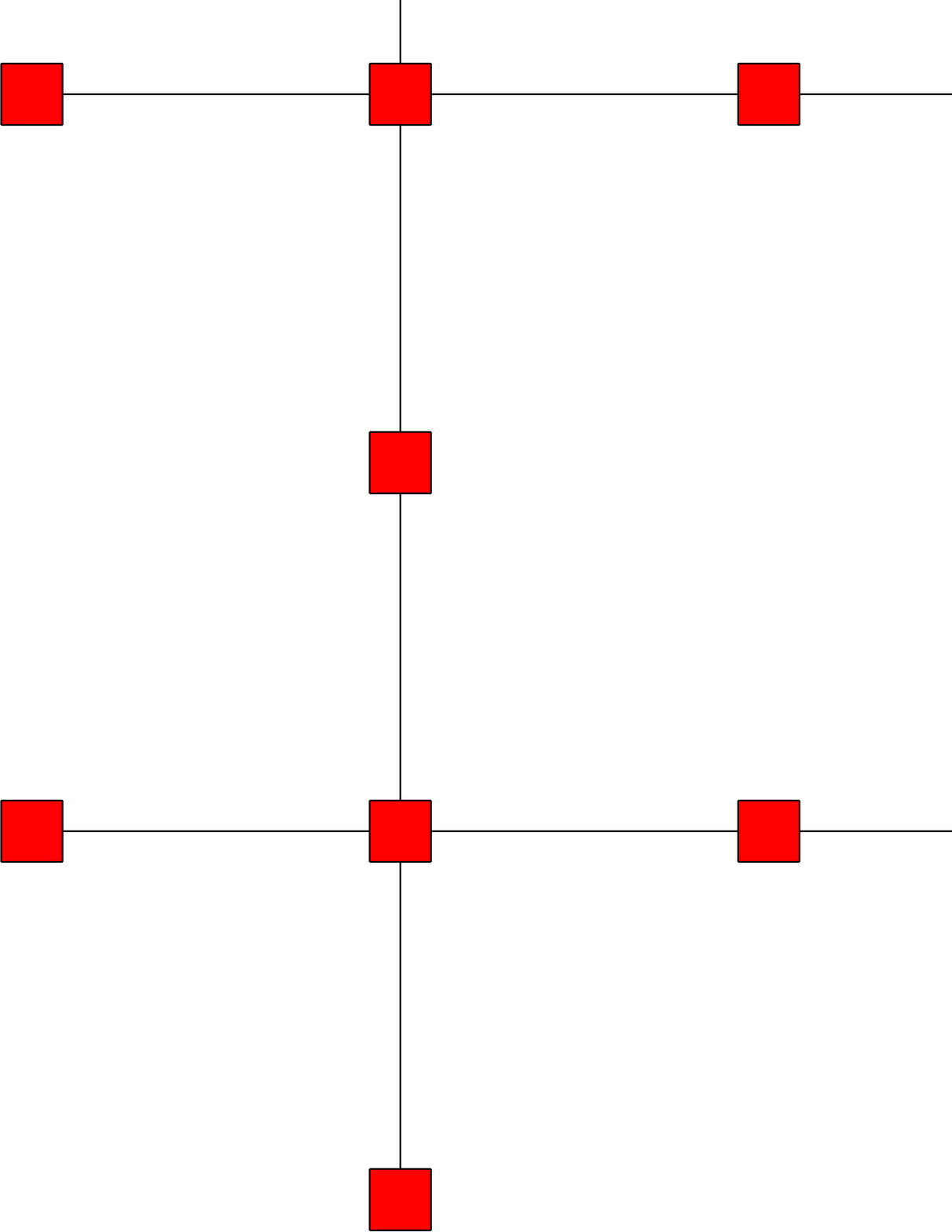}
}
\\
\subfloat[]{
  \includegraphics[width=0.15\columnwidth]{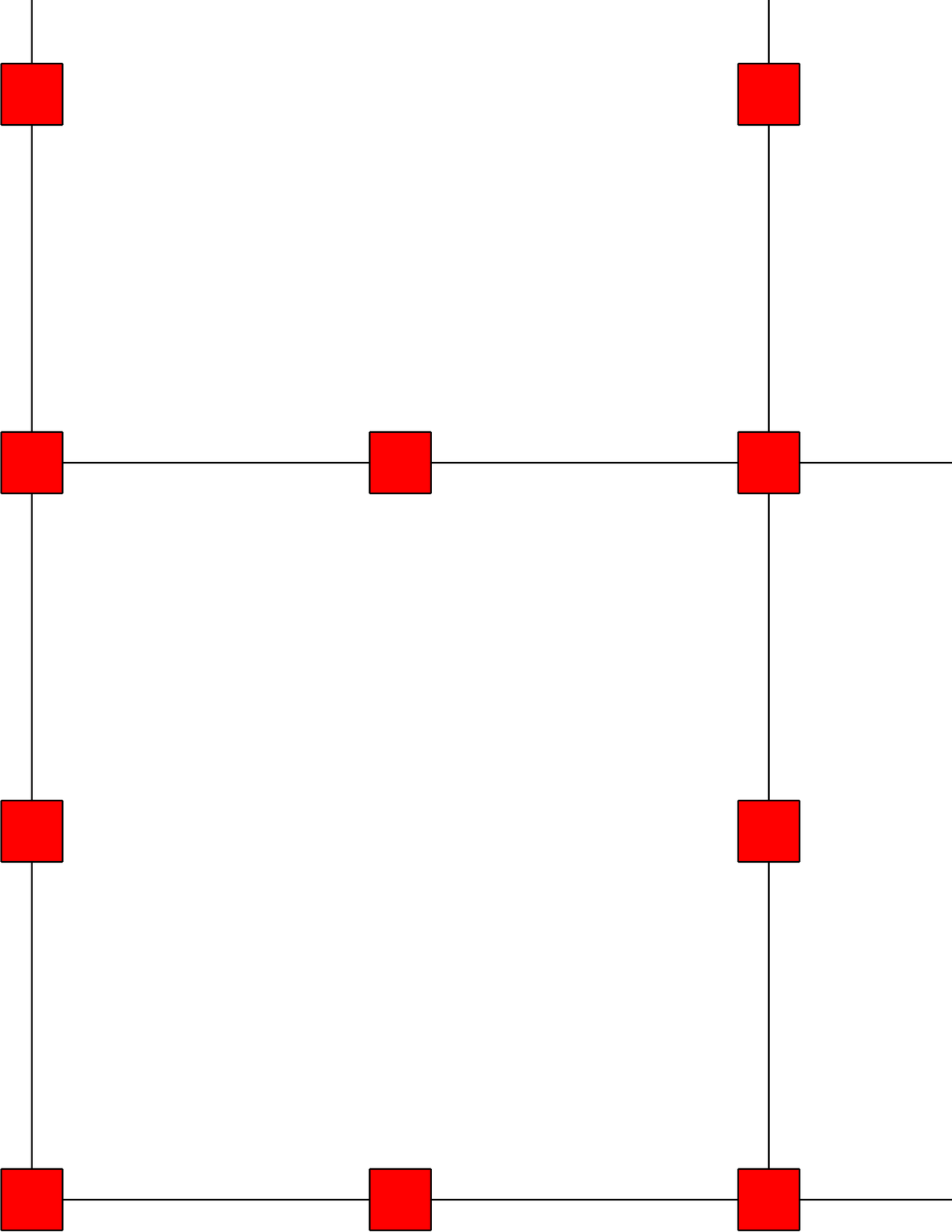}
}
\hfil
\subfloat[]{
  \includegraphics[width=0.15\columnwidth]{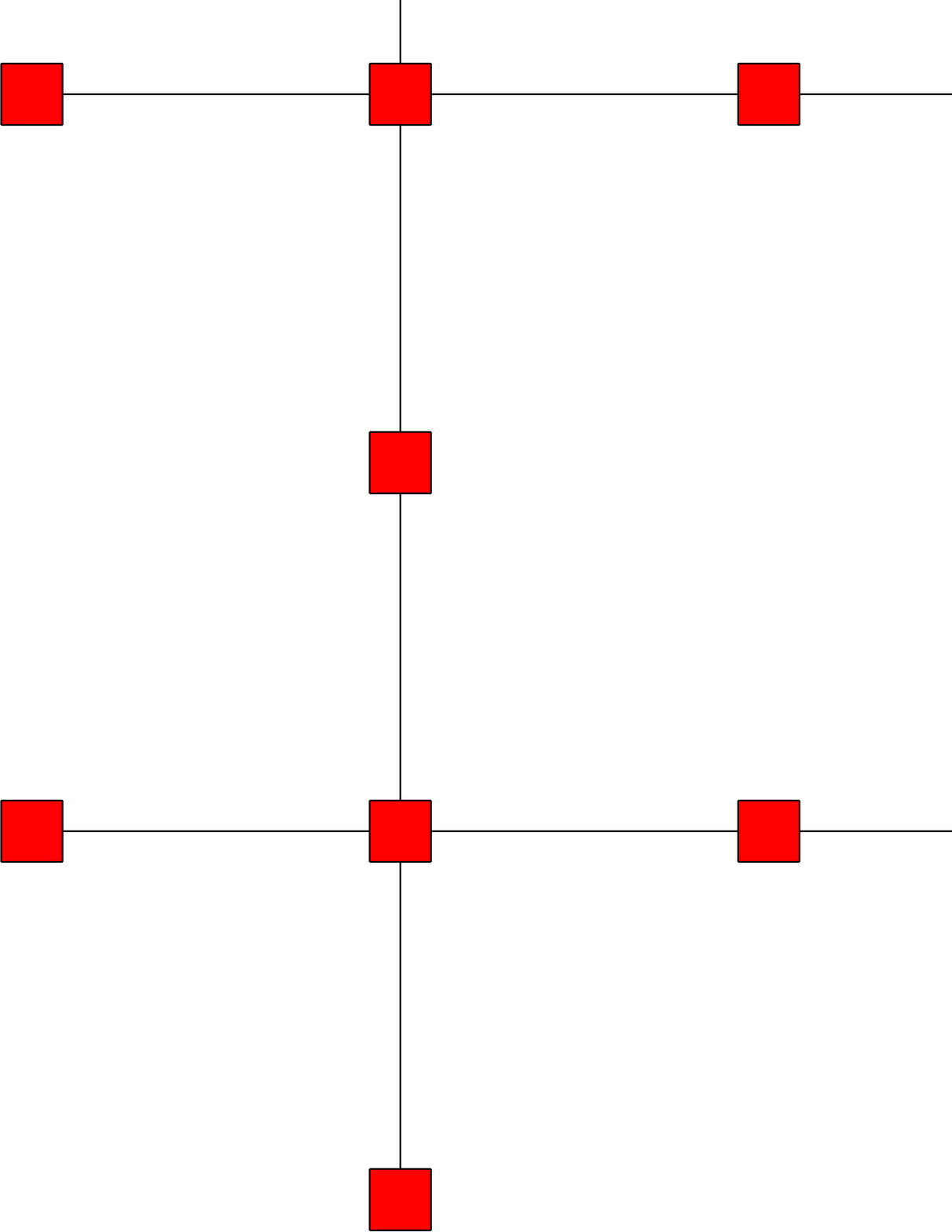}
}
\caption{A primal defect (green) braided with a dual defect (blue): a) defect  structure; b) defects superimposed on a lattice; c) j-axis view of the lattice having the slice positions indicated with parallel vertical lines; d-n) sequence of lattice slices.}
\label{fig:slices}
\end{figure}

The solution to the above problem uses a layered approach: the geometry is sliced along the t-axis into a sequence of primal and dual layers (Fig.~\ref{fig:slices}), and the number of required hardware entities is trimmed to just the one necessary to manipulate states from two subsequent layers. The highly regular structure of the lattice offers the support for obtaining the slices. Each logical qubit measurement is specified in terms of physical qubit coordinates (Sec.~\ref{sec:geomdesc}): the i- and j- axes coordinates are quantum hardware entity addresses and the t-axis coordinate indicates the time when the hardware is initialised, entangled with neighbouring entities and measured.

\begin{figure}[t!]
\centering
\includegraphics[width=0.45\columnwidth]{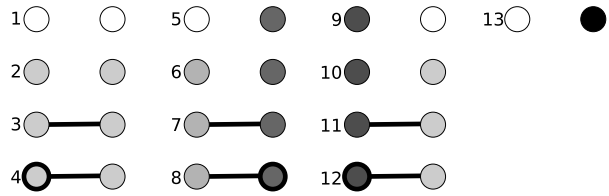}
\caption{Implementing an arbitrary rotation gate decomposed into three single qubit rotated measurements using two qubits. Colour coding: white - uninitialised state; black - output state; very light gray - initialised qubit; intermediate levels of gray - intermediate output states obtained after each rotated measurement of one of the qubits; measured qubit - thick black outline. At time step 1 both qubits are not initialised. At time steps 2 and 3, the qubits are initialised and entangled. The first measurement is performed on the leftmost qubit at step 4. The sequence is repeated again twice, until at time 13 the rightmost qubit has the output state of the rotation gate.}
\label{fig:graphm}
\end{figure}

Constructing and measuring the layers according to their t-axis coordinate will effectively enforce the ordering of the gates from the ICM circuit. The suitability of the layered approach is based on the observation that the execution of a TQEC circuit encoded into a topological cluster state is the execution of a measurement-based quantum computation \cite{childs2005unified}, where the measurement basis of individual physical qubits is a function of the computation implemented at the higher level (logical layer). The method is similar to how a linear graph state (e.g. Fig.~\ref{fig:graphm}) would be measured to implement a rotation gate: decompose the rotation in a time ordered measurement sequence (X,Z,X basis rotated measurements), and execute the measurement sequence.

Considering that a lattice is sliced then the number of primal layers is larger by one compared to the number of dual layers. Assuming that the layer sequence is $\{p_0,d_0,p_1,d_1,\ldots,d_{n-1},p_n\}$, a quantum computer executing a TQEC geometric description will execute the following steps: 1) initialise and entangle $p_0$ and $d_0$, for $i=0$; 2) measure the primal layer $p_i$; 3) initialise the primal layer $p_{i+1}$  and entangle it to the dual layer $d_i$; 4) measure the dual layer $d_i$; 5) reinitialise the dual layer $d_{i+1}$ and entangle it to the primal layer $p_{i+1}$; 6) increase $i$ by one; 7) go to step 2 until no primal layer is left ($i = n$).

The structure of the layers is illustrated in Fig.~\ref{fig:slices}. Primal layers have the physical qubits arranged according to Fig.~\ref{fig:sliceprimal}, and dual layers similar to Fig.~\ref{fig:slicedual}. In the process of entangling layers the physical qubits with the same i-, j-coordinates and sequential t-coordinates are entangled (e.g. the layer from Fig.~\ref{fig:sliceprimal} will be entangled to the one from Fig.~\ref{fig:slicedual}). As a consequence, the gradual construction and measurement of layers is respecting the structure imposed by tiling unit cells along three dimensions.

\section*{Glossary}

\textbf{Box}: the bounding box of the \emph{geometric description} of a \emph{distillation circuit}. The type of the box is identical to the state distilled by the distillation circuit. The term is used to also reference distillation sub circuits.

\textbf{Computational universality}: capability of a computing model to capture all of the power of quantum computation.

\textbf{Configurable input/output}: a circuit \emph{input/output} which can be initialised/measured in various bases (e.g. $X$, $Z$).

\textbf{Connection}: \emph{defect} spanned between a distillation box \emph{pin} and a circuit pin.

\textbf{Cube}: a subdivision of a \emph{volume unit} representing all the \emph{unit cells} necessary for constructing \emph{defects} with a diameter enabling \emph{topological error-correction} with a specific \emph{distance}.

\textbf{Defect (geometry)}: the geometric abstraction of a \emph{lattice} defect. It contains at least one \emph{segment}.

\textbf{Defect (lattice)}: a set of physical \emph{face qubits} arranged on neighbouring \emph{unit cells} missing from the \emph{lattice}. The qubits are removed without affecting the connectivity of their neighbourhood by performing $Z$-basis measurements.

\textbf{Distance (code)}: a measure of error-detection and -correction capabilities of a specific code.

\textbf{Dual (face qubit)}: \emph{face qubit} of a dual unit cell.

\textbf{Dual (lattice)}: the self-similar \emph{lattice} after geometrically translating the one constructed by tiling \emph{unit cells}.

\textbf{Dual (logical qubit)}: a \emph{logical qubit} defined in the \emph{dual lattice}.

\textbf{Dual (unit  cell)}: the unit cell resulting after tiling two primal unit cells along each of the three axis (Fig.~\ref{fig:cell}).

\textbf{Face qubit}: any qubit located at the center of one of the faces of the cubic \emph{unit cell}.

\textbf{Failure probability (box)}: the probability that a \emph{distillation circuit} will not output a state with a higher \emph{fidelity}.

\textbf{Failure probability (gate)}: the probability that a quantum gate (physical or logical) will not deliver the expected output.

\textbf{Fault-tolerant threshold}: physical gate failure probability below which it is possible to use error correction to perform an arbitrarily large quantum computation arbitrarily reliably.

\textbf{Fidelity}: for our purposes a distance measure between two states $u$ and $v$ both expressed as complex column vectors. A high fidelity is characteristic of states that are close. The fidelity $F$ of a state $u$ is computed by comparing it with an ideal version $v$ using $F = u^\dagger v$.

\textbf{Geometric description}: captures all the necessary elements for describing a \emph{TQEC circuit} at the \emph{logical layer}. It consists of the three dimensional coordinates of the \emph{defect segments} and \emph{inputs/outputs}.

\textbf{Ghost pin}: a nonexistent pin, but whose existence is simulated just for allowing the construction of a \emph{schedule}. \emph{Spare boxes} are scheduled using ghost pins, but are connected when necessary to \emph{real pins}.

\textbf{Graph state}: a specific quantum state derived from a graph of vertices and edges by associating a qubit prepared in $\ket{+}$ with each vertex and applying a CZ gate to every pair of qubits associated with the end points of every edge.

\textbf{Heterogeneous (schedule)}: a \emph{schedule} computed for \emph{boxes} of different types.

\textbf{Homogeneous (schedule)}: a \emph{schedule} computed for \emph{boxes} of the same type.

\textbf{ICM circuit}: a quantum circuit consisting only of qubit initialisation, CNOT gates and qubit measurements. The circuits are useful for representing fault-tolerant \emph{universal computations}.

\textbf{Input/output (circuit)}: the interface of a circuit that is used to either input a state to be processed or to output a resulting state. Accordingly, each \emph{logical qubit} has an input (used for initialisation) and an output (used for measurement). For \emph{TQEC circuits} the inputs/outputs abstract \emph{physical qubit} coordinates from the \emph{lattice}.

\textbf{Lattice}: a highly regular structure resulting after tiling \emph{unit cells}) along the three axes of a three dimensional space. Its resulting structure is of a \emph{graph state} where \emph{physical qubits} are entangled according to the unit cell pattern. Each qubit can be associated with an integer three dimensional coordinate. The lattice includes two self-similar sub lattices: itself (\emph{primal}) and a version of itself which is geometrically translated by the vector $(1,1,1)$ (\emph{dual}). By convention, the unit cells forming an initial lattice are considered primal.

\textbf{Logical layer}: an abstraction of the \emph{physical layer} where multiple entities from the physical layer are used for defining a single logical entity with better characteristics. For example, \emph{logical qubits} with lower failure probabilities are constructed out of multiple physical qubits each with higher failure probabilities. Another example is the logical CNOT gate which is the result of braiding \emph{defects}, an abstraction involving many physical operations.

\textbf{Logical qubit}: two \emph{defects} each abstracted in the \emph{geometric description} by a sequences of \emph{segment} end points.

\textbf{Matrix-like circuit representation}: the result of mapping the circuit representation to an array of integers. Each integer represents a different element of the circuit (e.g. -100 for inputs, -101 for outputs, 1 for CNOT control and 2 for CNOT target).

\textbf{Optimisation (TQEC circuit)}: the process of reducing the bounding box of a \emph{geometric description} corresponding to a \emph{TQEC circuit}.

\textbf{Physical layer}: one of the layers at which error corrected computational models operate. The physical layer is an abstraction of the hardware, where entities existing at this layer mostly have a direct hardware equivalent. For example, \emph{physical qubits} are direct abstractions of the hardware representing and/or operating a qubit (e.g. photon).

\textbf{Physical qubit}: the quantum state represented and manipulated by a single physical quantum hardware entity.

\textbf{Pin}: the three dimensional coordinate of a \emph{box} or circuit interface. Two pins exist for each box or circuit \emph{input/output} because their coordinates indicate where a connection has to be attached. A \emph{connection} is a geometric abstraction of a \emph{defect}, and because information is manipulated as \emph{logical qubits} there are two \emph{connections} required for each box or circuit input/output.

\textbf{Pin pair}: the two \emph{pins} associated to the same input/output.

\textbf{Primal (face qubit)}: \emph{face qubit} of a \emph{primal unit cell}.

\textbf{Primal (lattice)}: \emph{lattice}.

\textbf{Primal (logical qubit)}: a \emph{logical qubit} defined in the \emph{primal lattice}.

\textbf{Primal (unit cell)}: a \emph{unit cell} used for constructing the lattice necessary for \emph{TQEC circuits}.

\textbf{Real pin}: \emph{pin}.

\textbf{Region}: a rectangular two dimensional space extracted from a three dimensional space by constraining two coordinates to intervals and the third coordinate to a particular value. Regions are used for placing the \emph{geometric description} and the \emph{schedules}.

\textbf{Rotation gate}: arbitrary single qubit gate. The most utilised rotation gates are $T$, $P$ and $V$ because they are sufficient for achieving \emph{computational universality}. In TQEC these are implemented through teleportation circuits which use injected states.

\textbf{Schedule}: list of three dimensional space coordinates where \emph{boxes} are to be placed.

\textbf{Scheduler}: algorithm that places \emph{boxes} in a three dimensional space.

\textbf{Segment (connection)}: part of a \emph{connection}. It is geometrically abstracted by a line segment.

\textbf{Segment (defect)}: sub set of \emph{physical qubits}, from the \emph{defect}, which would exist on co-linear \emph{unit cells} if the \emph{lattice} is drawn in a three dimensional space.

\textbf{Spare box}: bounding box of a potential additionally required \emph{distillation circuit}. Spares are used because distillation circuits have an associated \emph{failure probability} and a failed box cannot be used for the implementation of a TQEC \emph{rotation gate}.

\textbf{State distillation}: circuit that takes multiple low \emph{fidelity} instances of the same state and outputs a single higher fidelity state.

\textbf{State injection}: process of encoding a specific rotated state of a lattice \emph{physical qubit} ($\ket{A}$ or $\ket{Y}$) at the \emph{logical layer}. The resulting \emph{logical qubit} has two \emph{defects} originating from the physical qubit. The latter does not belong to any of the defects. For this reason, both defects have a pyramid structure with its tip at the three dimensional coordinate of the physical qubit.

\textbf{Synthesis (circuit)}: process of transforming an algorithmic description into a circuit description. For the case of TQEC, the \emph{logical layer} circuit description is the \emph{geometric description}.

\textbf{Topological assembly}: the set of elements representing a \emph{TQEC circuit} specified by a \emph{geometric description}.

\textbf{Topological cluster state}: the quantum state represented as \emph{graph state} having the structure of a \emph{lattice}.

\textbf{Topological error-correction}: a class of quantum error-correction procedures based on \emph{topological cluster states}.

\textbf{TQEC circuit}: Topological Quantum Error Corrected circuit. For this work, an \emph{ICM circuit}.

\textbf{Unit cell}: a particular \emph{graph state} shown in \ref{fig:cell}. A single cell contains 18 \emph{physical qubits} such that six are entangled with four neighbours and 12 qubits are entangled with two neighbours. In an infinite three dimensional tiling of unit cells, every qubit is entangled with four neighbours and each unit cell contributes six qubits.

\textbf{Verification (TQEC circuit)}: process of checking if a specified \emph{geometric description} has the structure required for implementing a particular quantum computation.

\textbf{Volume unit}: a cubic structure consisting of 125 \emph{cubes} used for measuring the volume of a \emph{geometric description} in a manner which is error-correction code independent.

\end{document}